\documentclass[12pt, draftclsnofoot, perreview, onecolumn]{IEEEtran}

\usepackage{amsfonts}
\usepackage{amsmath}
\usepackage{amssymb}
\usepackage{color}
\usepackage{epsfig}
\usepackage{float}
\usepackage{graphics}
\usepackage{hhline}
\usepackage{multirow}
\usepackage{rotating}
\usepackage{stfloats}
\usepackage{subfigure}
\usepackage{tabularx}
\usepackage{textcomp}
\usepackage{url}
\usepackage{vector}
\usepackage{wasysym}

\IEEEoverridecommandlockouts

\newcolumntype{L}[1]{>{\raggedright\arraybackslash}p{#1}}
\newcolumntype{C}[1]{>{\centering\arraybackslash}p{#1}}
\newcolumntype{R}[1]{>{\raggedleft\arraybackslash}p{#1}}

\makeatletter

\newcommand{\Rmnum}[1]{\rm\expandafter\@slowromancap\romannumeral #1@}
\makeatother

\def\Rmone{\Rmnum 1}
\def\Rmtwo{\Rmnum 2}

\def\GFq{\text{GF}(q)}

\def\bfB{\mathbf B}
\def\bfBSC{\bfB_{\rm SC}}

\def\B{\mathcal B} 
\def\C{\mathcal C} 

\def\Bjk{\B(J,K,m)}
\def\Cjkmsonetypep{\C_1(J,K,m,p)}
\def\Cjkmsonetypeone{\C_1(J,K,m,1)}
\def\Cjkmsonetypetwo{\C_1(J,K,m,2)}
\def\Cjkmsonetypethree{\C_1(J,K,m,3)}
\def\Cjkmsjminusone{\C_{J-1}(J,K,m)}

\def\Bjj{\B(J,2J,m)}
\def\Cjjmsonetypeone{\C_1(J,2J,m,1)}
\def\Cjjmsonetypetwo{\C_1(J,2J,m,2)}
\def\Cjjmsonetypethree{\C_1(J,2J,m,3)}
\def\Cjjmsjminusone{\C_{J-1}(J,2J,m)}

\def\Cjjjmsjminusone{\C_{J-1}(J,3J,m)}

\def\Cjjjjmsjminusone{\C_{J-1}(J,4J,m)}

\def\Btwofour{\B(2,4,m)}
\def\Ctwofourmsone{\C_1(2,4,m)}

\def\Bthreesix{\B(3,6,m)}
\def\Cthreesixmsonetypeone{\C_1(3,6,m,1)}
\def\Cthreesixmsonetypetwo{\C_1(3,6,m,2)}
\def\Cthreesixmsonetypethree{\C_1(3,6,m,3)}
\def\Cthreesixmstwo{\C_2(3,6,m)}

\def\Cfoureightmsonetypeone{\C_1(4,8,m,1)}
\def\Cfoureightmsonetypetwo{\C_1(4,8,m,2)}
\def\Cfoureightmsonetypethree{\C_1(4,8,m,3)}
\def\Cfoureightmsthree{\C_3(4,8,m)}

\def\Cfivetenmsonetypeone{\C_1(5,10,m,1)}
\def\Cfivetenmsonetypetwo{\C_1(5,10,m,2)}
\def\Cfivetenmsonetypethree{\C_1(5,10,m,3)}
\def\Cfivetenmsfour{\C_4(5,10,m)}

\def\Bthreenine{\B(3,9,m)}
\def\Cthreeninemsonetypeone{\C_1(3,9,m,1)}
\def\Cthreeninemsonetypetwo{\C_1(3,9,m,2)}
\def\Cthreeninemsonetypethree{\C_1(3,9,m,3)}
\def\Cthreeninemsonetypefour{\C_1(3,9,m,4)}
\def\Cthreeninemstwo{\C_2(3,9,m)}

\def\Cthreetwelvemsonetypetwo{\C_1(3,12,m,2)}
\def\Cthreetwelvemsonetypethree{\C_1(3,12,m,3)}
\def\Cthreetwelvemsonetypefour{\C_1(3,12,m,4)}
\def\Cthreetwelvemsonetypefive{\C_1(3,12,m,5)}

\def\Ni{N(i) \backslash j}
\def\Mj{M(j) \backslash i}

\def\bfp{{\boldsymbol p}}
\def\pzero{\bfp^{(0)}}
\def\pcl{\bfp_\text{C}^{(l)}} 
\def\pvl{\bfp_\text{V}^{(l)}}
\def\pvlm{\bfp_\text{V}^{(l-1)}}
\def\pvli{\bfp_\text{V}^{(0)}}
\def\pvlAPP{\bfp_\text{V, APP}^{(l)}}

\def\EJms{E(J,w)}
\def\bfBzeroms{\bfB_{0}^{w}}
\def\bfEA{{\mathbf E}_A}
\def\bfEB{{\mathbf E}_B}

\def\efsd{\epsilon^*}
\def\efsdm{\efsd(m)}

\def\ewd{\epsilon^*_\text{WD}}
\def\ewdmw{\ewd(m,W)}
\def\ewdmax{\hat{\epsilon}_\text{WD}}
\def\ewdmaxm{\hat{\epsilon}_\text{WD}(m)}

\def\Lm{L^{*}(m)}
\def\W{W^{*}}
\def\Wm{W^{*}(m)}

\def\O{\mathcal O}
\def\TSC{T_{\rm SC}}
\def\TBC{T_{\rm BC}}

\newtheorem{theorem}{Theorem}

\begin{document}

\title
{
    Design of Spatially Coupled LDPC Codes\\ over GF$(q)$ for Windowed Decoding
}

\author
{
    \IEEEauthorblockN
    {
        Lai Wei, \emph{Student Member, IEEE,} David G. M. Mitchell, \emph{Member, IEEE,}\\
        Thomas E. Fuja, \emph{Fellow, IEEE,} and Daniel J. Costello, Jr., \emph{Life Fellow, IEEE}
    }
    \thanks
    {
        This work was supported by the U.S. National Science Foundation under grant CCF-1161754. Some of the material in this paper was presented at the Information Theory and Applications Workshop, San Diego, CA, Feb. 2014, and at the IEEE International Symposium on Information Theory, Honolulu, HI, July 2014.

        L.~Wei, D.~G.~M.~Mitchell, T.~E.~Fuja, and D.~J.~Costello, Jr. are with the Department of Electrical Engineering, University of Notre Dame, Notre Dame, IN, 46556, U.S. (e-mail: \{lwei1, david.mitchell, tfuja, dcostel1\}@nd.edu).
    }
}

\markboth{IEEE transactions on information theory (submitted paper)}%
{Shell \MakeLowercase{\textit{et al.}}: Bare Demo of IEEEtran.cls for Journals}

\maketitle

\begin{abstract}
\label{Abstract}
In this paper we consider the generalization of binary spatially coupled low-density parity-check (SC-LDPC) codes to finite fields GF$(q)$, $q\geq 2$, and develop design rules for $q$-ary SC-LDPC code ensembles based on their iterative belief propagation (BP) decoding thresholds, with particular emphasis on low-latency windowed decoding (WD). We consider transmission over both the binary erasure channel (BEC) and the binary-input additive white Gaussian noise channel (BIAWGNC) and present results for a variety of $(J,K)$-regular SC-LDPC code ensembles constructed over GF$(q)$ using protographs. Thresholds are calculated using protograph versions of $q$-ary density evolution (for the BEC) and $q$-ary extrinsic information transfer analysis (for the BIAWGNC). We show that WD of $q$-ary SC-LDPC codes provides significant threshold gains compared to corresponding (uncoupled) $q$-ary LDPC block code (LDPC-BC) ensembles when the window size $W$ is large enough and that these gains increase as the finite field size $q=2^m$ increases. Moreover, we demonstrate that the new design rules provide WD thresholds that are close to capacity, even when both $m$ and $W$ are relatively small (thereby reducing decoding complexity and latency). The analysis further shows that, compared to standard flooding-schedule decoding, WD of $q$-ary SC-LDPC code ensembles results in significant reductions in both decoding complexity and decoding latency, and that these reductions increase as $m$ increases. For applications with a near-threshold performance requirement and a constraint on decoding latency, we show that using $q$-ary SC-LDPC code ensembles, with moderate $q>2$, instead of their binary counterparts results in reduced decoding complexity.
\end{abstract}


\section{Introduction}
\label{Sec: Introduction}
Low-density parity-check block codes (LDPC-BCs) constructed over finite fields GF$(q)$ of size $q>2$ outperform comparable binary LDPC-BCs~\cite{Davey98-LDPC-GFq}, in particular when the block length is short to moderate. However, this performance gain comes at the cost of an increase in decoding complexity. A direct implementation of the $q$-ary belief-propagation (BP) decoder, originally proposed by Davey and MacKay in~\cite{Davey98-LDPC-GFq}, has complexity $\O(q^2)$ per symbol. More recently, an implementation based on the fast Fourier transform~\cite{Barnault03-LDPC-GF2^q-FFT} was shown to reduce the complexity to $\O(q \log q)$. Beyond that, a variety of simple but sub-optimal decoding algorithms have been proposed in the literature, such as the extended min-sum (EMS) algorithm~\cite{Voicila10-EMS} and the trellis-based EMS algorithm~\cite{Li13-T-EMS}. For computing iterative BP decoding thresholds, a $q$-ary extrinsic information transfer (EXIT) analysis was proposed in~\cite{Bennatan06-NB-EXIT} and was later developed into a version suitable for \emph{protograph-based} code ensembles in~\cite{Dolecek14-P-NB-EXIT}.

\begin{figure}[!t]
\begin{center}
    \includegraphics{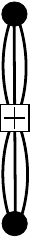}
    \hspace{4mm}\raisebox{11.5mm}
    {
        $\leftrightsquigarrow \hspace{2mm}
        \bfB =
        [
            \begin{array}{c c}
            3 & 3  \\
            \end{array}
        ]$
    }
    \caption{A $(3,6)$-regular protograph and its corresponding base-matrix representation. Black circles correspond to variable nodes and crossed boxes correspond to check nodes.}
    \label{Fig: protograph}
    \end{center}
\end{figure}

A protograph~\cite{Thorpe03-protograph} is a small Tanner graph, which can be used to produce a \emph{structured} LDPC code ensemble by applying a graph lifting procedure~\cite{Pusane11-protograph-lifting} with \emph{lifting factor} $M$, such that every code in the ensemble is $M$ times larger and maintains the structure of the protograph, i.e., it has the same degree distribution and the same type of edge connections. In this way, the computation graph~\cite{Richardson08-modern-coding-theory} is maintained in the lifted graph~\cite{Thorpe03-protograph}, so BP threshold analysis can be performed on the protograph. A protograph consisting of $(c-b)$ check nodes and $c$ variable nodes has \emph{design rate} $R=b/c$ and can be represented equivalently by a $(c-b) \times c$ \emph{base (parity-check) matrix} $\bfB$ consisting of non-negative integers, in which the $(i,j)$-th entry ($1 \leq i \leq c-b$ and $1 \leq j \leq c$) is the number of edges connecting check node $i$ and variable node $j$. Fig.~\ref{Fig: protograph} illustrates a $(3,6)$-regular protograph and its corresponding base matrix, which can be used to represent a $(3,6)$-regular LDPC-BC ensemble. To calculate the BP threshold of a protograph-based code ensemble, conventional tools are adapted to take the edge connections into account~\cite{Thorpe03-protograph, Divsalar09-protograph}. Although some freedom is lost in the code design when the protograph structure is adopted, one can use these modified protograph-based analysis tools to find ``good'' protograph-based ensembles with better BP thresholds than corresponding unstructured ensembles with the same degree distribution~\cite{Divsalar09-protograph, Liva07-P-EXIT}.

Spatially coupled LDPC (SC-LDPC) codes, also known as terminated LDPC convolutional codes~\cite{Felstrom99-LDPCC}, are constructed by coupling together a series of $L$ disjoint, or uncoupled, LDPC-BC Tanner graphs. Binary SC-LDPC code ensembles have been shown to exhibit a phenomenon called ``threshold saturation''~\cite{Lentmaier10-terminated-LDPCC, Kudekar11-th-sat, Kudekar13-th-sat}, in which, as the coupling length $L$ grows, the BP decoding threshold saturates to the maximum \emph{a-posteriori} (MAP) threshold of the corresponding uncoupled LDPC-BC ensemble, which, for the $(J,K)$-regular code ensembles considered in this paper, approaches channel capacity as the density of the parity-check matrix increases~\cite{Miller01-MAP}. This threshold saturation phenomenon has been reported for a variety of code ensembles (e.g., $(J,K)$-regular SC-LDPC code ensembles~\cite{Lentmaier09-edge-spreading}, accumulate-repeat-by-4-jagged-accumulate (AR4JA) irregular SC-LDPC code ensembles~\cite{Mitchell10-irregular-LDPCC-AR4JA}, bilayer SC-LDPC code ensembles~\cite{Si13-bilayer-LDPCC-relay}, and MacKay-Neal and Hsu-Anastasopoulos spatially-coupled code ensembles~\cite{Kasai2013-SC-MN-HA}) and channel models (e.g., channels with memory~\cite{Kudekar11-th-sat-memory}, multiple access channels~\cite{Kudekar11-th-sat-MAC}, intersymbol-interference channels~\cite{Nguyen12-th-sat-interfer}, and erasure relay channels~\cite{Uchikawa11-three-nodes-relay}), thus making SC-LDPC codes attractive candidates for practical applications requiring near-capacity performance. For a more comprehensive survey of the literature on SC-LDPC codes, refer to the introduction of~\cite{Mitchell14-SC-LDPC-protograph}.

BP decoding threshold results on the BEC for $q$-ary SC-LDPC code ensembles have been reported by Uchikawa~\emph{et al.}~\cite{Uchikawa11-RC-NB-LDPCC} and Piemontese~\emph{et al.}~\cite{Piemontese13-NB-SC-LDPC-BEC}, and the corresponding threshold saturation was proved by Andriyanova~\emph{et al.}~\cite{Andriyanova13-NB-SC-LDPC-th-sat-BEC}. In each of these papers, the authors assumed that decoding was simultaneously carried out across the entire parity-check matrix of the code; for simplicity, this will be referred to as flooding schedule decoding (FSD) in this paper. Employing FSD for SC-LDPC codes can result in large latency, since a large coupling length $L$ is needed to achieve near-capacity thresholds~\cite{Mitchell14-SC-LDPC-protograph}. To resolve this issue, a more efficient technique, called windowed decoding (WD), was proposed in~\cite{Iyengar12-WD, Lentmaier11-WD} for binary SC-LDPC codes. Compared to FSD, WD exploits the convolutional nature of the SC parity-check matrix to localize decoding and thereby reduce latency. Under WD, the decoding window contains only a small portion of the parity-check matrix, and within that window, BP decoding is performed.

In this paper, assuming that the binary image of a codeword is transmitted, we analyze the WD thresholds of a variety of $(J,K)$-regular protograph-based $q$-ary SC-LDPC code ensembles constructed from the corresponding uncoupled $q$-ary $(J,K)$-regular LDPC-BC ensembles via the \emph{edge-spreading} procedure~\cite{Lentmaier09-edge-spreading, Mitchell14-SC-LDPC-protograph}, where the finite field size is $q=2^m$ and $m$ is a positive integer. In particular,
\begin{enumerate}
    \item For the BEC, we extend the $q$-ary density evolution (DE) analysis proposed in~\cite{Rathi05-NB-DE-BEC} to a protograph version and apply this analysis in conjunction with WD to obtain \emph{windowed decoding thresholds} for $q$-ary SC-LDPC code ensembles;
    \item For the binary-input additive white Gaussian noise channel (BIAWGNC) with binary phase-shift keying (BPSK) modulation, we obtain windowed decoding thresholds for $q$-ary SC-LDPC code ensembles by applying a protograph-based EXIT analysis (originally proposed for $q$-ary LDPC-BC ensembles~\cite{Dolecek14-P-NB-EXIT}) in conjunction with WD.
\end{enumerate}

In both cases, our primary contribution is to determine how much the decoding latency of WD can be reduced without suffering a loss in threshold. We observe that
\begin{enumerate}
    \item Compared to FSD of the corresponding uncoupled $q$-ary LDPC-BC ensembles, WD of $q$-ary SC-LDPC code ensembles provides a threshold gain. This gain increases as the finite field size increases.
    \item Compared to FSD of a given $q$-ary SC-LDPC code ensemble, WD provides significant reductions in both decoding latency and decoding complexity, and these reductions increase as the finite field size increases.
    \item By carefully designing the protograph structure, using what we call a ``type $2$'' edge-spreading format, WD provides near-capacity thresholds for $q$-ary SC-LDPC code ensembles, even when both the finite field size and the window size are relatively small.
    \item When there is a constraint on decoding latency and operation close to the threshold of a binary SC-LDPC code ensemble is required, using the non-binary counterpart can provide a significant reduction in decoding complexity.
\end{enumerate}

The rest of the paper is organized as follows. Section~\ref{Sec: Ensembles} describes the construction of protograph-based $q$-ary SC-LDPC code ensembles and reviews the structure of WD. Then Sections~\ref{Sec: TH BEC} and~\ref{Sec: TH BIAWGNC} present the WD thresholds of various $q$-ary SC-LDPC code ensembles for the BEC and the BIAWGNC, respectively, as the finite field size and/or the window size vary. The WD threshold is evaluated from two perspectives: first, as the window size increases, whether it achieves its best numerical value when the window size is small to moderate; second, as the finite field size increases, whether this achievable value approaches capacity. Also, the effects of different protograph constructions on the WD threshold are evaluated and discussed. Finally, Section~\ref{Sec: Dec cmplx} studies the decoding latency and complexity of $q$-ary SC-LDPC code ensembles and examines the latency, complexity, and performance tradeoffs of WD.

In summary, by examining various $q$-ary SC-LDPC code ensembles, we bring additional insight to three questions:
\begin{enumerate}
    \item Why spatially coupled codes perform better than the corresponding uncoupled block codes,
    \item Why windowed decoding is preferred to flooding schedule decoding, and
    \item When non-binary codes should be used instead of binary codes.
\end{enumerate}
The results of this paper provide theoretical guidance for designing and implementing practical $q$-ary spatially coupled LDPC codes suitable for windowed decoding~\cite{Huang14-NB-SC-LDPC}.

\section{Windowed Decoding of Protograph-Based $q$-ary SC-LDPC Code Ensembles}
\label{Sec: Ensembles}

\subsection{Protograph-based $q$-ary SC-LDPC Code Ensembles}
\label{Sec: Ensembles; Subsec: Protograph}
\begin{figure}[!t]
    \centerline{\includegraphics[width=\columnwidth]{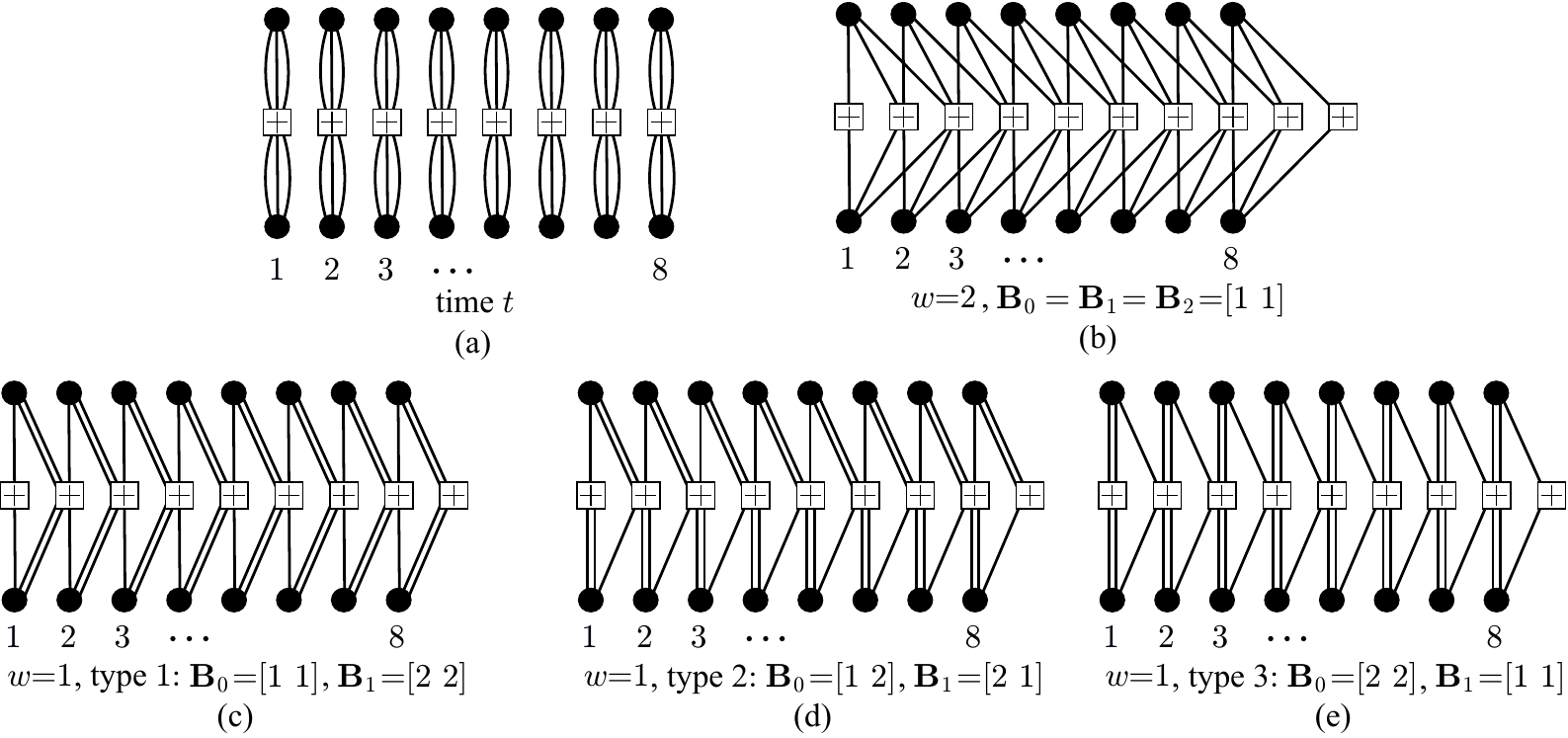}}
    \caption{(a) A sequence of $L=8$ uncoupled $(3,6)$-regular LDPC-BC protographs, and (b)-(e) various $(3,6)$-regular SC-LDPC protographs constructed following the edge-spreading procedure with coupling length $L=8$.}
    \label{Fig: protograph edge spreading}
\end{figure}

A $(J,K)$-regular SC-LDPC code ensemble can be constructed from a $(J,K)$-regular LDPC-BC ensemble using the edge-spreading procedure~\cite{Lentmaier09-edge-spreading, Mitchell14-SC-LDPC-protograph}, described here in terms of protograph representations of the code ensembles. Take $J=3$, $K=6$ as an example. As shown in Fig.~\ref{Fig: protograph edge spreading}, instead of transmitting a sequence of codewords from the $(3,6)$-regular LDPC-BC ensemble independently at time instants $t=1,2, \ldots, L$, edges from the variable nodes at time instant $t$, originally connected only to the check node at time instant $t$, are now ``spread'' to also connect to check nodes at time instants $t, t+1, \ldots, t+w$; in this way, memory is introduced and the different time instants are ``coupled'' together, i.e., a terminated convolutional, or spatially coupled, coding structure is introduced. The parameter $w$ is referred to as the \emph{coupling width}, and $L$ is called the \emph{coupling length}. Fig.~\ref{Fig: protograph edge spreading} shows three different types of edge-spreading formats for $w=1$ and one type for $w=2$, all for the case $J=3$, $K=6$, and $L=8$.

The above edge-spreading procedure can be described in terms of the base (parity-check) matrix representation of protographs as well. Let $\bfB$ be a $(c-b)\times c$ block base matrix representing an LDPC-BC ensemble with design rate $R=b/c$. Then the base matrix of an SC-LDPC code ensemble can be constructed from $\bfB$ as follows. First, $\bfB$ is ``spread'' into a set of $(w+1)$ \emph{component base matrices} following the rule
\begin{equation}
    \sum_{i=0}^{w} \bfB_i = \bfB,
\end{equation}
so that each $\bfB_i$ has the same size as $\bfB$. Next, an SC base matrix $\bfBSC$ is generated by ``stacking and shifting'' the base component matrices $\left\{ \bfB_i \right\}_{i=0}^{w}$ at each time instant $t=1,2,\ldots,L$, thereby forming a convolutional structure:
\begin{equation}
\label{Eq: SC base mx}
    \bfBSC =
    \begin{bmatrix}
        \bfB_0 &        &        &        &        &        \\
        \bfB_1 & \bfB_0 &        &        &        &        \\
        \vdots & \bfB_1 &        & \ddots &        &        \\
        \bfB_w & \vdots &        & \ddots &        & \bfB_0 \\
               & \bfB_w &        &        &        & \bfB_1 \\
               &        &        & \ddots &        & \vdots \\
               &        &        &        &        & \bfB_w \\
    \end{bmatrix}_{(L+w)(c-b) \times L c},
\end{equation}
where the design rate of $\bfBSC$ is
\begin{eqnarray}
\label{Eq: SC rate}
    R_L = 1 - \dfrac{(L+w)(c-b)}{Lc}=\dfrac{Lb-w(c-b)}{Lc}.
\end{eqnarray}
Due to the termination of $\bfBSC$ after $Lc$ columns, there is a loss in the SC-LDPC code ensemble design rate $R_L$ compared to the rate $R=b/c$ of $\bfB$. However, this rate loss diminishes as $L$ increases and vanishes as $L \rightarrow \infty$, i.e., $\lim_{L \rightarrow \infty}R_L = R = b/c$.

Next, a finite-length $q$-ary SC-LDPC code is constructed from $\bfBSC = \left[ b_{i,j} \right]$ by following the procedure for constructing a finite-length $q$-ary LDPC-BC from $\bfB$:
\begin{enumerate}
    \item ``Lifting''~\cite{Thorpe03-protograph}: Replace the nonzero entries $b_{i,j}$ in $\bfBSC$ with an $M \times M$ permutation matrix (or a sum of $b_{i,j}$ non-overlapping $M \times M$ permutation matrices if $b_{i,j}>1$), and replace the zero entries with the $M \times M$ all-zero matrix, where $M$ is called the lifting factor.
    \item ``Labeling'': Randomly assign to each non-zero entry in the lifted parity-check matrix a non-zero element uniformly selected from $\GFq$, where $q = 2^{m}$ is the finite field size.
\end{enumerate}
After the lifting step, the parity-check matrix is still binary, i.e., the non-binary feature does not arise until the labeling step.\footnote{Note that ``labeling'' can come before ``lifting'', resulting in a ``constrained'' protograph-based $q$-ary code as defined in~\cite{Dolecek14-P-NB-EXIT}.} The total code length is $n=LcM$, and we define the \emph{constraint length} as the maximum width of the non-zero portion of the parity-check matrix $\nu = (w+1)cM$. Both the permutation matrices and the $q$-ary labels can be carefully chosen to obtain good codes with desirable properties. But constructing specific codes is not the emphasis of this paper; rather, we are interested in a threshold analysis of general $q$-ary ensembles consisting of all possible combinations of liftings and labelings of a given protograph, where the dimension of the message model used in the analysis depends on the size of the finite field~\cite{Bennatan06-NB-EXIT, Rathi05-NB-DE-BEC}.

\subsection{Windowed Decoding (WD)}
\label{Sec: Ensembles; Subsec: WD}
\begin{figure}[!t]
    \centering
    \includegraphics[width=3.5in]{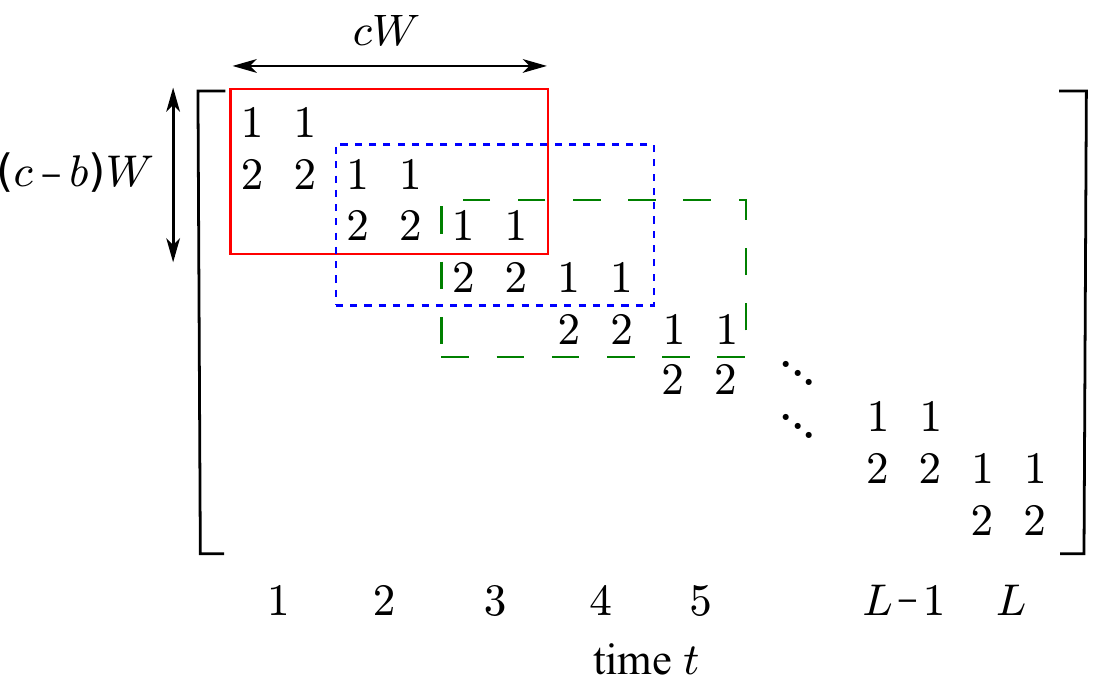}
    \caption{WD example with window size $W=3$: at $t=1$ (solid red), $t=2$ (dotted blue), and $t=3$ (dashed green). $J=3$, $K=6$, $w=1$; $\bfB_0 = [1,\,1]$ and $\bfB_1 = [2,\,2]$, both of size $(c-b) \times c = 1 \times 2$, for the $\bfBSC$ given by the protograph construction of Fig.~\ref{Fig: protograph edge spreading}(c). For each window position/time instant, the first $c=2$ column blocks are target symbols.}
    \label{Fig: WD example}
\end{figure}

In this subsection, we briefly review the structure of WD. By construction, any two variable nodes (columns of the parity-check matrix) in the graph of an SC-LDPC code cannot be connected to the same check node if they are more than a constraint length $\nu = (w+1)cM$ (of columns) apart. As previously mentioned, compared to FSD, where iterative decoding is carried out on the \emph{entire} parity-check matrix, WD of SC-LDPC code ensembles takes advantage of the convolutional structure of the parity-check matrix and localizes the decoding process to a small portion of the matrix, i.e., the BP algorithm is carried out only for those checks and variables covered by a ``window''. Consequently, WD is an efficient way to reduce the memory and latency requirements of SC-LDPC codes~\cite{Iyengar12-WD, Lentmaier11-WD}. The WD algorithm can be described as follows (see \cite{Iyengar12-WD} for further details):
\begin{itemize}
    \item In terms of the SC base matrix $\bfBSC$, the window is of fixed size $(c-b)W \times cW$ (recall that the size of the component base matrices $\bfB_i$'s in $\bfBSC$ is $(c-b) \times c$) measured in symbols, and slides from time instant $t=1$ to time instant $t=L$, where $W$, called the \emph{window size}, is defined as the number of column blocks of size $c$ in the window. An example of WD with $W=3$ is illustrated in Fig.~\ref{Fig: WD example} for the SC-LDPC code ensemble whose protograph is shown in Figure~\ref{Fig: protograph edge spreading}(c).

    \item At each time instant/window position, the BP algorithm runs until a fixed number of iterations has been performed or some stopping rule~\cite{Iyengar12-WD, Lentmaier11-WD, Huang14-NB-SC-LDPC} is satisfied, after which the window shifts $c$ column blocks and those $c$ column block symbols shifted out of the window are decoded. The first $c$ column blocks in a window are called the \emph{target symbols}. We assume that all the variables and checks in a window are updated during each iteration and that, after the window shifts, the final messages from the previously decoded target symbols are passed to the new window.

    \item Clearly, the largest possible $W$ is equal to $(L+w)$, in which case the whole parity check matrix is covered and makes WD equivalent to FSD, and the smallest possible $W$ is $(w+1)$, i.e., the window length (measured in variables) when decoding an SC-LDPC code must be at least one constraint length. We are interested in searching for $q$-ary SC-LDPC code ensembles for which a small window size $W$ can provide WD with a good threshold, which implies that the coupling width $w$ should be kept small. Indeed, our results for $q$-ary SC-LDPC codes together with those in the literature for binary SC-LDPC codes \cite{Iyengar12-WD, Lentmaier11-WD} show that ensembles with $w=1$ provide the best latency-constrained performance with WD.

\end{itemize}

\subsection{Code Ensemble Construction}
\label{Sec: Ensembles; Subsec: Construction}
In this paper, we restrict our attention to $(J,K)$-regular LDPC code ensembles.

\subsubsection{$(J,K)$-regular LDPC-BC ensembles}
Let
\begin{eqnarray}
\label{Eq: block base mx J K}
    \bfB =
    \begin{bmatrix}
        J & J & \cdots & J \\
    \end{bmatrix}_{1 \times k}
\end{eqnarray}
denote the block base matrix corresponding to the protograph representation of a $(J,K)$-regular LDPC-BC ensemble, where $K=kJ$, $k=1,2,\ldots$, and the design rate of the code ensemble is $R = \left( k-1 \right) / k$. That is, in the remainder of the paper, we let $c - b = 1$ and $c = k$. We denote the $(J,K)$-regular LDPC-BC ensemble constructed over GF$(2^m)$ as $\Bjk$.

\subsubsection{Edge spreadings of $\bfB$}
Given a variable node degree $J$, for a particular coupling width $w$, define
\begin{eqnarray}
\label{Eq: partition J w}
    \EJms =
    \left\{
        \begin{bmatrix}
            J_0 & J_1 & \cdots & J_w
        \end{bmatrix}^\intercal
        \left\rvert
            \sum\limits_{i=0}^{w} J_i = J, \,J_i \in \left\{1, 2, \ldots, J-w\right\}
        \right.
    \right\},
\end{eqnarray}
i.e., $\EJms$ is the set of all possible column vectors of length $(w+1)$ satisfying the constraint $\sum_{i=0}^{w} J_i = J$, where $J_i \in \{1, 2, \ldots, J-w\}$. Moreover, define $\bfBzeroms$ as
\begin{eqnarray}
\label{Eq: comp base mx stack}
    \bfBzeroms =
    \begin{bmatrix}
    \bfB_0 \\
    \bfB_1 \\
    \vdots \\
    \bfB_w \\
    \end{bmatrix}_{(w+1)\times k},
\end{eqnarray}
i.e., $\bfBzeroms$ is the ``stack'' of all the component base matrices $\left\{ \bfB_i \right\}_{i=0}^w$. Then an edge-spreading format can be generated by selecting $k$ elements (with replacement) from $\EJms$ as the $k$ columns of $\bfBzeroms$. Recall from Section~\ref{Sec: Ensembles; Subsec: WD} that our major interest lies in $q$-ary SC-LDPC code ensembles for which windowed decoding (WD) achieves good thresholds under tight latency constraints, i.e., for a small window size $W$, which implies that the coupling width $w$ should be small. Therefore, we do not allow $w$ to exceed $(J-1)$, i.e., the block base matrix $\bfB$ should be spread into at most $J$ component base matrices $\bfB_i$. In other words, for $\EJms$ in~\eqref{Eq: partition J w}, we consider only values of $w$ in the range $1 \leq w \leq J-1$.

The edge-spreading format $\bfBzeroms$ determines the SC base matrix $\bfBSC$, and the $q$-ary WD thresholds depend on $\bfBSC$. For a given $\bfBzeroms$, column permutations do not affect the WD threshold, but row permutations do. Consequently, for each combination of $J$ and $w$, there will be $\left|\EJms \right| \cdot \left( 1 + \left| \EJms \right| \right) / 2$ possible edge-spreading formats that can result in diffferent WD thresholds. For example, consider the $(4,8)$-regular degree distribution with $J=4$ and $w=2$. Then
\begin{eqnarray}
\label{Eq: partition J4 w2}
    E(4,2) =
    \left\{
    \begin{bmatrix}
    1 & 1 & 2\\
    \end{bmatrix}^\intercal,\,
    \begin{bmatrix}
    1 & 2 & 1\\
    \end{bmatrix}^\intercal,\,
    \begin{bmatrix}
    2 & 1 & 1\\
    \end{bmatrix}^\intercal
    \right\},
\end{eqnarray}
and the $\left|E(4, 2)\right|\cdot \left( 1 + \left|E(4, 2 )\right| \right) / 2 = 6$ possible edge-spreading formats that can give different WD thresholds are given by
\begin{eqnarray}
\label{Eq: spreading J4 w2}
    \bfBzeroms \in \left\{
        \begin{bmatrix}
            1 & 1 \\
            1 & 1 \\
            2 & 2 \\
        \end{bmatrix},\,
        \begin{bmatrix}
            1 & 1 \\
            1 & 2 \\
            2 & 1 \\
        \end{bmatrix},\,
        \begin{bmatrix}
            1 & 2 \\
            1 & 1 \\
            2 & 1 \\
        \end{bmatrix},\,
        \begin{bmatrix}
            1 & 1 \\
            2 & 2 \\
            1 & 1 \\
        \end{bmatrix},\,
        \begin{bmatrix}
            1 & 2 \\
            2 & 1 \\
            1 & 1 \\
        \end{bmatrix},\,
        \begin{bmatrix}
            2 & 2 \\
            1 & 1 \\
            1 & 1 \\
        \end{bmatrix}
    \right\}.
\end{eqnarray}

\subsubsection{$(J,K)$-regular SC-LDPC code ensembles}
We now detail the particular constructions of SC-LDPC code ensembles considered in the remainder of the paper. The first construction we consider is the ``classical'' edge spreading~\cite{Lentmaier10-terminated-LDPCC} of the $(J,K)$-regular LDPC-BC base matrix $\bfB$ given by~\eqref{Eq: block base mx J K}, where $K=kJ$ and $w=J-1$:
\begin{equation}
\label{Eq: full spreading}
\bfB_0 = \bfB_1 = \cdots = \bfB_{w} =
        \begin{bmatrix}
            1 & 1 & \cdots & 1\\
        \end{bmatrix}_{1 \times k}.
\end{equation}
Unless noted otherwise, the coupling length for all the $q$-ary SC-LDPC code ensembles in this paper is taken to be $L=100$, in order to keep the rate loss small. Consequently, we do not include $L$ in the ensemble notation, and we denote as $\Cjkmsjminusone$ the SC-LDPC code ensemble constructed over GF$(2^m)$ using the component matrices $\bfB_0$, $\bfB_1$, $\dots$, $\bfB_w$ given by~\eqref{Eq: full spreading} in the base matrix $\bfBSC$ given by~\eqref{Eq: SC base mx}, with coupling width $w=J-1$.

As noted previously, under tight latency constraints, the WD threshold can be improved by using small $w$; in fact, excellent WD performance has been shown for binary SC-LDPC code ensembles using repeated edges in the protograph and $w=1$~\cite{Iyengar12-WD, Lentmaier11-WD}. In the case of $q$-ary SC-LDPC code ensembles, we have also found that the case $w=1$, i.e., the set of edge spreadings
\begin{eqnarray}
\label{Eq: partition J w(J-1)}
    E(J,w=1) =
    \left\{
    \begin{bmatrix}
        1 \\
        J-1
    \end{bmatrix},\,
    \begin{bmatrix}
        2 \\
        J-2
    \end{bmatrix},\,
    \ldots,
    \begin{bmatrix}
        J-1 \\
        1
    \end{bmatrix}
    \right\},
\end{eqnarray}
results in the best thresholds for low latency WD. Moreover, if we further restrict our attention to the edge-spreading pair
\begin{equation}
\label{Eq: EA EB}
    \bfEA =
    \begin{bmatrix}
        1 \\
        J-1 \\
    \end{bmatrix},\,
    \bfEB
    \begin{bmatrix}
        J-1 \\
        1 \\
    \end{bmatrix}
    \in E(J,1),
\end{equation}
we obtain the most interesting and representative constructions compared to the other possible selections of column vectors from $E(J,1)$.

Combining $\bfEA$ and $\bfEB$, there are $(k+1)$ possible choices for $\bfB_{0}^{w=1}$. An edge-spreading format is called ``type-$p$'' if there are $(k-p+1)$ columns of $\bfEA$ in $\bfB_{0}^{1}$ followed by $(p-1)$ columns of $\bfEB$, i.e.,
\begin{align}
\label{Eq: spreading J w1}
    \bfB_{0}^{1} &=
    \underbrace
    {
    \left[
        \begin{matrix}
            \bfEA & \cdots & \bfEA \\
        \end{matrix}
    \right.
    }_{k-p+1}\,
    \underbrace
    {
    \left.
        \begin{matrix}
            \bfEB & \cdots & \bfEB \\
        \end{matrix}
    \right]
    }_{p-1}\nonumber\\
    &=
    \underbrace
    {
    \left[
        \begin{matrix}
            1 & \cdots & 1 \\
            J-1 & \cdots & J-1 \\
        \end{matrix}
    \right.
    }_{k-p+1}\,
    \underbrace
    {
    \left.
        \begin{matrix}
            J-1 & \cdots & J-1 \\
            1 & \cdots & 1 \\
        \end{matrix}
    \right]
    }_{p-1} =
    \left[
        \begin{matrix}
            \bfB_0 \\
            \bfB_1 \\
        \end{matrix}
    \right],
\end{align}
where $1 \leq p \leq k+1$. Again, note that the ordering of columns is not important, because this simply results in column permutations of the resulting base matrix $\bfBSC$ and does not change the code or graph properties. We again omit $L$ from the ensemble notation and denote as $\Cjkmsonetypep$ the type-$p$ SC-LDPC code ensemble constructed over GF$(2^m)$ using component matrices $\bfB_0$ and $\bfB_1$ to form $\bfBSC$, with coupling width $w=1$, where $1 \leq p \leq k+1$.

For a particular $(J,K)$ pair and Galois field GF$(2^m)$, we refer informally to the collection of ensembles
\begin{equation}
\label{Eq: ensembles}
    \left\{
        \Bjk,
        \>\Cjkmsjminusone,
        \>\Cjkmsonetypep
        \>|\>p=1,2,...,k+1
    \right\}
\end{equation}
as ``the $(J,K,m)$ ensembles'', and we further refer to the collection of ensembles
\begin{equation}
\label{Eq: SC ensembles}
    \left\{
        \>\Cjkmsjminusone,
        \>\Cjkmsonetypep
        \>|\>p=1,2,...,k+1
    \right\}
\end{equation}
as ``the $(J,K,m)$ SC ensembles''. For example, for an arbitrary $m$, let $(J,K)=(3,6)$. In this case $k=2$, and we consider the ``classical'' edge spreading with $w=J-1=2$ along with $k+1=3$ types of edge spreading with $w=1$, viz.:
\begin{itemize}
    \item $\Cthreesixmstwo$:
        $\bfB_0 = \bfB_1 = \bfB_2 =
        \begin{bmatrix}
            1 & 1 \\
        \end{bmatrix}$;
    \item $\Cthreesixmsonetypeone$:
        $\bfB_0 =
        \begin{bmatrix}
            1 & 1 \\
        \end{bmatrix}$,
        $\bfB_1 =
        \begin{bmatrix}
            2 & 2 \\
        \end{bmatrix}$;
    \item $\Cthreesixmsonetypetwo$:
        $\bfB_0 =
        \begin{bmatrix}
            1 & 2 \\
        \end{bmatrix}$,
        $\bfB_1 =
        \begin{bmatrix}
            2 & 1 \\
        \end{bmatrix}$;
    \item $\Cthreesixmsonetypethree$:
        $\bfB_0 =
        \begin{bmatrix}
            2 & 2 \\
        \end{bmatrix}$,
        $\bfB_1 =
        \begin{bmatrix}
            1 & 1 \\
        \end{bmatrix}$.
\end{itemize}
These four ensembles form the $(3,6,m)$ SC ensembles, and together with $\Bthreesix$ they form the $(3,6,m)$ ensembles. Fig.~\ref{Fig: protograph edge spreading} shows each of the $(3,6,m)$ ensembles with coupling length $L=8$ and arbitrary $m$.

\section{Threshold Analysis of $q$-ary SC-LDPC Code Ensembles on the BEC}
\label{Sec: TH BEC}
\subsection{Protograph Density Evolution (DE) for $q$-ary LDPC Code Ensembles on the BEC}
\label{Sec: TH BEC; Subsec: q-ary PDE}
The $q$-ary DE algorithm presented in~\cite{Rathi05-NB-DE-BEC} was originally derived for randomized uncoupled $q$-ary LDPC-BC ensembles where 1) the symbol set is the vector space GF$_2^m$ of dimension $m$ over the binary field, and 2) the edge labeling set is the general linear group GL$_2^m$ over the binary field, which is the set of all $m \times m$ invertible matrices whose entries are in $\{0,1\}$. The thresholds of these code ensembles, as pointed out by the authors of~\cite{Rathi05-NB-DE-BEC}, are very good approximations to those of $q$-ary LDPC-BC ensembles defined over GF$(2^m)$, since the numerical difference is on the order of $10^{-4}$.

Consider an ordered list of the elements of GF$_2^m$, and assume that the zero element is in the $0$th position of the list. For a specific code, a probability domain message vector in $q$-ary BP decoding is of length $2^m$, where the entry at position $i$ corresponds to the \emph{a posteriori} probability that the symbol is the $i$-th element from GF$_2^m$. Since transmission is on the BEC and it can be assumed that the all-zero codeword is transmitted without affecting decoding performance~\cite{Rathi05-NB-DE-BEC}, all the non-zero elements in the message vector must be equal; in fact, the set of symbols (elements from GF$_2^m$) whose \emph{a posteriori} probabilities are non-zero forms a subspace of GF$_2^m$, and the message vector is said to have dimension $n$ if it contains $2^n$ non-zero elements, $n=0,1,\ldots,m$. Consequently, for the purpose of $q$-ary DE, which is concerned only with asymptotic ensemble-average properties rather than decoding a specific finite-length code, only the dimension of the BP decoding message vector needs to be tracked by the algorithm. As a result, a $q$-ary DE message vector for the BEC can be represented by a vector of length $(m+1)$, whose $n$-th entry, $n=0,1,\ldots,m$, indicates the \emph{a posteriori} probability that the BP decoding message vector has dimension $n$.

Similar to the procedure used to extend $q$-ary EXIT analysis to a protograph version in~\cite{Dolecek14-P-NB-EXIT}, we now extend the $q$-ary DE algorithm to a protograph version, which we refer to as $q$-ary protograph DE (PDE). Since the edge connections are taken into account and the computation graph is equal for all members of the ensemble, PDE reduces to the BP algorithm performed on the protograph. We use notation similar to that in~\cite{Dolecek14-P-NB-EXIT} and~\cite{Andriyanova13-NB-SC-LDPC-th-sat-BEC}. Let $b_{i,j}$ denote a non-zero entry in the base matrix and recall that, from the perspective of the protograph, the value of $b_{i,j}$ is the number of edges connecting check node $i$ (the row index in the matrix) to variable node $j$ (the column index), rather than an edge label. Let $N(i)$ (resp. $M(j)$) denote the neighboring variables (resp. checks) of check $i$ (resp. variable $j$). Let $\pcl(i,j)$ (resp. $\pvl(i,j)$) denote the check-$i$-to-variable-$j$ (resp. variable-$j$-to-check-$i$) $q$-ary DE message vector during iteration $l$. Finally, let the erasure probability of the BEC be $\epsilon$. Then the $q$-ary PDE algorithm consists of four steps as follows:
\begin{itemize}
    \item Initialization: for each $b_{i,j}>0$, let
        \begin{eqnarray}
            \pvli(i,j) = \pvli(j) = \pzero(\epsilon),
        \end{eqnarray}
        where $\pzero(x)$ is a vector of length $(m+1)$ in the probability domain, whose $n$-th entry is defined as
        \begin{eqnarray}
            \binom{m}{n}x^{n}(1-x)^{m-n}.
        \end{eqnarray}
    \item Check-to-variable update: the message vector from check $i$ to variable $j$ is
        \begin{eqnarray}
            \pcl(i,j) =
            \left[
                \boxtimes_{s \in \Ni}
                \left(
                    \boxtimes^{b_{i,s}}\pvlm(i,s)
                \right)
            \right]
            \boxtimes
            \left(
                \boxtimes^{b_{i,j}-1}\pvlm(i,s)
            \right),
        \end{eqnarray}
        where the ``$\boxtimes$'' notation (see Appendix A of~\cite{Andriyanova13-NB-SC-LDPC-th-sat-BEC} for details) is described as follows. For two $q$-ary DE message vectors $\bfp_1$ and $\bfp_2$, $\bfp_1 \boxtimes \bfp_2$ has $n$-th element
        \begin{eqnarray}
            \sum\limits_{i=0}^{n} \sum\limits_{j=n-i}^{n} C_{i,j,n}^m p_{1,i} p_{2,j},
        \end{eqnarray}
        where $p_{1,i}$ is the $i$-th element of $\bfp_1$, $p_{2,j}$ is the $j$-th element of $\bfp_2$,
        \begin{eqnarray}
            C_{i,j,n}^m = \dfrac{ G_{m-i,m-n} G_{i,n-j} 2^{(n-i)(n-j)} }{ G_{m,m-j} }
        \end{eqnarray}
        is the probability of choosing a subspace (of GF$_2^m$) of dimension $j$ whose sum with a subspace of dimension $i$ has dimension $n$, and
        \begin{eqnarray}
            G_{m,k} =
            \begin{cases}
            1 &\text{if $k=m$ or $k=0$,}\\
            \prod\limits_{l=0}^{k-1}\dfrac{2^m - 2^l}{2^k - 2^l} &\text{if $0 < k < m$,}\\
            0 &\text{otherwise,}\\
            \end{cases}
        \end{eqnarray}
        is the Gaussian binomial coefficient, the number of different subspaces of dimension $k$ of GF$_2^m$. Finally, $\boxtimes^{b_{i,j}-1}\bfp = \bfp \boxtimes \bfp \boxtimes \ldots \boxtimes \bfp$, with $(b_{i,j}-1)$ occurrences of $\bfp$.
    \item Variable-to-check update: the message vector from variable $j$ to check $i$ is
        \begin{eqnarray}
            \pvl(i,j) = \pvli(j) \boxdot
            \left[
                \boxdot_{s \in \Mj}
                \left(
                    \boxdot^{b_{s,j}}\pcl(s,j)
                \right)
            \right]
            \boxdot
            \left(
                \boxdot^{b_{s,j}-1}\pcl(s,j)
            \right),
        \end{eqnarray}
        where $\bfp_1 \boxdot \bfp_2$ has $n$-th element
        \begin{eqnarray}
            \sum\limits_{i=n}^{m} \sum\limits_{j=n}^{m-i+n} V_{i,j,n}^m p_{1,i} p_{2,j},
        \end{eqnarray}
        and
        \begin{eqnarray}
            V_{i,j,n}^m = \dfrac{ G_{i,n} G_{m-i,j-n} 2^{(i-n)(j-n)} }{ G_{m,j} }
        \end{eqnarray}
        is the probability of choosing a subspace of dimension $j$ whose intersection with a subspace of dimension $i$ has dimension $n$ (again, see Appendix A of~\cite{Andriyanova13-NB-SC-LDPC-th-sat-BEC} for details).
    \item Convergence check: the \emph{a-posteriori} message vector for variable $j$ is
        \begin{eqnarray}
            \pvlAPP(j) = \pvli(j) \boxdot
            \left[
                \boxdot_{i \in M(j)}
                \left(
                    \boxdot^{b_{i,j}}\pcl(i,j)
                \right)
            \right].
        \end{eqnarray}
        The $q$-ary PDE algorithm ends when
        \begin{itemize}
            \item Either a decoding success is declared: for all the variables to be decoded, the $0$th entry of each $\pvlAPP(j)$ (denoted as $\pvlAPP(j)[0]$) is at least $(1-\delta)$, i.e., $\pvlAPP(j)[0] \geq 1 - \delta$, where $\delta \in [0,1]$ is a preset erasure rate,
            \item Or a decoding failure is declared: the algorithm reaches some maximum number of iterations.
        \end{itemize}
        The parameter $\delta$ should be chosen small enough so that it is essentially certain that $q$-ary PDE has converged if the condition is satisfied.
\end{itemize}

\subsubsection{Flooding-Schedule Decoding (FSD) Thresholds for $q$-ary SC-LDPC Code Ensembles}
Given $m$ characterizing the symbol set and $\epsilon$ characterizing the BEC, if $q$-ary PDE is performed over the entire base matrix $\bfBSC$ of an SC-LDPC code ensemble, then the algorithm determines asymptotically (i.e., for coupling length $L \to \infty$ and lifting factor $M \to \infty$) whether FSD can be successful on an ensemble average basis for that specific BEC. Thus, $q$-ary PDE can be used to calculate the FSD threshold, denoted $\efsd(m,\delta)$, which is the largest channel erasure rate such that all transmitted symbols can be recovered successfully with probability at least $(1-\delta)$, as the number of iterations $l$ goes to infinity, i.e.,
\begin{eqnarray}
    \efsd(m,\delta) =
    \sup
    \left\{
        \epsilon \in [0,1]
        \left\rvert
        \pvlAPP(j)[0] \geq 1-\delta
        \text{ for } 1 \leq j \leq kL,
        \text{ as }l \to \infty
        \right.
    \right\}.
\end{eqnarray}

The following numerical FSD threshold results on the BEC are obtained for $\delta = 10^{-6}$, and from this point forward $\efsd(m,\delta)$ will be denoted simply as $\efsdm$.

\subsubsection{Windowed Decoding (WD) Thresholds for $q$-ary SC-LDPC Code Ensembles}
We also apply $q$-ary PDE to WD in order to calculate the WD threshold of an SC-LDPC code ensemble defined over GF$(2^m)$.

The $q$-ary WD-PDE algorithm consists of performing $q$-ary PDE for all the window positions/time instants $t=1$, $2$, $\ldots$, $L$, as illustrated in Fig.~\ref{Fig: WD example}. For each window position, $q$-ary PDE is performed within the $W \times kW$ window; however, unlike the case of FSD, now the convergence check involves only the target symbols, i.e., the first $k$ symbols in the window. Starting from $t=1$, if $q$-ary PDE declares a decoding failure, then the whole $q$-ary WD-PDE terminates and declares a decoding failure; otherwise, the window slides forward and $q$-ary PDE is performed for the next window position. The $q$-ary WD-PDE algorithm declares a decoding success for a specific BEC if and only if its ``component'' $q$-ary PDE declares decoding successes for all the window positions. Thus, given $m$, $\epsilon$, and $W$, $q$-ary WD-PDE can be used to calculate the WD threshold of an SC-LDPC code ensemble.

We now define
\begin{eqnarray}
    \ewd(m,W,t,\delta) =
    \sup
    \left\{
        \epsilon \in [0,1]
        \left\rvert
        \begin{split}
            \pvlAPP(j)[0] \geq 1-\delta
            &\text{ for } tk-k+1 \leq j \leq tk,\\
            &\text{ as }l \to \infty
        \end{split}
        \right.
    \right\}
\end{eqnarray}
as the largest channel erasure rate such that all the target symbols in window position $t$ can be recovered successfully with probability at least $(1-\delta)$, as $l$ goes to infinity, given that all the target symbols in the previous $(t-1)$ windows have already been recovered successfully with probability at least $(1-\delta)$. Then the WD threshold $\ewd(m,W,\delta)$ is the infimum of $\left\{ \ewd(m,W,t,\delta) \right\}_{t=1}^L$, i.e.,
\begin{eqnarray}
    \ewd(m,W,\delta) = \inf_{1 \leq t \leq L}{\ewd(m,W,t,\delta)},
\end{eqnarray}
guaranteeing that all the transmitted symbols -- consisting of all the target symbols in all the windows -- can be recovered successfully with probability at least $(1-\delta)$, as $l$ goes to infinity.

It was proved in Proposition~$1$ of~\cite{Iyengar12-WD} that the WD thresholds of binary SC-LDPC code ensembles on the BEC are non-decreasing with increasing $W$, i.e., $\ewd(1,W,\delta) \leq \ewd(1,W+1,\delta)$ for any $\delta \in \left[0,1\right]$ and all $W$, $W = w+1$, $w+2$, $\dots$, $w+L$. By combining this proof with the monotonicity of $q$-ary variable and check node updates, proved in Appendix~B of~\cite{Andriyanova13-NB-SC-LDPC-th-sat-BEC}, we can state the following theorem.

\begin{theorem}[Monotonicity of $\ewd(m,W,\delta)$ with increasing $W$]
\label{Theorem: monotonicity}
For a fixed $m \geq 1$, any $\delta \in \left[0,1\right]$, and all $W$, $W = w+1$, $w+2$, $\dots$, $w+L$,
    \begin{equation}
        \ewd(m,W,\delta) \leq \ewd(m,W+1,\delta).
    \end{equation}
\end{theorem}

As in the case of FSD thresholds, we choose $\delta = 10^{-6}$, and from this point forward $\ewd(m,W,\delta)$ will be denoted simply as $\ewdmw$.

\subsection{Numerical results: $k=2$ ($R=1/2$)}
\label{Sec: TH BEC; Subsec: k=2}
In this subsection we focus on the BP thresholds of the rate $R=1/2$ $q$-ary SC-LDPC code ensembles with $k=2$: in particular, we consider the $(2,4)$-, $(3,6)$-, $(4,8)$-, and $(5,10)$-regular code ensembles. Our emphasis is on the scenario when WD is used, and the $q$-ary WD-PDE algorithm described in the previous subsection is adopted to calculate the corresponding BP thresholds.

Recall from Section~\ref{Sec: Ensembles; Subsec: Construction} that, for $k=2$, the SC-LDPC code ensembles we consider are the following:
\begin{eqnarray}
& \Cjkmsjminusone:
\label{Eq: C J K w(J-1)}
    &\bfB_0 = \bfB_1 = \cdots = \bfB_{J-1} =
    \begin{bmatrix}
        1 & 1 \\
    \end{bmatrix};\\
&\Cjkmsonetypeone:
\label{Eq: C J K w1 type1}
    &\bfB_0 =
    \begin{bmatrix}
        1 & 1 \\
    \end{bmatrix},\,
    \bfB_1 =
    \begin{bmatrix}
        J-1 & J-1 \\
    \end{bmatrix};\\
&\Cjkmsonetypetwo:
\label{Eq: C J K w1 type2}
    &\bfB_0 =
    \begin{bmatrix}
        1 & J-1 \\
    \end{bmatrix},\,
    \bfB_1 =
    \begin{bmatrix}
        J-1 & 1 \\
    \end{bmatrix};\\
&\Cjkmsonetypethree:
\label{Eq: C J K w1 type3}
    &\bfB_0 =
    \begin{bmatrix}
        J-1 & J-1 \\
    \end{bmatrix},\,
    \bfB_1 =
    \begin{bmatrix}
        1 & 1 \\
    \end{bmatrix}.
\end{eqnarray}

The classical edge spreading results in the maximum coupling width $w=J-1$ by choosing each $\bfB_i$ in $\bfBzeroms$ equal to $\begin{bmatrix} 1 & 1 \end{bmatrix}$. When $w=1$, the type $1$ and type $3$ ensembles, $\Cjkmsonetypeone$ and $\Cjkmsonetypethree$, will have the same FSD threshold $\efsd(m)$, since their SC base matrices are equal up to row permutations and the $q$-ary PDE algorithm is performed over the entire base matrix $\bfBSC$. However, their WD thresholds are different. Type $2$ has one column of $\bfBzeroms$ that is the same as type $1$ and the other column that is the same as type $3$, so it is expected that its WD threshold will be between those of types $1$ and $3$.

\subsubsection{The $(2,4)$ ensembles}
\begin{figure}[!t]
    \centering
    \includegraphics[width=0.475\columnwidth]{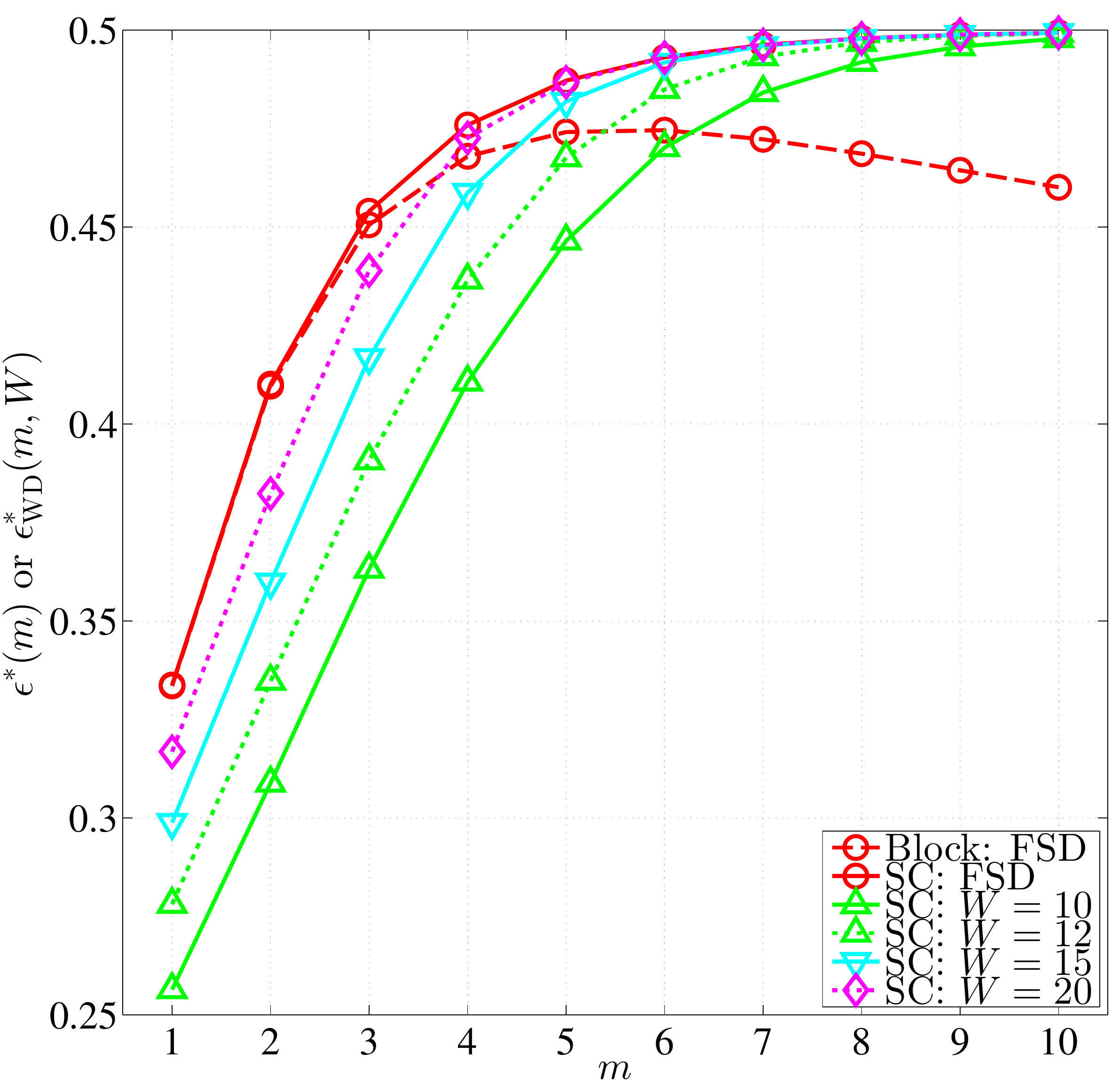}
    \caption{FSD thresholds $\efsdm$ and WD thresholds $\ewdmw$ of $\Ctwofourmsone$.}
    \label{Fig: TH BEC 2 4}
\end{figure}

When $(J,K)=(2,4)$, all four types of edge spreading for $q$-ary SC-LDPC code ensembles are the same. For $m=1$, $2$, $\ldots$, $10$, the FSD and WD thresholds are shown in Fig.~\ref{Fig: TH BEC 2 4}:
\begin{itemize}
    \item Comparing $\Ctwofourmsone$ to $\Btwofour$, the improvement in the FSD threshold $\efsdm$ introduced by the spatially coupled structure is negligible for small $m$. However, as $m$ increases, $\efsdm$ for $\Ctwofourmsone$ increases and approaches the BEC capacity of a rate $R=1/2$ code ensemble.\footnote{Since $L=100$, the design rate of $\Ctwofourmsone$ is $R_L = 0.495$ and capacity is $\epsilon_{\text{Sh}}=1-0.495 = 0.505$. This gap to capacity vanishes as $L \to \infty$, since the thresholds do not further decay and $R_L \to 1/2$.} This is consistent with the observations made in~\cite{Uchikawa11-RC-NB-LDPCC}. We note that the $\Btwofour$ ensembles do not display this behavior; in particular, their thresholds diverge from capacity as $m$ increases, $m\geq 5$.
    \item For WD of $\Ctwofourmsone$ with fixed $m$, the threshold $\ewdmw$ improves as the window size $W$ increases -- see Theorem~\ref{Theorem: monotonicity} in Section~\ref{Sec: TH BEC; Subsec: q-ary PDE} -- and saturates numerically to a (maximum) constant value $\ewdmaxm$. Thus, we define
        \begin{eqnarray}
            \Wm =
            \min
            \left\{
                W \>\left\rvert\> \ewdmw \cong \ewdmaxm \right.
            \right\}
        \end{eqnarray}
        as the smallest window size that provides the best threshold $\ewdmaxm$ for a fixed $m$; here, ``$\cong$'' is used to denote a numerically indistinguishable equality.\footnote{For our purposes, two thresholds are numerically indistinguishable if their absolute difference is no more than $10^{-6}$.} We now make three observations regarding the ensemble $\Ctwofourmsone$:
        \begin{itemize}
            \item For all $m$, $\ewdmaxm = \efsdm$, i.e., when the window size $W$ is large enough, the WD threshold equals the FSD threshold.
            \item As $m$ increases, $\Wm$ is non-increasing, i.e., increasing the finite field size can ``speed up'' the saturation of $\ewdmw$ to $\ewdmaxm$ as $W$ increases.
            \item The saturation of $\ewdmw$ to $\ewdmaxm$ is relatively slow as $W$ increases, especially when $m$ is small. For example, when $m=1$, we need a window size of $\W(1) = 30$ to obtain the threshold $\ewdmax(1)$. Moreover, even for a fairly large window, say $W=20$, the WD threshold of $\Ctwofourmsone$ is worse than the FSD threshold of $\Btwofour$ for $m=1$, $2$, and $3$. This indicates that $\Ctwofourmsone$ does not perform well unless $W$ and/or $m$ are large.
        \end{itemize}
\end{itemize}

As a result, we conclude that $\Ctwofourmsone$ is not a good candidate for WD, since a desirable $q$-ary SC-LDPC code ensemble should provide a near-capacity threshold when both the finite field size and the window size are relatively small, resulting in both small decoding latency and small decoding complexity -- details will be discussed later in Section~\ref{Sec: Dec cmplx}. We will see in the remainder of this section, however, that increasing the node degrees in the code graph speeds up the saturation of $\ewdmw$ to $\ewdmaxm$.

\subsubsection{The $(3,6)$ ensembles}
\begin{figure}[!t]
    \centering
    \includegraphics[width=0.475\linewidth]{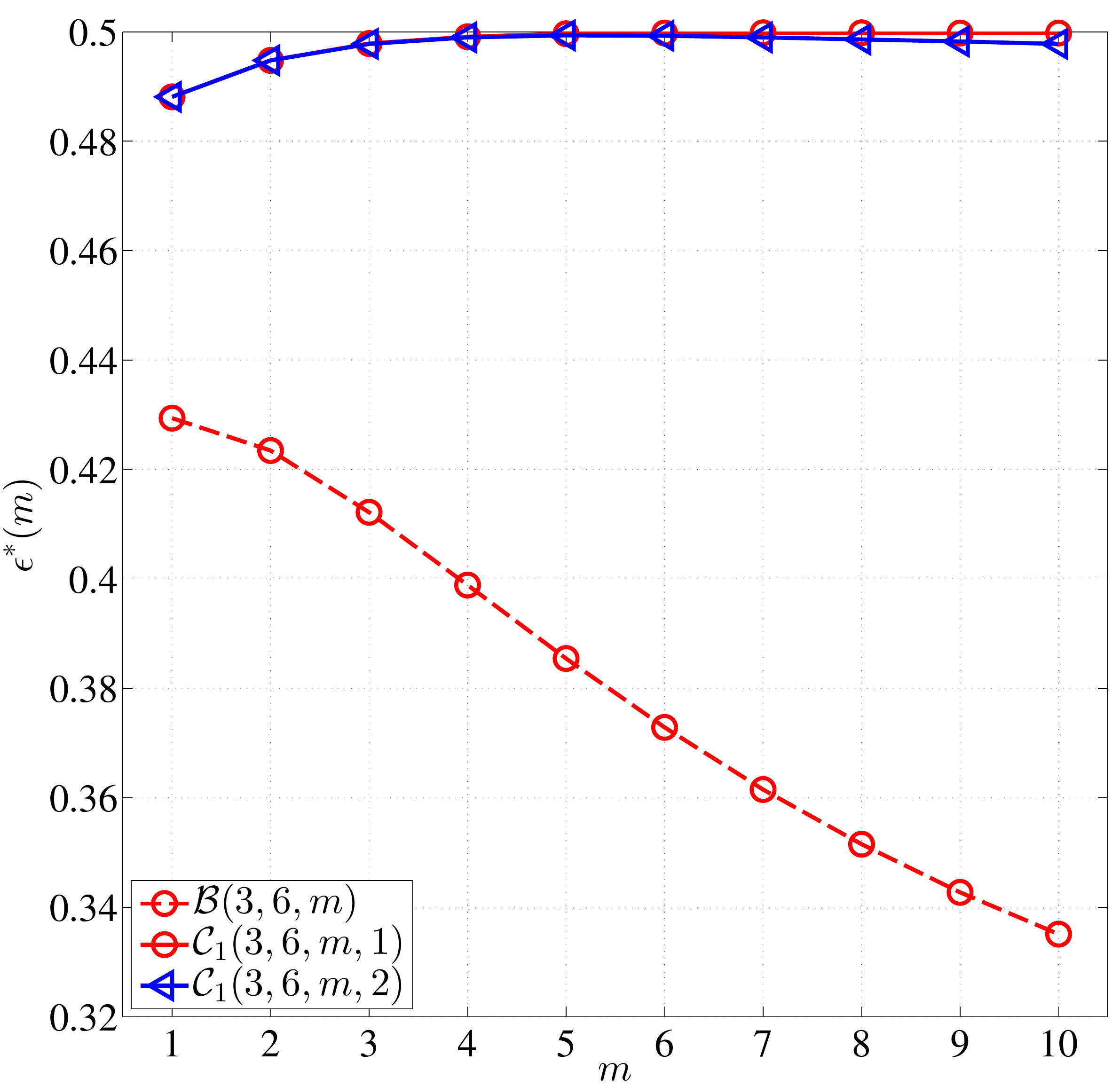}
    \caption{FSD thresholds $\efsdm$ comparison of the $(3,6,m)$ ensembles.}
    \label{Fig: TH BEC FSD 3 6 blk vs SC}
\end{figure}

As a benchmark, Fig.~\ref{Fig: TH BEC FSD 3 6 blk vs SC} compares the FSD thresholds of ensembles $\Cthreesixmsonetypeone$ (and thus $\Cthreesixmsonetypethree$) and $\Cthreesixmsonetypetwo$ to that of $\Bthreesix$ for various $m$.\footnote{Code ensembles $\Cthreesixmstwo$ are not included in Fig.~\ref{Fig: TH BEC FSD 3 6 blk vs SC} because their thresholds are almost indistinguishable (although slightly different) from those of $\Cthreesixmsonetypeone$.} It is observed that:
\begin{itemize}
    \item For all finite field sizes $2^m$, the introduction of the spatially coupled structure provides all four $q$-ary SC-LDPC code ensembles with significant improvement in the FSD threshold compared to the corresponding $q$-ary LDPC-BC ensemble. In fact, the gap between $\Bthreesix$ and the $(3,6,m)$ SC ensembles increases as $m$ increases. Again, this is consistent with the observations made in~\cite{Uchikawa11-RC-NB-LDPCC} and~\cite{Piemontese13-NB-SC-LDPC-BEC}.
    \item Note that, like the $\Btwofour$ ensembles discussed above, the $\Bthreesix$ thresholds diverge from capacity as $m$ increases, while the FSD thresholds of $\Cthreesixmsonetypeone$ and $\Cthreesixmstwo$ increase and approach the BEC capacity for rate $R=1/2$ as $m$ increases. Surprisingly, this is not the case for $\Cthreesixmsonetypetwo$, whose FSD threshold increases and approaches capacity for $m=5$, but then decreases slowly and thus diverges slightly from capacity as $m$ increases further. As a result, in Fig.~\ref{Fig: TH BEC FSD 3 6 blk vs SC}, there exists a small gap between the thresholds of $\Cthreesixmsonetypeone$ and $\Cthreesixmsonetypetwo$ for large $m$.
\end{itemize}

\begin{figure}[!t]
    \centering
    \subfigure[$\Cthreesixmsonetypetwo$.]
    {
        \includegraphics[width=0.475\linewidth]{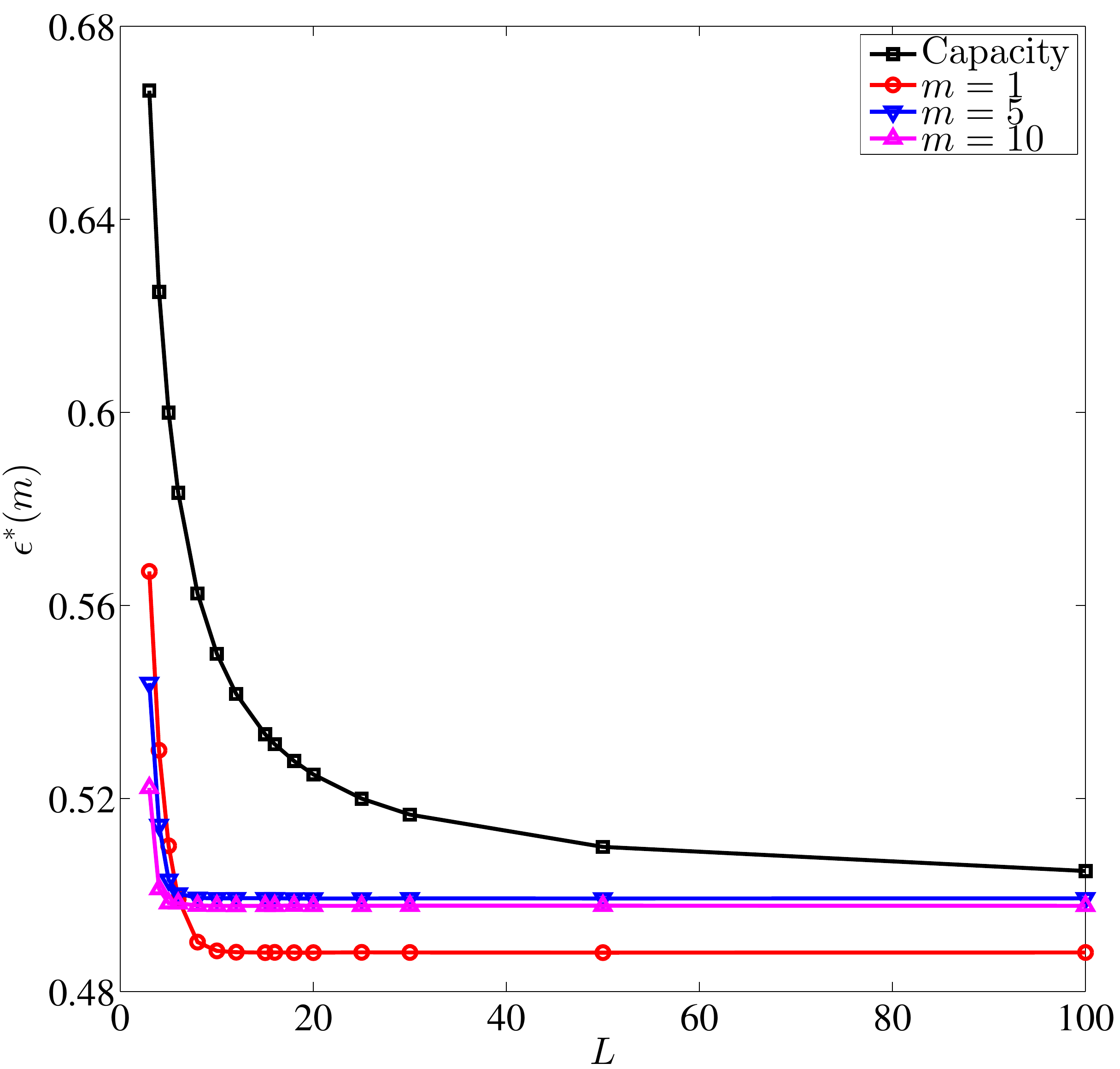}
        \label{Fig: TH BEC FSD 3 6 ms1 type 2 diffL}
    }
    \subfigure[$\Cthreesixmstwo$.]
    {
        \includegraphics[width=0.475\linewidth]{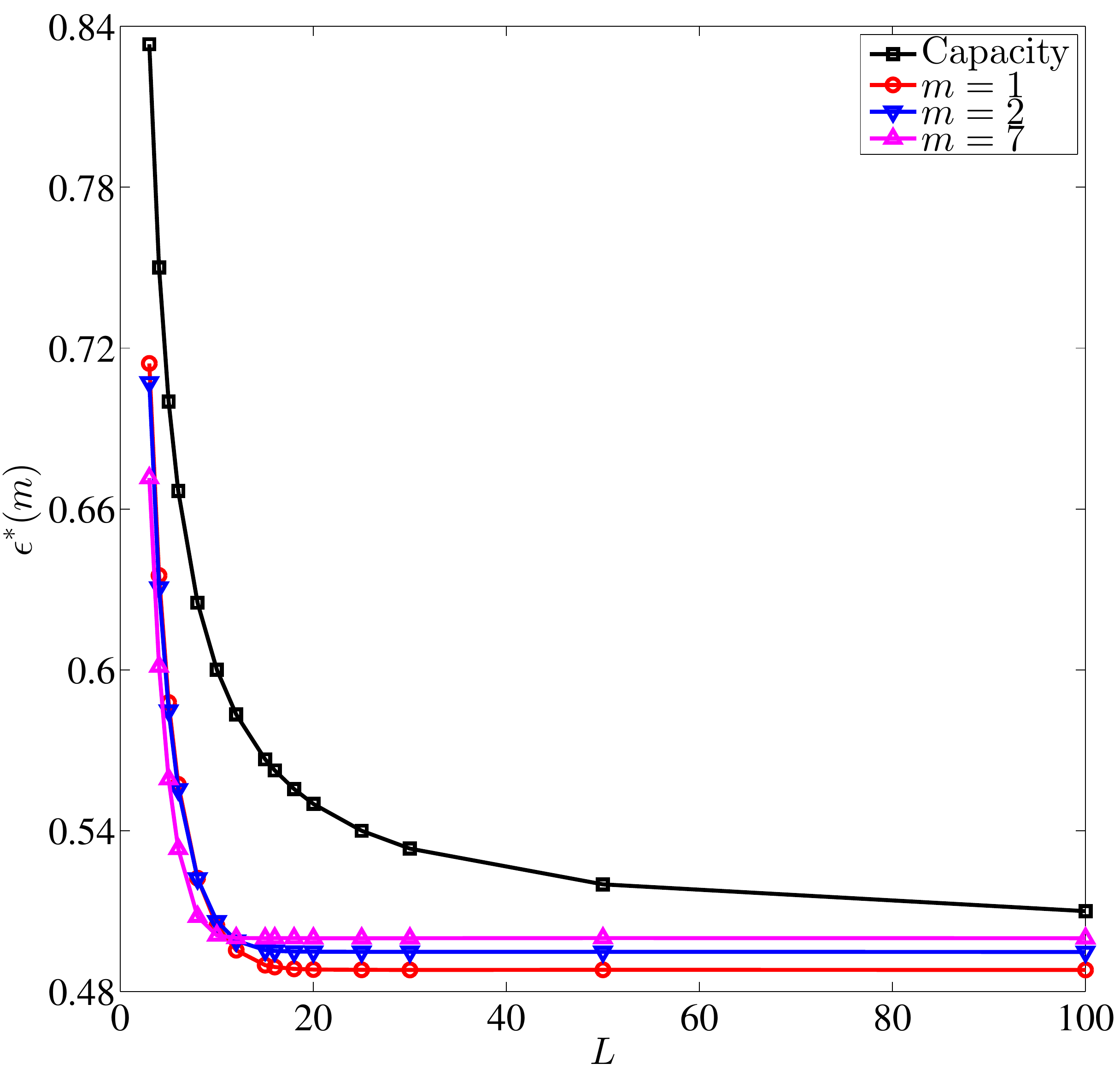}
        \label{Fig: TH BEC FSD 3 6 ms2 diffL}
    }
    \caption{FSD thresholds $\efsdm$ of the SC-LDPC code ensembles with different coupling lengths $L$.}
    \label{Fig: TH BEC FSD 3 6 diffL}
\end{figure}

We now briefly demonstrate the FSD threshold behavior of SC-LDPC code ensembles for varying coupling lengths $L$. Fig.~\ref{Fig: TH BEC FSD 3 6 diffL} shows the FSD thresholds $\efsdm$ for ensembles $\Cthreesixmsonetypetwo$ and $\Cthreesixmstwo$ with increasing $L$. For fixed $m$ and increasing $L$, the FSD thresholds initially decrease and then saturate to a constant value for sufficiently large $L$, which is consistent with results for binary protograph-based SC-LDPC code ensembles~\cite{Lentmaier10-terminated-LDPCC, Mitchell14-SC-LDPC-protograph}.

Note that Figure~\ref{Fig: TH BEC FSD 3 6 diffL} also illustrates the point made above that the $\Cthreesixmsonetypetwo$ ensemble does not have monotonically increasing thresholds with $m$. Specifically, in Fig.~\ref{Fig: TH BEC FSD 3 6 ms1 type 2 diffL}, for $\Cthreesixmsonetypetwo$, we have $\efsd(1) < \efsd(5)$ but $\efsd(10) < \efsd(5)$, while in Fig.~\ref{Fig: TH BEC FSD 3 6 ms2 diffL}, for $\Cthreesixmstwo$, $\efsd(m)$ increases uniformly as $m$ increases: this confirms our observation of the small gap between the FSD thresholds of $\Cthreesixmsonetypeone$ (almost indistinguishable from $\Cthreesixmstwo$) and $\Cthreesixmsonetypetwo$ for large $m$ noted in Fig.~\ref{Fig: TH BEC FSD 3 6 blk vs SC}.

With reference to Fig~\ref{Fig: TH BEC FSD 3 6 diffL}, given $m$, let $\Lm$ be the minimum $L$ such that the threshold has saturated to its constant value, i.e.,
\begin{eqnarray}
    \Lm =
    \min
    \left\{
        L \>\left\rvert\> \efsd(m,L) \cong \efsd(m,L'),\, L'=L+1, L+2, \ldots \right.
    \right\}.
\end{eqnarray}
As shown in Figs.~\ref{Fig: TH BEC FSD 3 6 ms1 type 2 diffL} and~\ref{Fig: TH BEC FSD 3 6 ms2 diffL}, $\Lm$ is non-increasing as $m$ increases; for example, for $\Cthreesixmsonetypetwo$, $L^{*}(1)=15$, $L^{*}(3)=10$, and $\Lm=8$ when $m \geq 6$. Thus, we see that increasing the finite field size speeds up the saturation of the FSD threshold as $L$ increases. To avoid repetition, we omit the FSD thresholds obtained for other $(J,K)$-regular SC-LDPC code ensembles with varying $L$; however, it should be noted that the threshold saturation behavior described above is consistent over all considered code ensembles.

\begin{figure*}[!t]
    \centering
    \subfigure[$\Cthreesixmstwo$]
    {
        \includegraphics[width=0.475\linewidth]{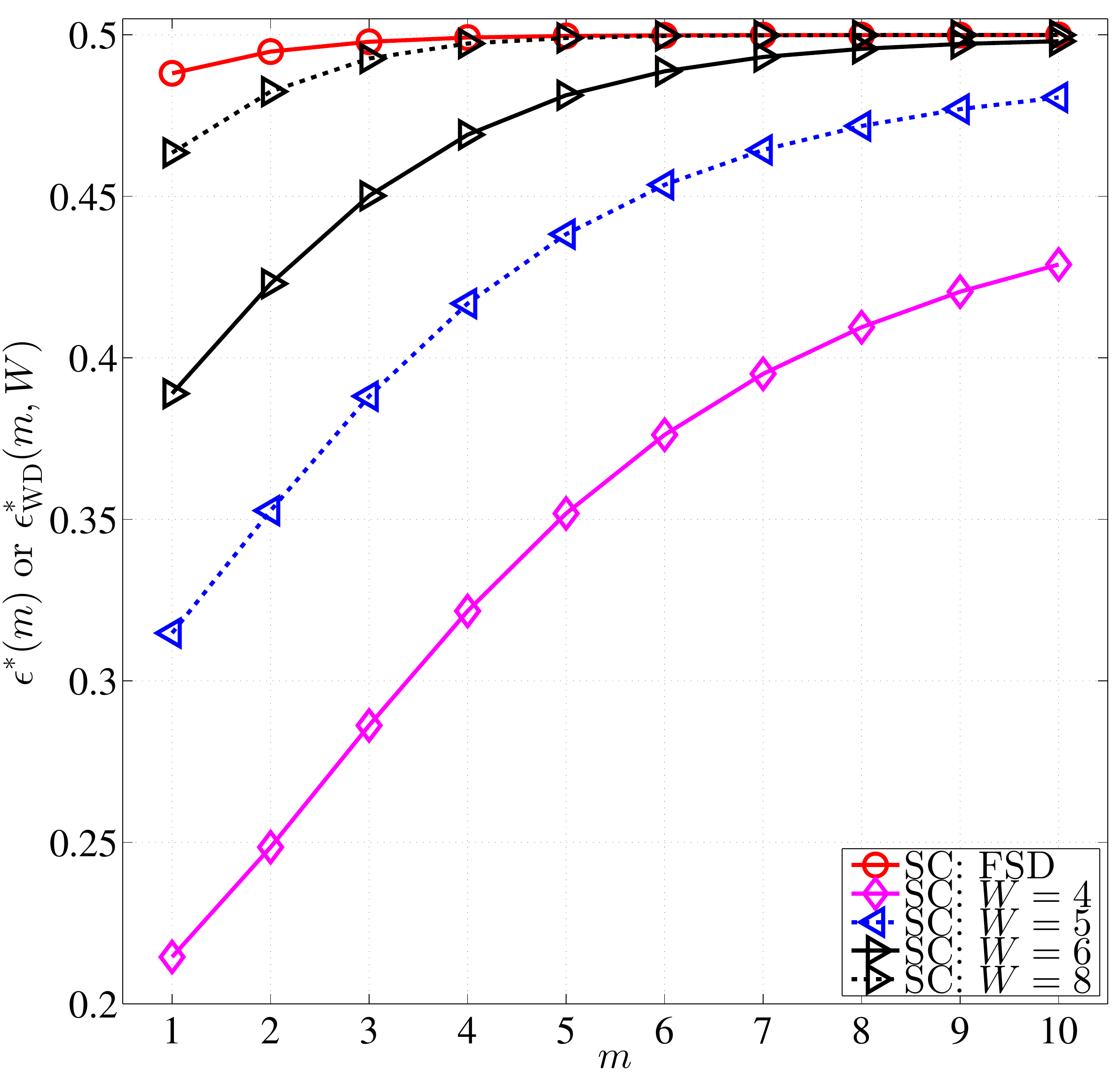}
        \label{Fig: TH BEC WD 3 6 ms2}
    }
    \subfigure[$\Cthreesixmsonetypeone$]
    {
        \includegraphics[width=0.475\linewidth]{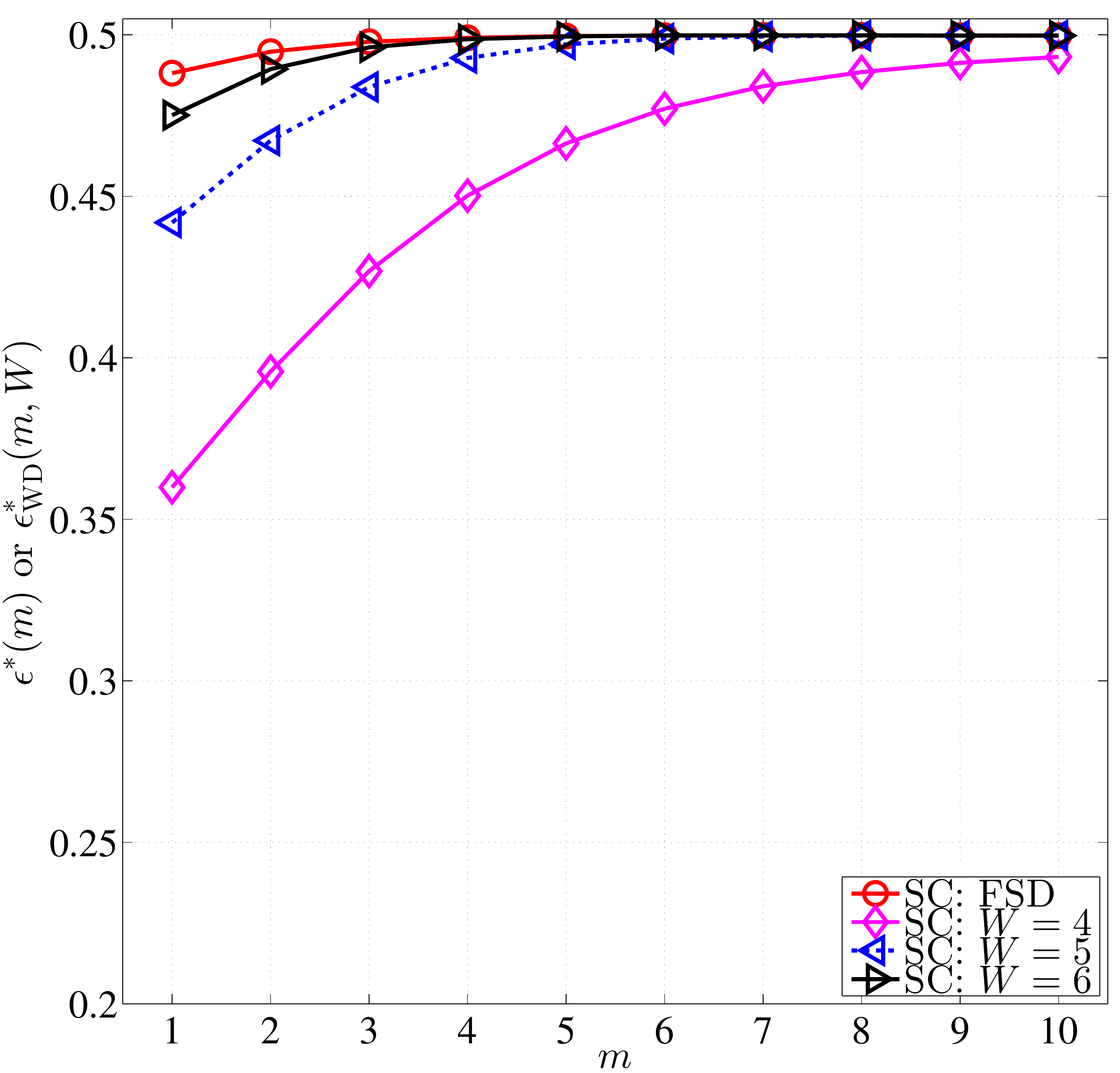}
        \label{Fig: TH BEC WD 3 6 ms1 type 1}
    }
    \subfigure[$\Cthreesixmsonetypetwo$]
    {
        \includegraphics[width=0.475\linewidth]{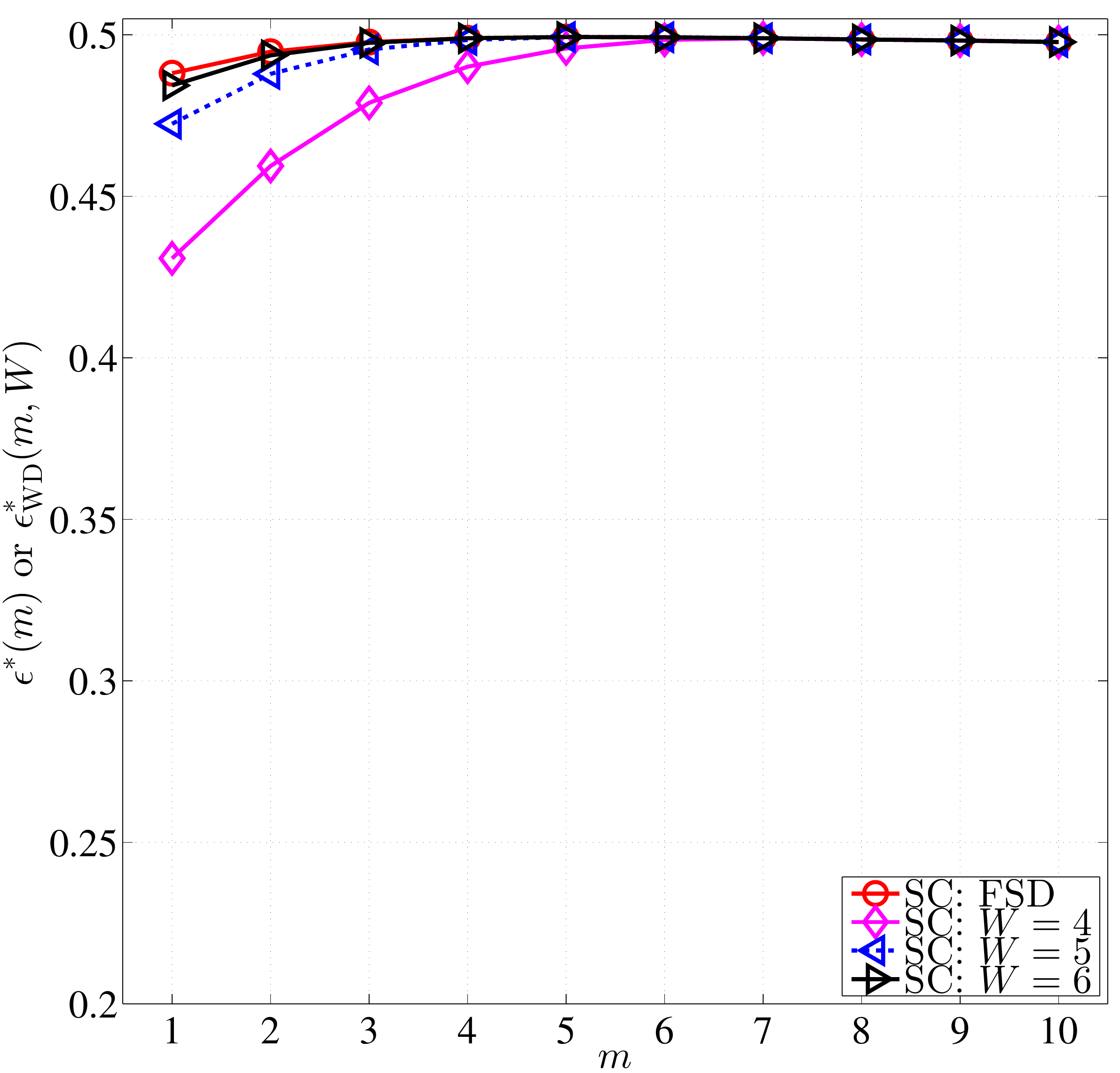}
        \label{Fig: TH BEC WD 3 6 ms1 type 2}
    }
    \subfigure[$\Cthreesixmsonetypethree$]
    {
        \includegraphics[width=0.475\linewidth]{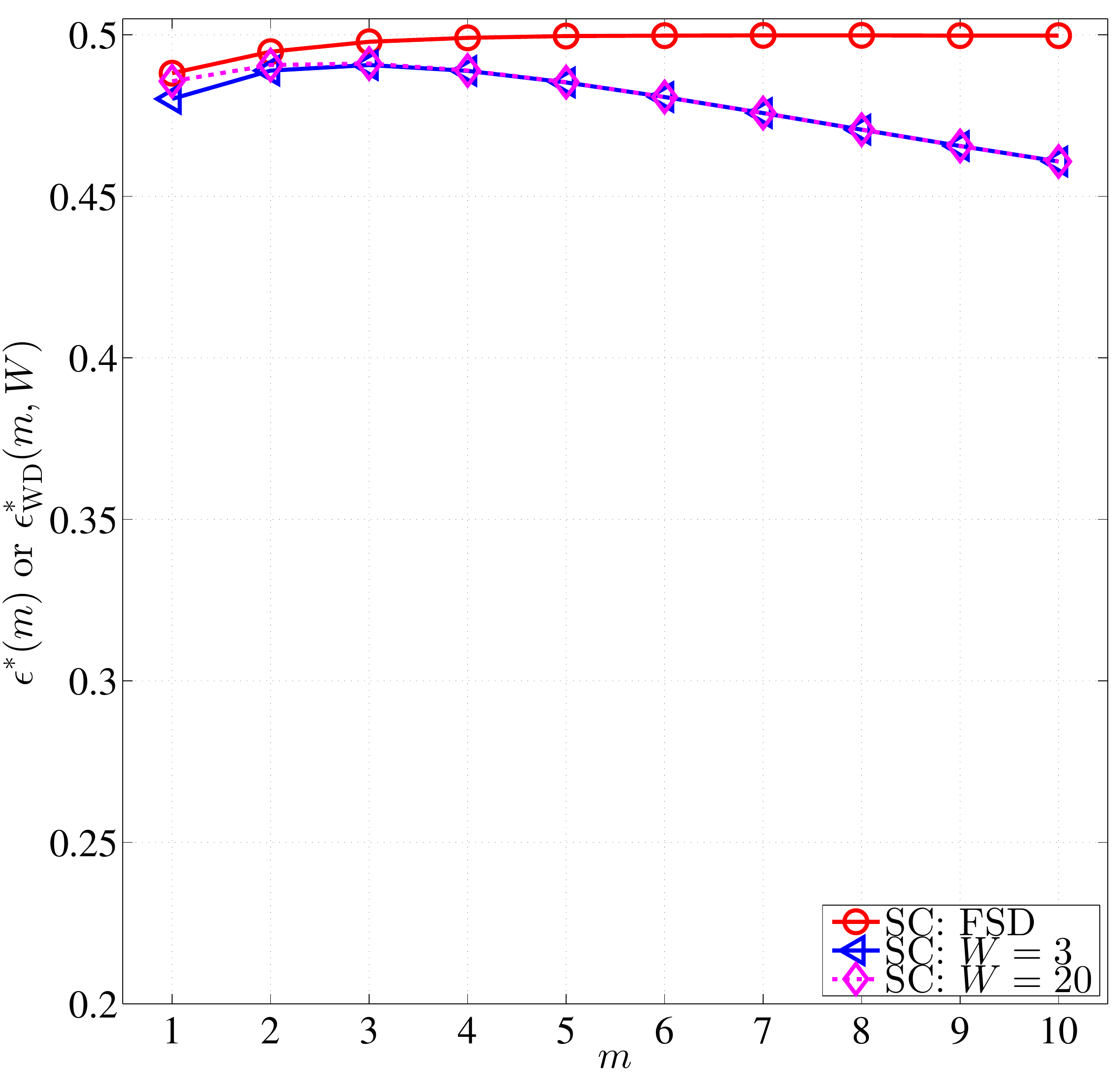}
        \label{Fig: TH BEC WD 3 6 ms1 type 3}
    }
    \caption{WD thresholds $\ewdmw$ of the $(3,6,m)$ SC-LDPC code ensembles. FSD thresholds $\efsdm$ are included as benchmarks.}
    \label{Fig: TH BEC WD 3 6}
\end{figure*}

We now consider the WD thresholds of the $(3,6,m)$ SC-LDPC code ensembles, again with $L=100$. The WD thresholds of $\Cthreesixmstwo$ with the classical edge-spreading format are shown in Fig.~\ref{Fig: TH BEC WD 3 6 ms2}. As expected, for fixed $m$, the WD thresholds improve with increasing $W$, and we find that $\ewdmaxm = \efsd(m)$ for $W \geq \Wm$, i.e., for a sufficiently large window, the WD threshold is equal to the FSD threshold for all $m$. We note that $\Wm$ is non-increasing as $m$ increases, i.e., the saturation of the WD thresholds $\ewdmw$ to $\ewdmaxm$ is faster for larger $m$. For example, $\W(2) = 15$, $\W(4) = 12$, and for $m\geq 7$, $\W(m)= 8$. Due to a combination of the existence of degree-$1$ variable nodes in the window and the larger coupling width $w=2$, $\Cthreesixmstwo$ does not perform well using WD with a relatively small window.

Next, we consider the cases when $w=1$: $\Cthreesixmsonetypeone$, $\Cthreesixmsonetypetwo$, and $\Cthreesixmsonetypethree$, shown in Figs.~\ref{Fig: TH BEC WD 3 6 ms1 type 1}, \ref{Fig: TH BEC WD 3 6 ms1 type 2}, and~\ref{Fig: TH BEC WD 3 6 ms1 type 3}, respectively. We observe that
\begin{itemize}
    \item Similar to the $\Cthreesixmstwo$ ensemble, for each of the three ensembles, at a particular $m$, the WD threshold $\ewdmw$ improves as $W$ increases and saturates numerically to a constant value $\ewdmaxm$. Again, increasing the finite field size speeds up the saturation as $W$ increases; for example, $\W(2)=10$, $\W(4)=8$, and $\W(6)=6$ for $\Cthreesixmsonetypeone$.
    \item Simply choosing $W \geq \Wm$ does not necessarily guarantee good WD thresholds, since $\ewdmaxm$ may not equal $\efsdm$ even when $W$ is large.\footnote{Of course, as mentioned earlier, by selecting $W = L+w$ in WD, the decoding window covers the whole parity-check matrix and WD is equivalent to FSD. However, we are not considering this extreme case here.} In fact, $\ewdmaxm = \efsdm$ for all $m$ only for $\Cthreesixmsonetypeone$ and $\Cthreesixmsonetypetwo$; for $\Cthreesixmsonetypethree$, on the other hand, $\ewdmaxm$ diverges from $\efsdm$ as $m$ increases, as shown in Fig.~\ref{Fig: TH BEC WD 3 6 ms1 type 3}.
\end{itemize}

\begin{figure}[!t]
    \centering
    \subfigure[$\Cthreesixmsonetypeone$]
    {
        \includegraphics[width=0.4\linewidth]{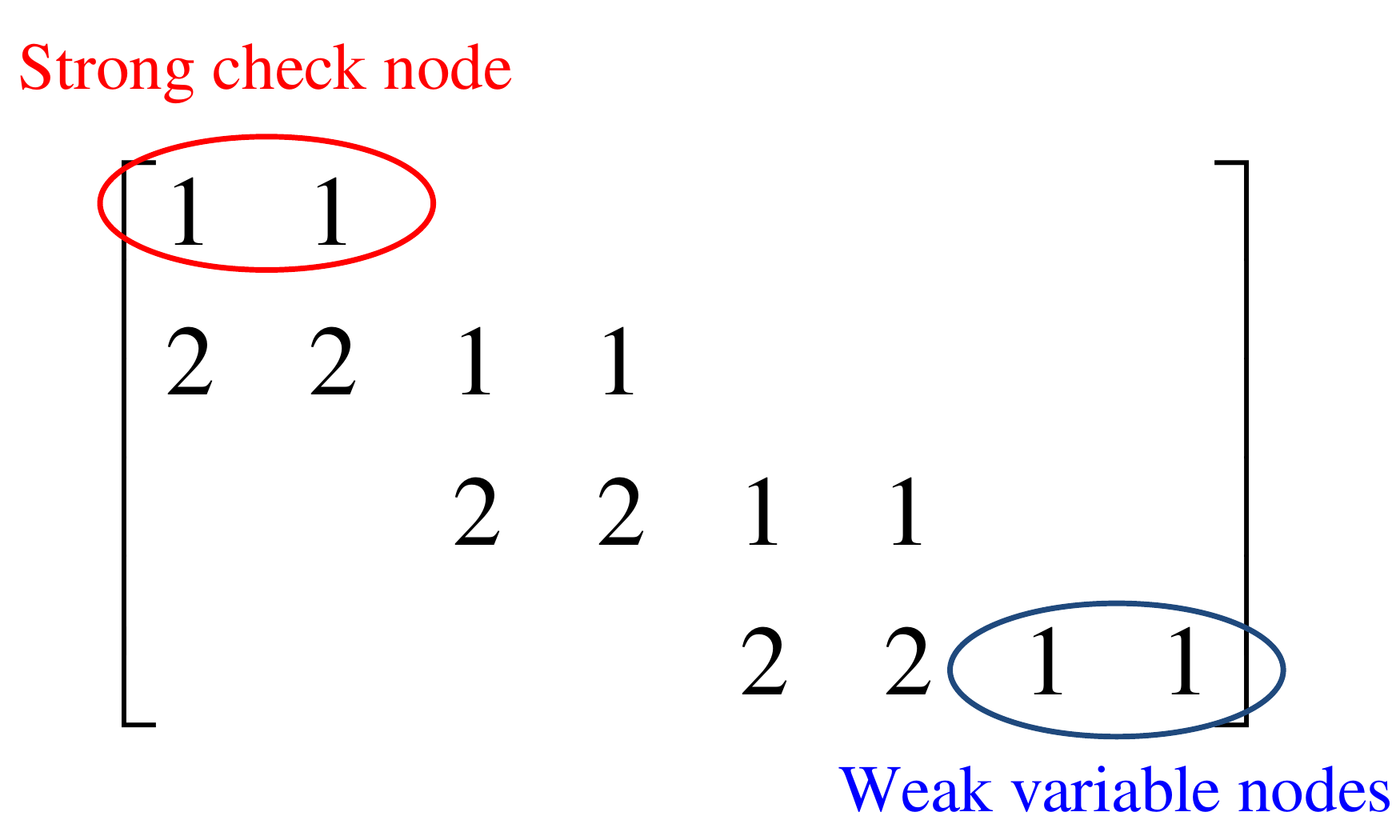}
        \label{Fig: mx WD 3 6 ms1 type1 W4}
    }
    \subfigure[$\Cthreesixmsonetypethree$]
    {
        \includegraphics[width=0.4\linewidth]{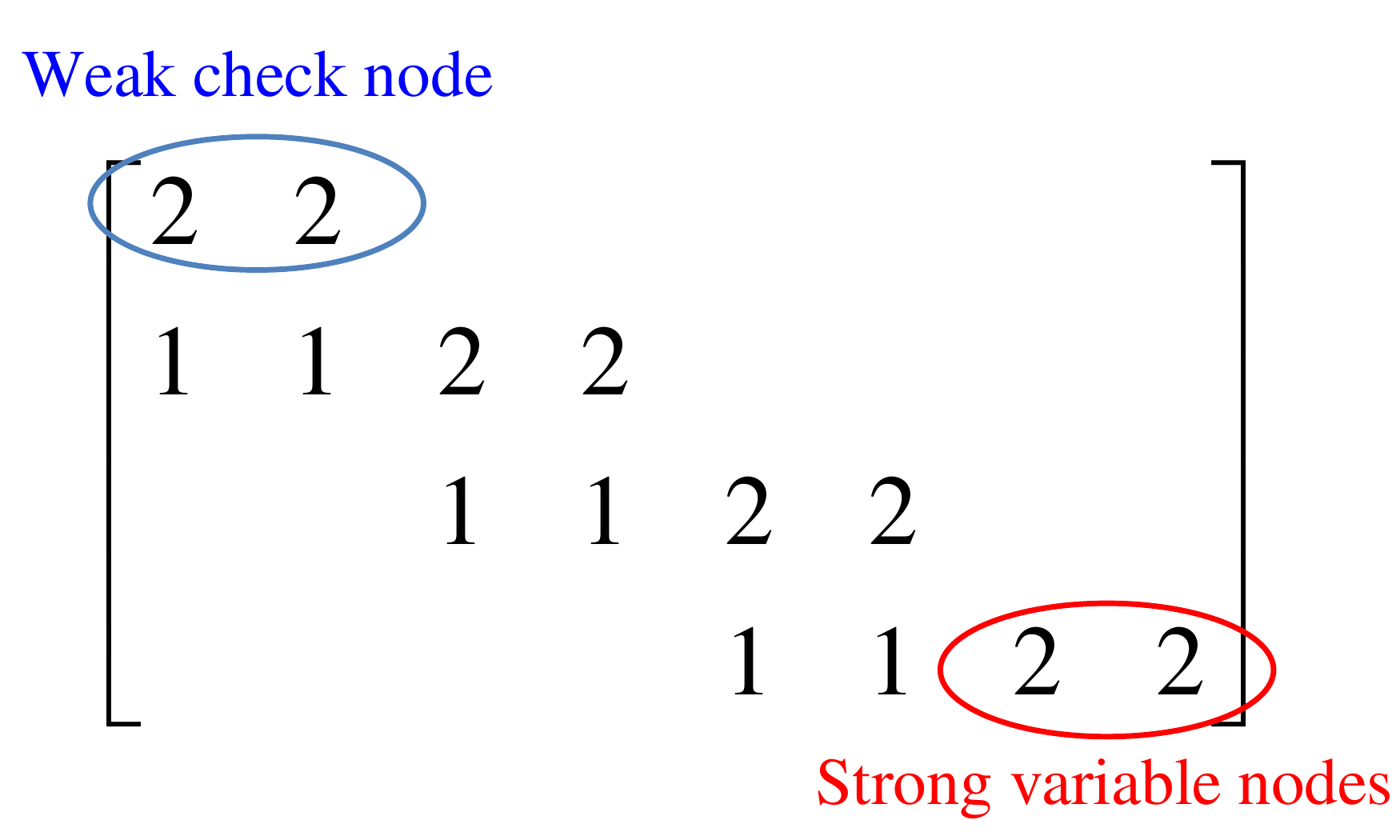}
        \label{Fig: mx WD 3 6 ms1 type3 W4}
    }
    \caption{The portion of the base matrix covered by the window when $W=4$.}
    \label{Fig: mx WD 3 6 ms1 W4}
\end{figure}

We turn our attention now to the implications of the WD thresholds on protograph design. Recall the three types of edge-spreading formats of the $(3,6,m)$ SC ensembles with $w=1$ defined in \eqref{Eq: C J K w1 type1}-\eqref{Eq: C J K w1 type3}, where $\bfB_0^1$ is given as
$\begin{bmatrix} \bfEA & \bfEA \end{bmatrix}$,
$\begin{bmatrix} \bfEA & \bfEB \end{bmatrix}$, and
$\begin{bmatrix} \bfEB & \bfEB \end{bmatrix}$, respectively.
As we move from type $1$ to type $2$ to type $3$, the $q$-ary SC-LDPC code ensemble includes more
$\bfEB = \begin{bmatrix} 2 & 1 \end{bmatrix}^\intercal$
spreading and less
$\bfEA = \begin{bmatrix} 1 & 2 \end{bmatrix}^\intercal$ spreading.
As illustrated in Fig.~\ref{Fig: mx WD 3 6 ms1 type1 W4} for $\Cthreesixmsonetypeone$ with a window size $W=4$, $\bfEA$ spreading has a strong (lower degree) check node at the beginning of the window and weak variable nodes (with degree $1$) at the end of the window. As a result, for all $m$, $\ewdmaxm = \efsdm$ when $W$ is large enough, but $\ewdmw$ is not very good when $W$ is relatively small -- for example, $W=4$ in Fig.~\ref{Fig: TH BEC WD 3 6 ms1 type 1}. (See also the threshold behavior of the $\Cthreesixmstwo$ ensembles in Fig.~\ref{Fig: TH BEC WD 3 6 ms2} which have a similar structure but larger $w$.)

On the other hand, as illustrated in Fig.~\ref{Fig: mx WD 3 6 ms1 type3 W4} for $\Cthreesixmsonetypethree$, $\bfEB$ spreading provides strong (higher degree) variable nodes at the end of the window and a weak (higher degree) check node at the beginning of the window. As a result, compared to $\Cthreesixmsonetypeone$ and $\Cthreesixmsonetypetwo$, $\Cthreesixmsonetypethree$ has the smallest $\Wm$ when $m$ is fixed, i.e., threshold saturation to $\ewdmaxm$ is fastest as $W$ increases, but $\ewdmaxm$ itself does not converge to $\efsdm$, resulting in unsatisfactory WD thresholds, especially when $m$ is large. In fact, comparing Fig.~\ref{Fig: TH BEC WD 3 6 ms1 type 3} to Fig.~\ref{Fig: TH BEC FSD 3 6 blk vs SC}, we observe that the WD threshold of $\Cthreesixmsonetypethree$ becomes more ``block-like'' as $m$ increases, i.e., the curve for the WD threshold of $\Cthreesixmsonetypethree$ behaves similarly to the curve for the FSD threshold of $\Bthreesix$ for $m \geq 4$. This ``block-like'' behavior occurs for type $3$ spreading because the edges of the block protograph have not been sufficiently spread, i.e., only one edge from each variable node in a block protograph is spread to the adjacent block protograph.

We summarize the above observations for WD thresholds with respect to the advantages and disadvantages of $\bfEA$ and $\bfEB$ spreading based on their effects on the portion of the parity-check matrix covered by the window:
\begin{enumerate}
    \item The advantage of $\bfEA$: Due to the strong check node at the start of the window, for a sufficiently large window size, the WD threshold saturates to the FSD threshold, which in turn approaches the channel capacity as the finite field size increases.
    \item The disadvantage of $\bfEA$: Due to the weak variable nodes at the end of the window, WD does not perform well when the window size is relatively small, so for small finite field sizes, there are large gaps between the WD threshold and the FSD threshold.
    \item The advantage of $\bfEB$: Due to the strong variable nodes at the start of the window, for relatively small window sizes, the WD threshold quickly saturates to its best achievable value, even for relatively small finite field sizes.
    \item The disadvantage of $\bfEB$: Due to the weak check node at the end of the window, WD tends to provide more ``block-like'' behavior, so that as the finite field size increases, the WD threshold diverges from the FSD threshold of the $q$-ary SC-LDPC code ensemble and approaches the FSD threshold of the corresponding uncoupled $q$-ary LDPC-BC ensemble.
\end{enumerate}

Based on the advantages and disadvantages of these two antipolar spreading formats, we can develop design rules that combine fast saturation and FSD-achieving thresholds by mixing $\bfEB$ spreading and $\bfEA$ spreading, resulting in the type $2$ spreading $\Cthreesixmsonetypetwo$. For example, as shown in Fig.~\ref{Fig: TH BEC WD 3 6 ms1 type 2}, we see that $\Cthreesixmsonetypetwo$ has good WD thresholds even when both $m$ and $W$ are relatively small, i.e., with $m=5$ and $W=5$, the best performance is already achieved and lies within $0.15\%$ of channel capacity. These design rules are consistent with the design rules proposed in~\cite{Iyengar12-WD} for the binary case, but they are more general in the sense that the effect of non-binary code alphabets is included.

To summarize, given the $(3,6)$-regular degree distribution, to achieve near-capacity thresholds with small decoding latency and small decoding complexity (see Section~\ref{Sec: Dec cmplx} for further details), the $q$-ary SC-LDPC code ensemble $\Cthreesixmsonetypetwo$ is recommended due to its excellent thresholds when the window size $W$ and the finite field size $q$ are both relatively small.

\subsubsection{The $(4,8)$ and $(5,10)$ ensembles}
\begin{figure}[!t]
    \centering
    \subfigure[$(3,6,m)$ ensembles, $W=3$]
    {
        \includegraphics[width=0.475\linewidth]{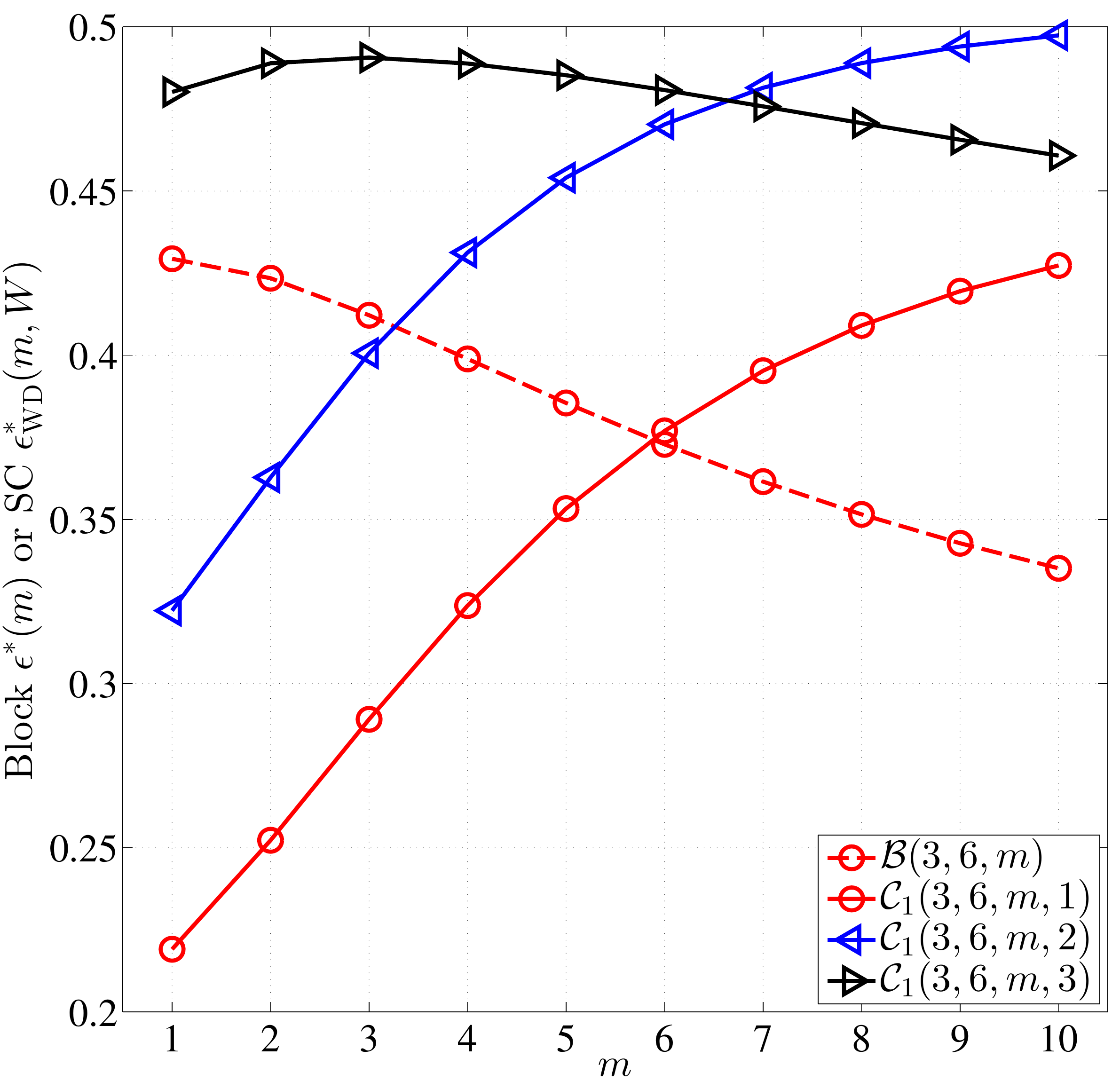}
        \label{Fig: TH BEC WD 3 6 ms1 W3}
    }
    \subfigure[$(4,8,m)$ ensembles, $W=3$]
    {
        \includegraphics[width=0.475\linewidth]{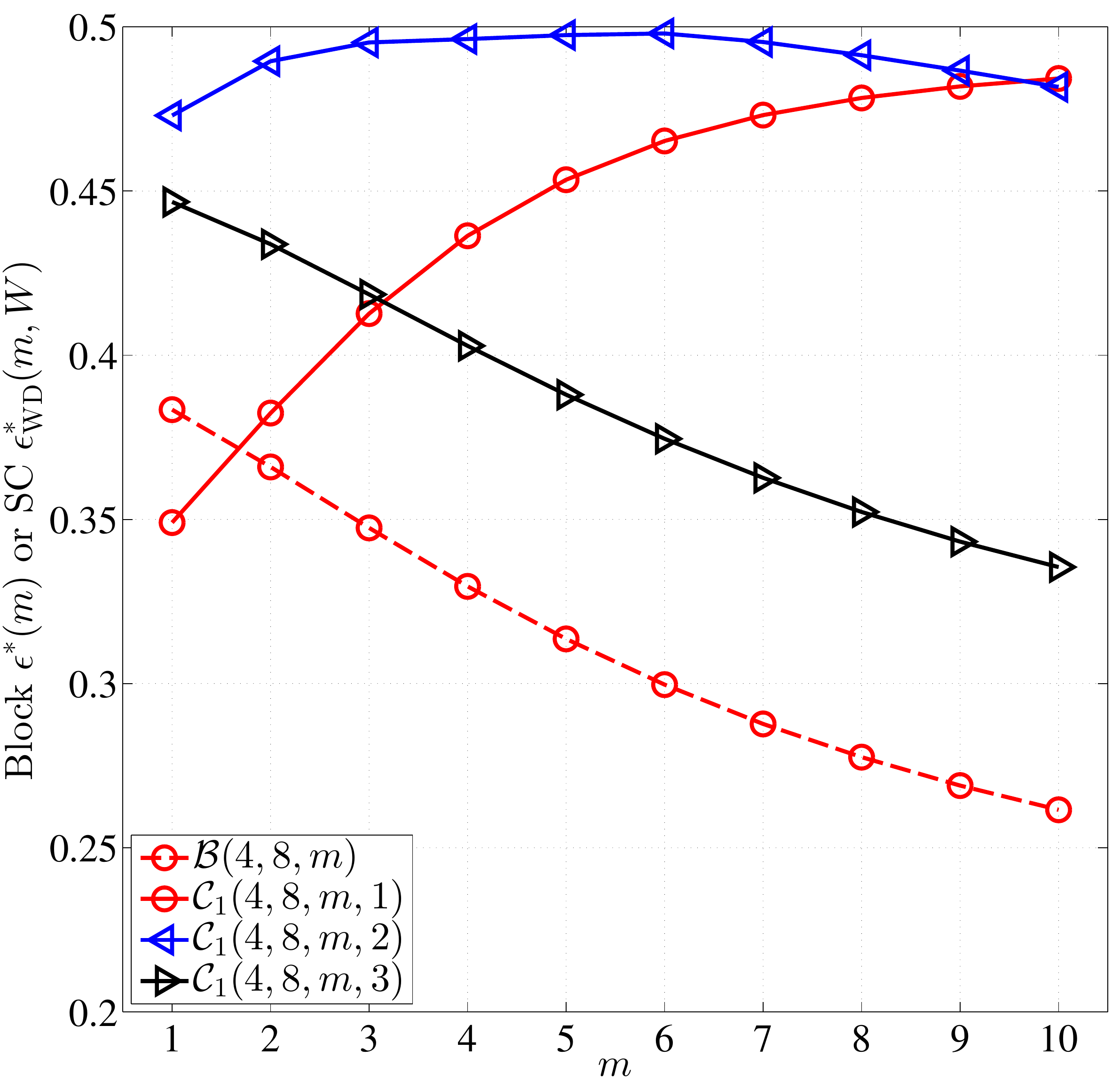}
        \label{Fig: TH BEC WD 4 8 ms1 W3}
    }
    \subfigure[$(5,10,m)$ ensembles, $W=3$]
    {
        \includegraphics[width=0.475\linewidth]{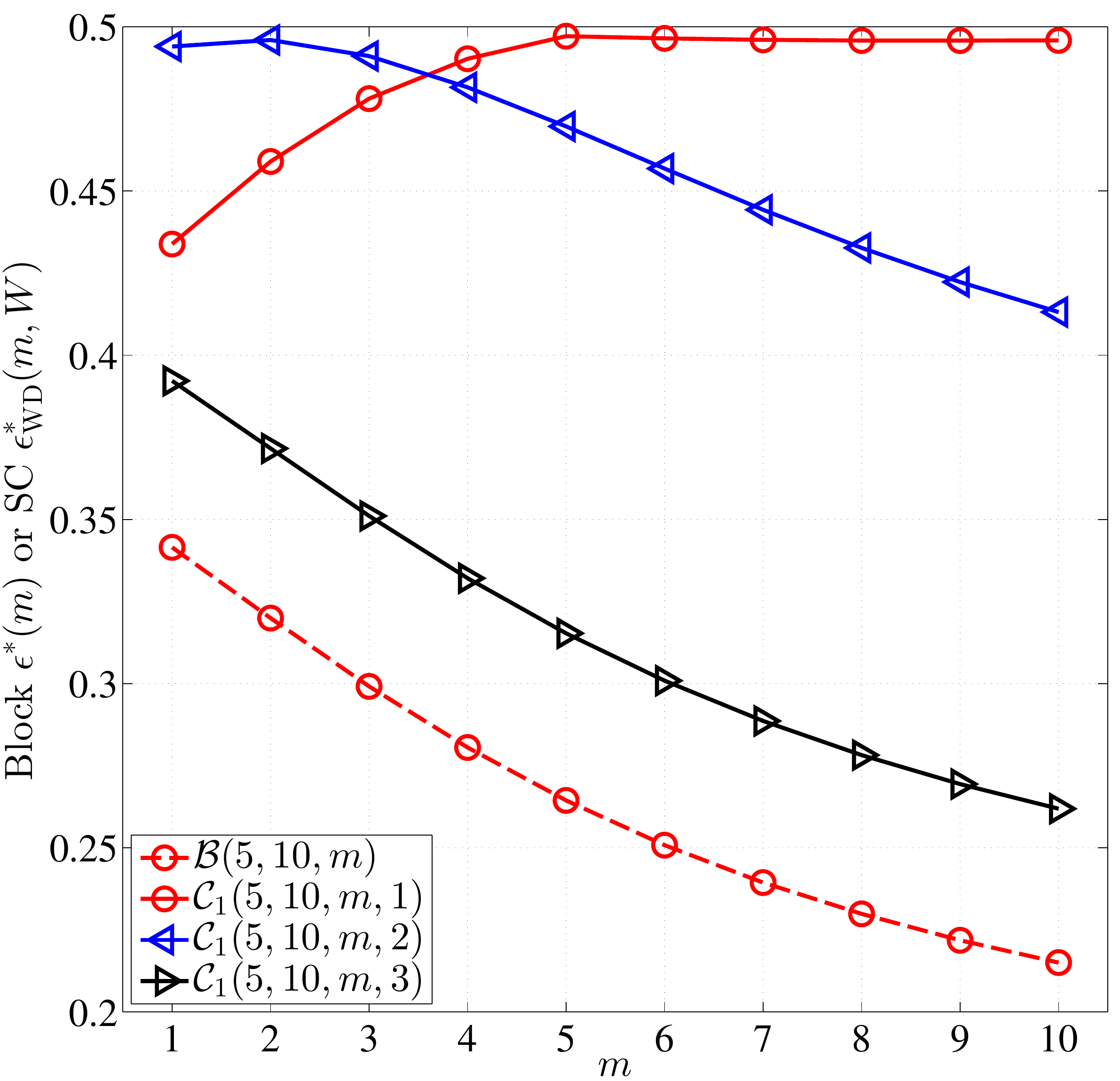}
        \label{Fig: TH BEC WD 5 10 ms1 W3}
    }
    \subfigure[Type $2$ spreading, $W=5$]
    {
        \includegraphics[width=0.475\linewidth]{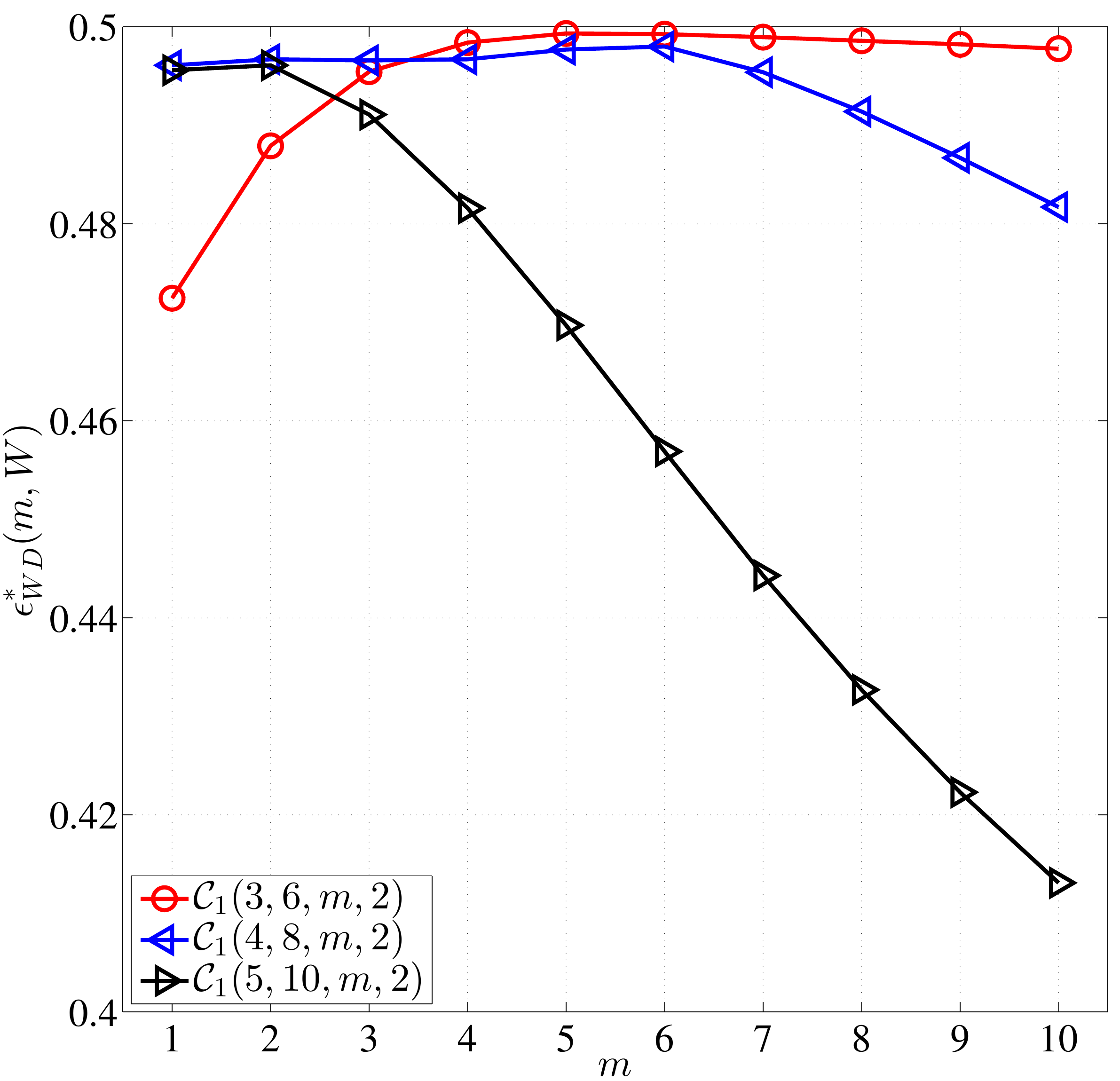}
        \label{Fig: TH BEC WD 3 6 4 8 5 10 ms1 type2 W5}
    }
    \caption{WD thresholds $\ewdmw$ of $q$-ary SC-LDPC code ensembles with $w=1$ and $W=3$: (a) the $(3,6,m)$ ensembles, (b) the $(4,8,m)$ ensembles, and (c) the $(5,10,m)$ ensembles. FSD thresholds $\efsdm$ of the corresponding $q$-ary LDPC-BC ensembles are included for reference. WD thresholds $\ewdmw$ of the $\Cjjmsonetypetwo$ ensembles, $J=3$, $4$, and $5$, with $W=5$ are shown in (d).}
    \label{Fig: TH BEC WD 3 6 4 8 5 10}
\end{figure}

We now examine the WD thresholds of the $(4,8)$-regular $q$-ary SC-LDPC code ensembles with $w=1$ and the $(5,10)$-regular $q$-ary SC-LDPC code ensembles with $w=1$ to explore how the advantages and disadvantages of $\bfEA$ and $\bfEB$ spreading are affected by the density $(J,K)$ of the parity-check matrix, where we still have $k=K/J=2$.

\begin{figure}[!t]
    \centering
    \includegraphics[width=0.95\linewidth]{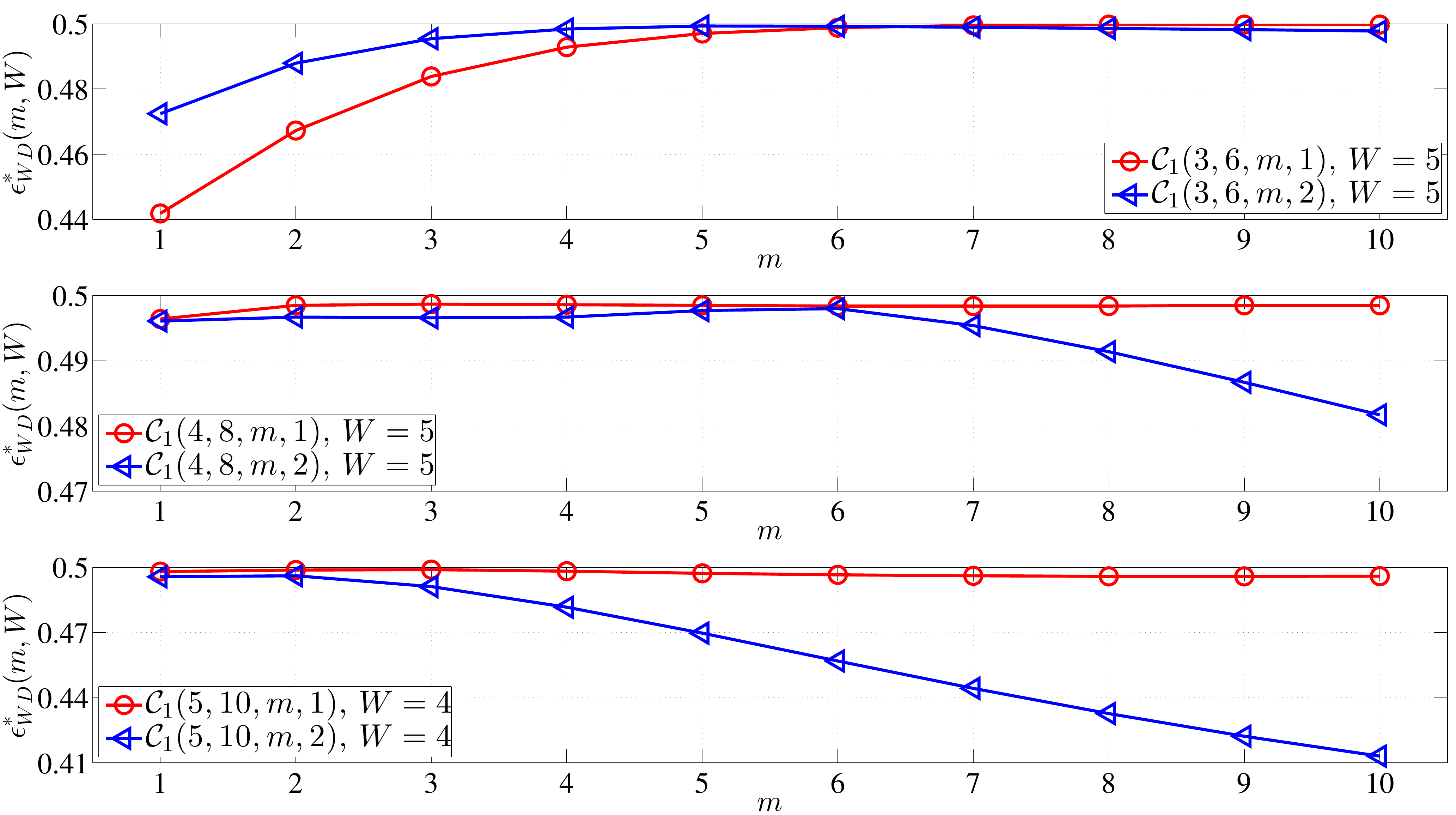}
    \caption{Comparison of WD thresholds $\ewdmw$: type-$\Rmone$ spreading vs. type-$\Rmtwo$ spreading for $J=3$ with $W=5$, for $J=4$ with $W=5$, and for $J=5$ with $W=4$. Note that for the latter two comparisons, $\Cjjmsonetypeone$ has better thresholds than $\Cjjmsonetypetwo$ for all $m$.}
    \label{Fig: TH BEC WD 3 6 ms1 W5 4 8 ms1 W5 5 10 ms1 W4}
\end{figure}

For comparison, the WD thresholds of the $(3,6,m)$ SC ensembles with $w=1$ and $W=3$ are shown in Fig.~\ref{Fig: TH BEC WD 3 6 ms1 W3}, and the WD thresholds of the $(4,8)$ and $(5,10)$ SC ensembles with $w=1$ and $W=3$ are shown in Figs.~\ref{Fig: TH BEC WD 4 8 ms1 W3} and~\ref{Fig: TH BEC WD 5 10 ms1 W3}, respectively. In addition to several features that are similar to the $(3,6)$ SC ensembles, some further observations can be made for the $(4,8)$ and $(5,10)$ SC ensembles:
\begin{itemize}
    \item Recall that for $\bfEB$ spreading, the advantage results from the strong variable nodes with degree $(J-1)$ at the end of the window, and the disadvantage results from the weak check node with degree $2(J-1)$ at the beginning of the window, as shown in Fig.~\ref{Fig: mx WD 3 6 ms1 type3 W4} for $J=3$. Thus, as the density $J$ increases, both the positive and the negative effects are strengthened. On the one hand, the saturation of the WD threshold $\ewdmw$ to its best achievable value $\ewdmaxm$ as $W$ increases is faster. For example, for $m=3$, we find that $\W(3)=4$ for $\Cthreesixmsonetypethree$, $\W(3)=4$ for $\Cfoureightmsonetypethree$, and $\W(3)=3$ for $\Cfivetenmsonetypethree$, i.e., for fixed $m$, $\Wm$ is non-increasing for $\Cjjmsonetypethree$ as $J$ increases. On the other hand, we observe from Fig.~\ref{Fig: TH BEC WD 3 6 4 8 5 10} that:
        \begin{itemize}
            \item The WD thresholds of $\Cjjmsonetypethree$ monotonically decrease as $m$ increases ($m \geq 3$ for $\Cthreesixmsonetypethree$),
            \item Their curves are almost parallel to the corresponding curves for the FSD thresholds of $\Bjj$ -- this effect is more apparent for $J=4$ and $5$, and
            \item The gap between these two curves decreases as $J$ increases.
        \end{itemize}
        Thus, the denser the parity-check matrix is, the more ``block-like'' the WD thresholds of type $3$ spreading
        $\begin{bmatrix} \bfEB & \bfEB \end{bmatrix}$
        become. As previously mentioned, this is because only one edge from each variable node in a block protograph is spread to the adjacent block protograph in type $3$ edge spreading.
    \item The disadvantage of $\bfEB$ spreading also affects the WD thresholds of type $2$ edge spreading. Fig.~\ref{Fig: TH BEC WD 3 6 4 8 5 10 ms1 type2 W5} compares the WD thresholds of the $\Cthreesixmsonetypetwo$, $\Cfoureightmsonetypetwo$, and $\Cfivetenmsonetypetwo$ ensembles with $W=5$. We see that, as $J$ increases, the thresholds of $\Cjjmsonetypetwo$ diverge more significantly from channel capacity as $m$ increases, consistent with the observation that the disadvantage of $\bfEB$ spreading is strengthened as $J$ increases. Moreover, the divergence occurs sooner as $J$ increases, e.g., the WD threshold of $\Cfivetenmsonetypetwo$ increases only up to $m=2$ and then starts to decrease as $m$ increases further, whereas the divergence for both $\Cthreesixmsonetypetwo$ and $\Cfoureightmsonetypetwo$ does not occur until $m=6$.
    \item For type $1$ edge spreading, where both columns of $\bfB_0^1$ equal $\bfEA$, the WD thresholds improve dramatically as $J$ increases for small $W$, as we see in Fig.~\ref{Fig: TH BEC WD 3 6 4 8 5 10} for $W=3$. In other words, to a certain extent, the negative effect of $\bfEA$ spreading due to the presence of the degree-$1$ variable nodes at the end of the window, which results in poor WD thresholds for small $W$, is compensated for by the increased density of the parity-check matrix. This observation is further supported by Fig.~\ref{Fig: TH BEC WD 3 6 ms1 W5 4 8 ms1 W5 5 10 ms1 W4}, which compares the WD thresholds of $\Cjjmsonetypeone$ and $\Cjjmsonetypetwo$ for $J=3$ with $W=5$, for $J=4$ with $W=5$, and for $J=5$ with $W=4$. We observe that for $J=4$ and $J=5$, $\Cjjmsonetypeone$ has better thresholds than $\Cjjmsonetypetwo$ for all finite field sizes, even with relatively small $W$. This indicates that, although $\Cthreesixmsonetypeone$ does not perform as well as $\Cthreesixmsonetypetwo$ with WD, $\Cfoureightmsonetypeone$ and $\Cfivetenmsonetypeone$ are both excellent choices for use with WD.
\end{itemize}

Based on the above observations, since the thresholds of type $2$ spreading
$\begin{bmatrix} \bfEA & \bfEB \end{bmatrix}$
deteriorate as $J$ increases (see Fig.~\ref{Fig: TH BEC WD 3 6 4 8 5 10 ms1 type2 W5}), while the thresholds of type $1$ spreading
$\begin{bmatrix} \bfEA & \bfEA \end{bmatrix}$
improve, we conclude for these two edge-spreading formats that
\begin{enumerate}
    \item When $J=3$, $\Cthreesixmsonetypetwo$ is better for WD than $\Cthreesixmsonetypeone$,
    \item When $J=4$, both $\Cfoureightmsonetypeone$ (for all $m$) and $\Cfoureightmsonetypetwo$ (for $m \leq 6$) give excellent performance with WD, and
    \item When $J=5$, $\Cfivetenmsonetypeone$ is a better choice for WD than $\Cfivetenmsonetypetwo$.
\end{enumerate}
Moreover, for $J=3$, if the code construction is restricted to a very small finite field size -- say $m=1$ ($q=2$) or $m=2$ ($q=4$) -- and the threshold requirement can be slightly relaxed, then $\Cthreesixmsonetypethree$ with
$\begin{bmatrix} \bfEB & \bfEB \end{bmatrix}$
spreading also performs well for WD (see Fig.~\ref{Fig: TH BEC WD 3 6 ms1 type 3}). Again, this is consistent with the design rules proposed in~\cite{Iyengar12-WD} for binary SC-LDPC code ensembles suitable for WD. Finally, $\Cfoureightmsonetypethree$ and $\Cfivetenmsonetypethree$ ensembles are clearly not suitable for WD, as shown in Figs.~\ref{Fig: TH BEC WD 4 8 ms1 W3} and~\ref{Fig: TH BEC WD 5 10 ms1 W3}.

The key point we wish to make throughout the paper is that desirable protograph-based $q$-ary SC-LDPC code ensembles for windowed decoding should achieve good thresholds when both the finite field size $q$ and the window size $W$ are small. To this end, we can summarize the above observations made for the $(3,6)$, $(4,8)$, and $(5,10)$ SC ensembles with $w=1$ into two design rules as follows:
\begin{itemize}
    \item Combining $\bfEA$ spreading and $\bfEB$ spreading, i.e., type $2$ edge spreading, is attractive when $J$, characterizing the density of the parity-check matrix, is small;
    \item As $J$ increases, $\bfEB$ spreading becomes less attractive, and it should be totally avoided in favor of $\bfEA$ spreading when $J \geq 5$.
\end{itemize}

\begin{figure}[!t]
    \centering
    \includegraphics[width=0.475\columnwidth]{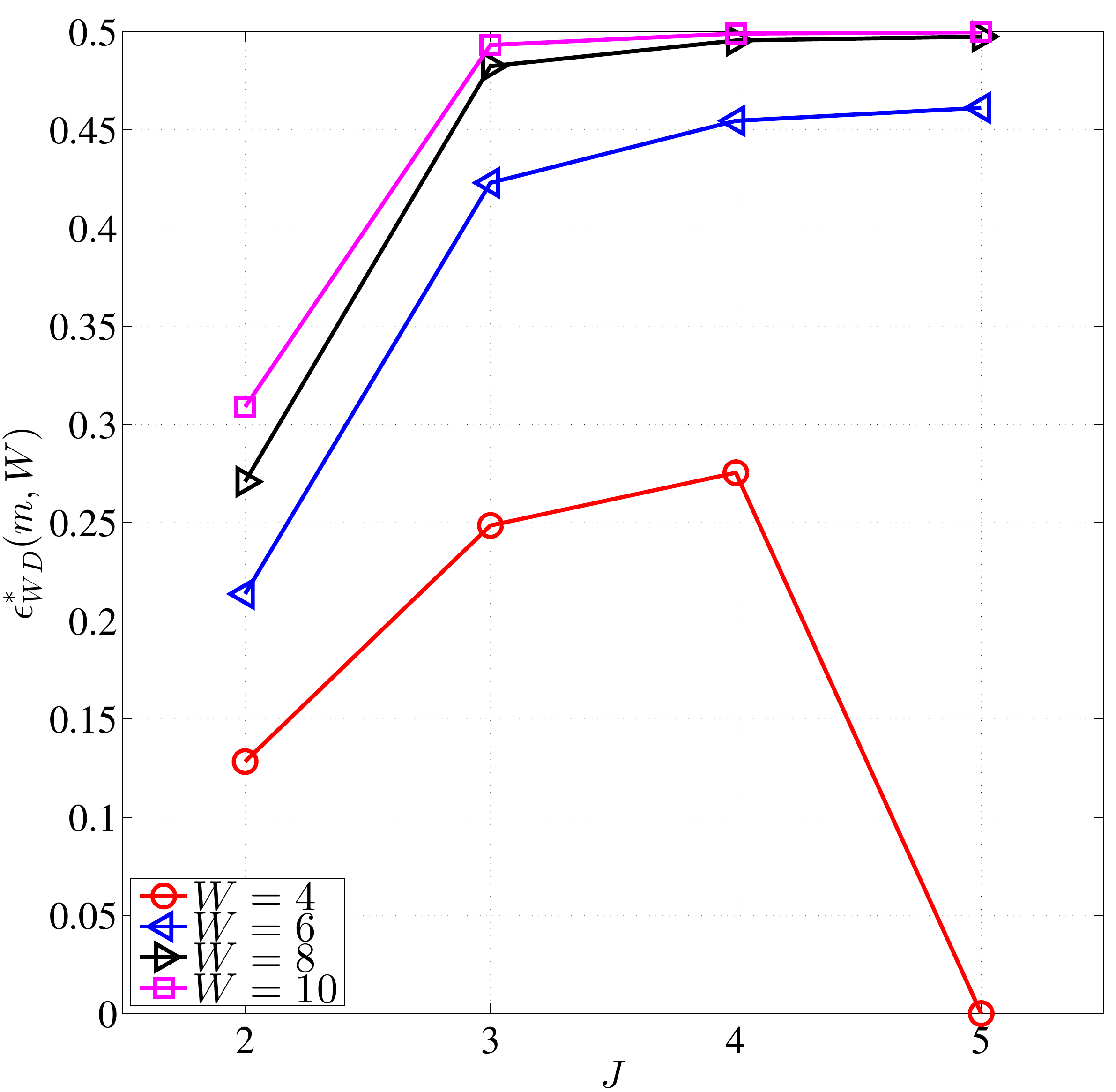}
    \caption{WD thresholds $\ewdmw$ of $\C_{J-1}(J,2J,m=2)$ ensembles as $J$ increases for window sizes $W=4$, $6$, $8$, and $10$. (Note that for $\C_4(5,10,m=2)$ with $W=4$, $\ewdmw=0$, because the minimum required $W$ is $5$.)}
    \label{Fig: TH BEC WD full spreading vs J}
\end{figure}

The classical $(4,8)$-regular and $(5,10)$-regular $q$-ary SC-LDPC code ensembles with $w=J-1$, i.e., $\Cfoureightmsthree$ and $\Cfivetenmsfour$ defined by \eqref{Eq: C J K w(J-1)}, provide WD thresholds analogous to $\Cthreesixmstwo$. To be more specific, for all $m$, the WD thresholds $\ewdmw$ improve with $W$ and saturate to $\ewdmaxm = \efsdm$ (the FSD thresholds), which are non-decreasing and numerically achieve capacity as $m$ increases. Further, when $m$ is fixed and $W$ is sufficiently large, the WD thresholds of the $\Cjjmsjminusone$ ensembles improve as $J$ increases, as shown in Fig.~\ref{Fig: TH BEC WD full spreading vs J} for $m=2$, $W=8$ and $10$. Nevertheless, when $W$ is small to moderate, the thresholds are not satisfactory; for example, when $W=6$, there is still significant space for threshold improvement by increasing $W$ further. In fact, since the minimum $W$ required for WD of $\Cjjmsjminusone$ is $(w+1) = J$, as $J$ increases the classical edge spreading format is even less attractive if there is a constraint on decoding latency, i.e., if a small $W$ must be adopted. For example, $\ewd(m,W=4)=0$ for $\Cfivetenmsfour$, as shown in Fig.~\ref{Fig: TH BEC WD full spreading vs J} for $m=2$, because the minimum required window size in this case is $W = 5$.

\subsection{Numerical results: $k=3$ and $k=4$ ($R>1/2$)}
\label{Sec: TH BEC; Subsec: k=3, k=4}
The previous subsection presented the advantages and disadvantages of using $\bfEA$ spreading and $\bfEB$ spreading in the construction of rate $R=1/2$ protograph-based $q$-ary SC-LDPC code ensembles suitable for WD, and results were presented on the influence of varying the density $J$ (and thus $K$) of the parity-check matrix on the WD thresholds. This subsection presents additional results on WD thresholds for higher rate ($R=2/3$ and $3/4$) protograph-based $q$-ary SC-LDPC code ensembles, with emphasis on how the particular mix of $\bfEA$ spreading and $\bfEB$ spreading affects the WD thresholds. We expect that the more a certain kind of spreading is used, the more its corresponding advantages and disadvantages will be observed. For simplicity, we fix $J=3$.

\subsubsection{$(J,K)=(3,9)$, $k=3$}
We consider the $(3,9,m)$ SC code ensembles over GF$(2^m)$ defined by~\eqref{Eq: ensembles}. The asymptotic rate of $(3,9)$-regular $q$-ary SC-LDPC code ensembles is $R=(k-1)/k = 2/3$, when the coupling length $L$ goes to infinity. Since $k=3$, the component matrices $\bfB_i$ used to construct $\bfBSC$ in~\eqref{Eq: SC base mx} are of size $1 \times 3$ and, in addition to the classical edge spreading with $w=2$, there are four types of $w=1$ spreading where, for types $1$ through $4$, $\bfB_0^1$ is given as
$\begin{bmatrix} \bfEA & \bfEA & \bfEA \end{bmatrix}$,
$\begin{bmatrix} \bfEA & \bfEA & \bfEB \end{bmatrix}$,
$\begin{bmatrix} \bfEA & \bfEB & \bfEB \end{bmatrix}$, and
$\begin{bmatrix} \bfEB & \bfEB & \bfEB \end{bmatrix}$,
respectively.

\begin{figure}[!t]
    \centering
    \subfigure[$W=4$]
    {
        \includegraphics[width=0.475\linewidth]{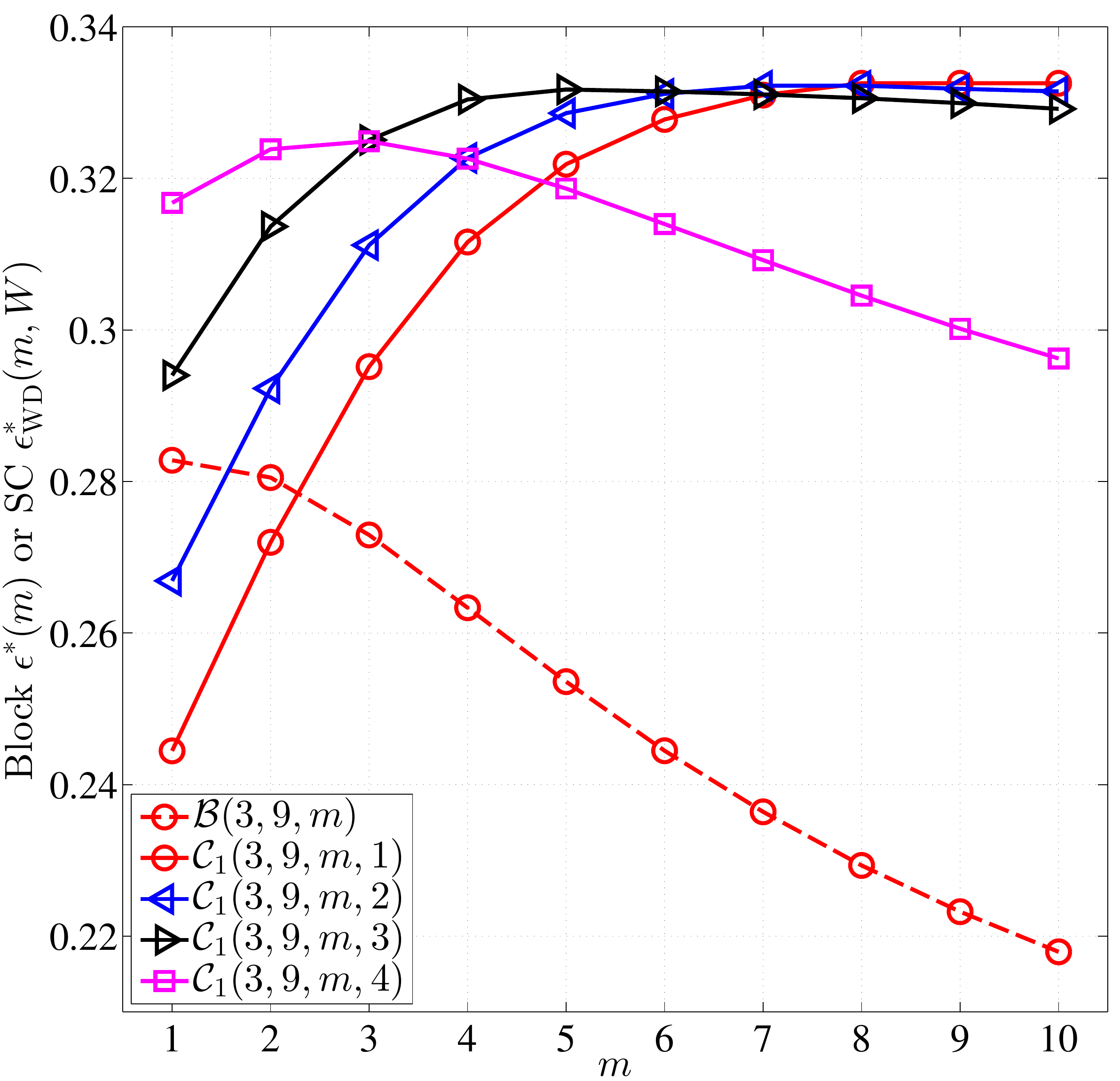}
        \label{Fig: TH BEC WD 3 9 ms1 W4}
    }
    \subfigure[$W=10$]
    {
        \includegraphics[width=0.475\linewidth]{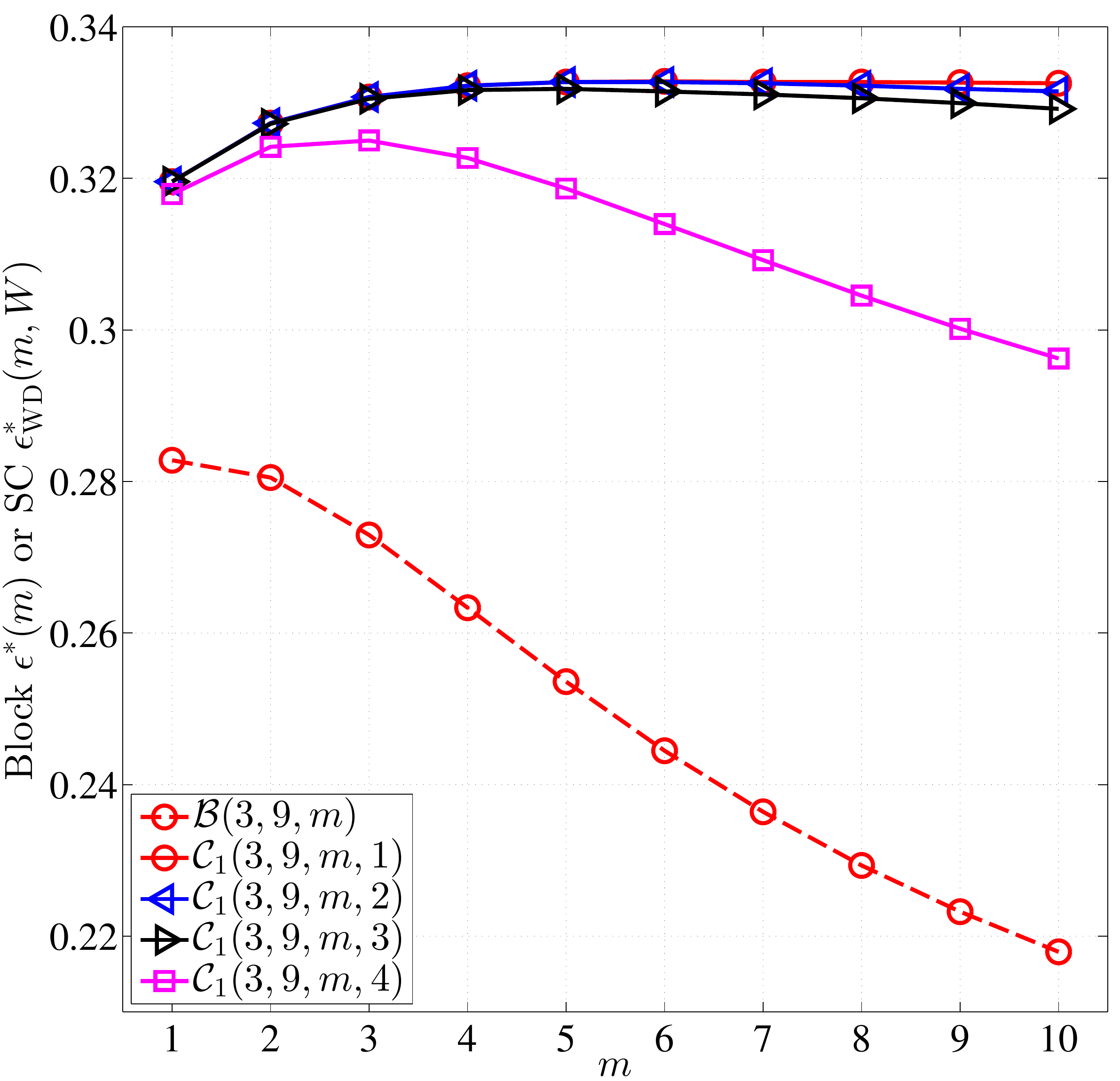}
        \label{Fig: TH BEC WD 3 9 ms1 W10}
    }
    \caption{WD thresholds $\ewdmw$ of the $(3,9,m)$ SC code ensembles with $w=1$: (a) $W=4$, and (b) $W=10$, a sufficiently large window size such that the best WD thresholds are achieved for all the SC-LDPC code ensembles. FSD thresholds $\efsdm$ of $\Bthreenine$ are included as a benchmark.}
    \label{Fig: TH BEC WD 3 9 ms1}
\end{figure}

We expect that if there are more $\bfEB$ spreadings, then, for fixed $m$, the WD threshold $\ewdmw$ will saturate faster to its best achievable value $\ewdmaxm$ as $W$ increases, and that if there are more $\bfEA$ spreadings, then $\ewdmaxm$ will diverge less from channel capacity as $m$ increases. These expectations are met, as illustrated in Figs.~\ref{Fig: TH BEC WD 3 9 ms1 W4} and~\ref{Fig: TH BEC WD 3 9 ms1 W10} when $W=4$ and $10$, respectively. Combined with other numerical results, it is observed that
\begin{itemize}
    \item When $m$ is fixed, the WD threshold $\ewdmw$ of $\Cthreeninemsonetypefour$, which contains all $\bfEB$ spreadings, has the fastest saturation to the corresponding $\ewdmaxm$ of all the $w=1$ code ensembles. This indicates that there is little room for threshold improvement for $\Cthreeninemsonetypefour$ by increasing $W$ to a large value; indeed, comparing the two $\Cthreeninemsonetypefour$ curves in Figs.~\ref{Fig: TH BEC WD 3 9 ms1 W4} and~\ref{Fig: TH BEC WD 3 9 ms1 W10}, we observe that, over the entire range of $m$, $\ewdmw$ hardly changes when the window size goes from small ($W=4$) to large ($W=10$). In fact, even in the case $m=1$ with the slowest saturation ($\Wm=6$) among all field sizes, $\ewd(m,W=4)$ for $\Cthreeninemsonetypefour$ already lies within $0.35\%$ of $\ewdmaxm$.
    \item On the other hand, this fast saturation of $\ewdmw$ to $\ewdmaxm$ for type $4$ edge spreading is accompanied by reduced threshold values. In Fig.~\ref{Fig: TH BEC WD 3 9 ms1 W4}, where $W$ is small, for $m \leq 3$ the ordering of the ensemble types from best to worst is $4$, $3$, $2$, $1$, and WD of $\Cthreeninemsonetypeone$ has even worse performance than FSD of the block code ensemble $\Bthreenine$ for $m \leq 2$. When $W < \Wm$, the WD thresholds are worse than $\ewdmaxm$ because the decoder performance is impaired. In this regime, any additional structural weakness, such as weak variable nodes arising from an $\bfEA$ spreading, further harm the threshold, especially for small $m$, where we observe that fewer $\bfEA$ spreadings result in better thresholds. However, for a larger window size, the decoder is more robust, in the sense that some weaker variable nodes can be included without significantly harming performance. This allows for a stronger check node at the start of the window to initiate the ``wave-like'' decoding that results in threshold saturation for SC-LDPC code ensembles. This effect is more obvious in Fig.~\ref{Fig: TH BEC WD 3 9 ms1 W10}, where the window size $W=10$ is chosen to be sufficiently large such that $\ewd(m,10) = \ewdmaxm$, i.e., $W = 10 \geq \Wm$, for each SC code ensemble. Compared to Fig~\ref{Fig: TH BEC WD 3 9 ms1 W4}, for $m \leq 3$, types $1$, $2$, and $3$ now provide almost identical WD thresholds, which are all better than type $4$. In this regime, we clearly favor an edge-spreading format with a mixture of $\bfEA$ and $\bfEB$.
    \item The introduction of $\bfEB$ spreadings causes a divergence from capacity $\epsilon_\text{Sh} =1/3$ of a rate-$R=2/3$ code ensemble as $m$ increases. This is observed whether the window size is small (Fig.~\ref{Fig: TH BEC WD 3 9 ms1 W4}) or large (Fig.~\ref{Fig: TH BEC WD 3 9 ms1 W10}), and the divergence becomes more significant as more $\bfEB$ spreadings are used. This behavior is similar to what was observed for $\Cthreesixmsonetypethree$ in Fig.~\ref{Fig: TH BEC WD 3 6 ms1 W3}, and again the ``block-like'' behavior as $m$ increases can be explained by insufficient edge spreading to the adjacent block protograph.
    \item Finally, the $\Cthreeninemsonetypeone$ ensembles, with all $\bfEA$ spreading, and the $\Cthreeninemstwo$ classical edge spreading ensembles with $w=2$ display non-decreasing maximum WD thresholds $\ewdmaxm$ that approach channel capacity as $m$ increases. However, the weak variable nodes at the end of the windows for these two ensembles imply that when $m$ is small, $\Wm$ should be large.
\end{itemize}

\subsubsection{$(J,K)=(3,12)$, $k=4$}
\begin{figure}[!t]
    \centering
    \subfigure[All $\bfEB$ spreading]
    {
        \includegraphics[width=0.475\linewidth]{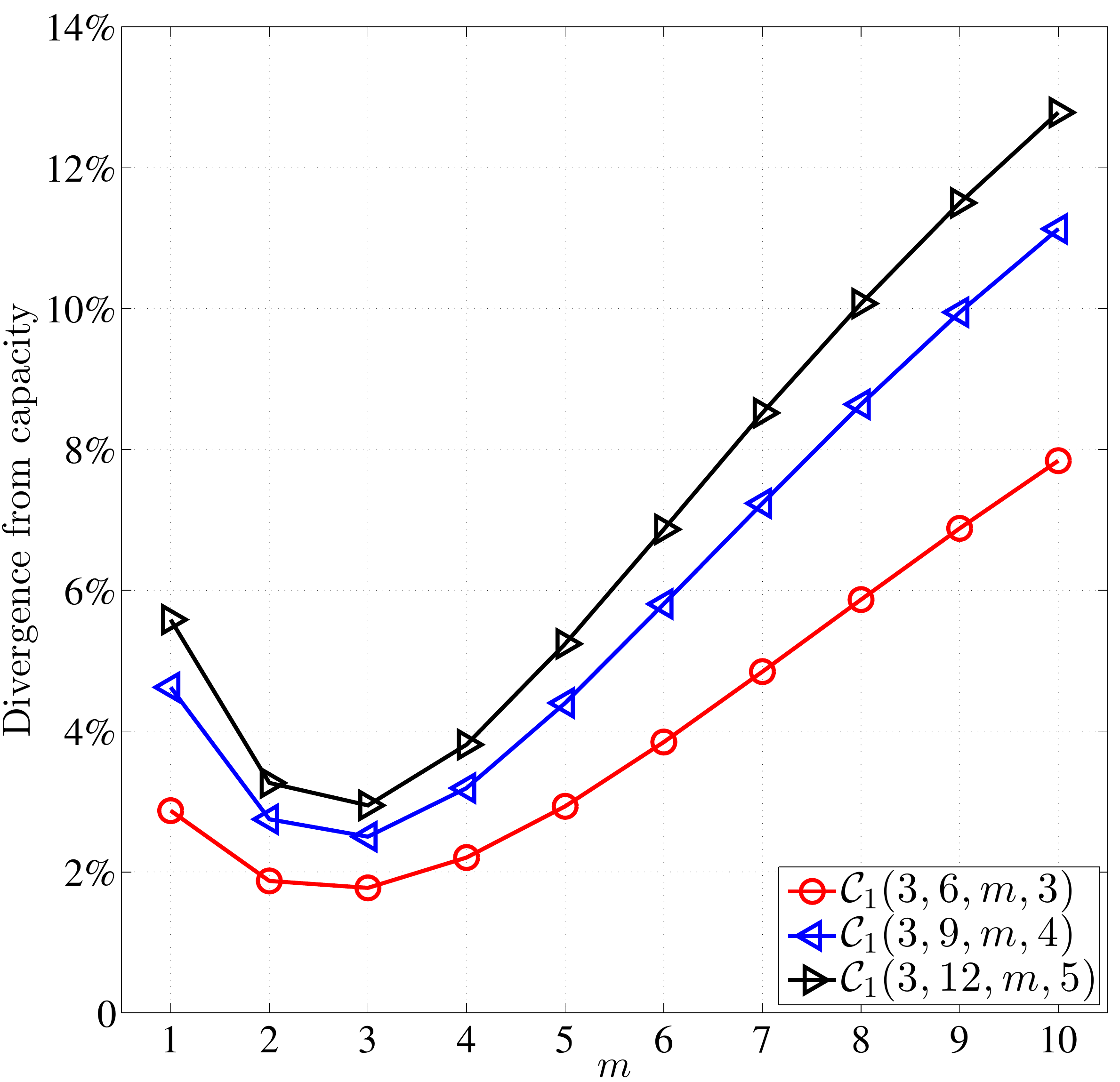}
        \label{Fig: TH BEC diverge capacity allEB}
    }
    \subfigure[One $\bfEA$ spreading]
    {
        \includegraphics[width=0.475\linewidth]{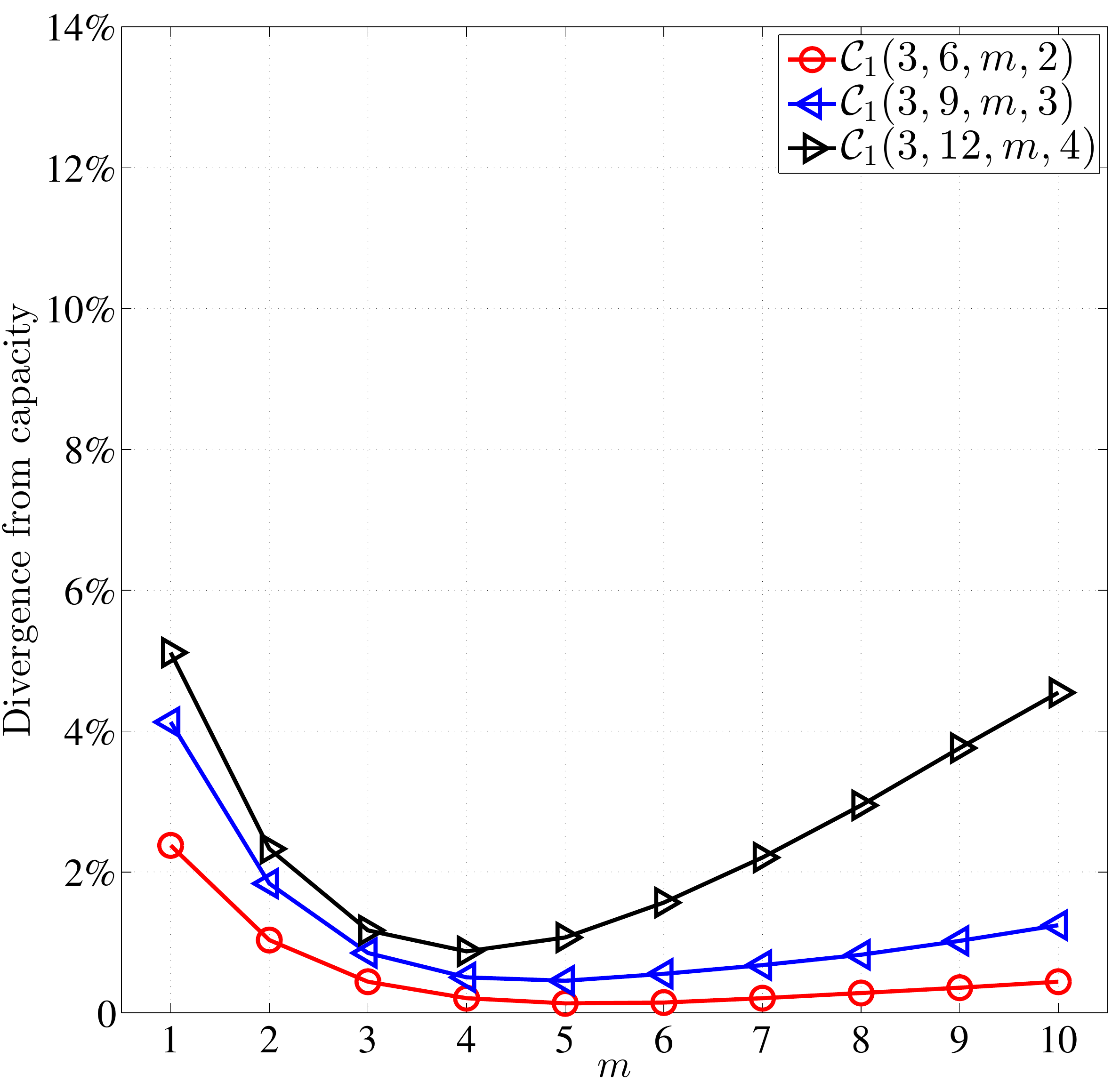}
        \label{Fig: TH BEC diverge capacity oneEA}
    }
    \caption{Percentage divergence of the best achievable WD thresholds $\ewdmaxm$ from the corresponding channel capacities for (a) the $\Cthreesixmsonetypethree$, $\Cthreeninemsonetypefour$, and $\Cthreetwelvemsonetypefive$ ensembles with all $\bfEB$ spreading, and (b) the $\Cthreesixmsonetypetwo$, $\Cthreeninemsonetypethree$, and $\Cthreetwelvemsonetypefour$ ensembles containing only one $\bfEA$ spreading.}
    \label{Fig: TH BEC diverge capacity}
\end{figure}

The asymptotic rate of $(3,12)$-regular $q$-ary SC-LDPC code ensembles is $R = (k-1)/k = 3/4$, when the coupling length $L$ goes to infinity. For $w=1$ and an arbitrary $m$, there are $k+1=5$ types of $(3,12,m)$ SC ensembles over GF$(2^m)$ defined in~\eqref{Eq: ensembles}, and the behavior of their thresholds is similar to the $(3,9,m)$ SC ensembles with $w=1$. Fig.~\ref{Fig: TH BEC diverge capacity allEB} shows the percentage divergence of the $\ewdmaxm$ from the corresponding channel capacities for the $\Cthreesixmsonetypethree$, $\Cthreeninemsonetypefour$, and $\Cthreetwelvemsonetypefive$ ensembles, where all $\bfEB$ spreading is adopted in each case. The results strengthen our observation that the more a particular spreading is used, the greater are its effects: the WD threshold of $\Cthreetwelvemsonetypefive$ shows the most significant divergence from the corresponding BEC capacity because it uses four $\bfEB$ spreadings, compared to three in $\Cthreeninemsonetypefour$ and two in $\Cthreesixmsonetypethree$.

Similar observations can be made in Fig.~\ref{Fig: TH BEC diverge capacity oneEA} as well, which compares the percentage divergence for the $\Cthreesixmsonetypetwo$, $\Cthreeninemsonetypethree$, and $\Cthreetwelvemsonetypefour$ ensembles, where one $\bfEA$ spreading and $(k-1)$ $\bfEB$ spreadings are adopted in each case. Again, the ensemble that uses the most $\bfEB$ spreadings -- in this case $\Cthreetwelvemsonetypefour$ with three $\bfEB$'s -- shows the most significant divergence as $m$ increases. However, compared to the thresholds in Fig.~\ref{Fig: TH BEC diverge capacity allEB} with the same degree distribution $(J=3,K=3k)$, we observe that introducing only one $\bfEA$ spreading can significantly alleviate the divergence effect of the $\bfEB$ spreading(s) and thus improve the WD thresholds, i.e., it is desirable to mix $\bfEA$ and $\bfEB$ spreadings in designing of WD-suitable code ensembles.

Finally, classical edge spreading of the $\Cjjjmsjminusone$ and $\Cjjjjmsjminusone$ code ensembles with $w = J-1$ is not suitable for WD, despite their excellent capacity-achieving thresholds when $m$ and $W$ are both large enough, as noted previously for the $\Cjjmsjminusone$ code ensembles.

We emphasize again the design rule that combining $\bfEB$ spreading and $\bfEA$ spreading is a good strategy for designing $(J,K)$-regular $q$-ary SC-LDPC code ensembles suitable for windowed decoding when $J$ is small for two reasons:
\begin{enumerate}
    \item The coupling width $w=1$ makes the minimum required window size only $W = w+1 = 2$, and
    \item The threshold can be near capacity when $m$ and $W$ are both small.
\end{enumerate}
The above conclusions are supported by WD threshold results for the $\Cthreesixmsonetypetwo$, $\Cfoureightmsonetypetwo$, $\Cthreeninemsonetypetwo$, $\Cthreeninemsonetypethree$, $\Cthreetwelvemsonetypetwo$, and $\Cthreetwelvemsonetypethree$ ensembles. For the cases when $J=3$, these conclusions are further reinforced by decoding performance simulations of finite-length codes with different rates; see~\cite{Huang14-NB-SC-LDPC} for details.

\section{Threshold Analysis of $q$-ary SC-LDPC Code Ensembles on the BIAWGNC}
\label{Sec: TH BIAWGNC}
\subsection{$q$-ary Protograph EXIT Analysis on the BIAWGNC}
\label{Sec: TH BIAWGNC; Subsec: P-NB-EXIT}
We use the $q$-ary protograph EXIT (PEXIT) algorithm presented in~\cite{Dolecek14-P-NB-EXIT} to analyze the FSD thresholds of protograph-based $q$-ary SC-LDPC code ensembles on the BIAWGNC, assuming that the binary image of a codeword is transmitted using BPSK modulation, and we extend it in a similar fashion to the $q$-ary WD-PDE algorithm to obtain WD thresholds for the BIAWGNC. Similar to the $q$-ary PDE analysis on the BEC, the $q$-ary PEXIT analysis is also a BP algorithm performed on a protograph, where the messages now represent \emph{mutual information} (MI) values, a model obtained by approximating the distribution of the log-likelihood ratio messages in BP decoding as (jointly) Gaussian. The thresholds are obtained by determining the smallest signal-to-noise ratio $E_b/N_0$ (in dB) for which decoding is successful, i.e., the smallest value of $E_b/N_0$ such that the \emph{a-posteriori} MI between each variable node and a corresponding codeword symbol goes to $1$ as the number of iterations goes to infinity.

\subsection{Numerical Results}
\label{Sec: TH BIAWGNC; Subsec: Sim}
Our observations and conclusions made regarding the WD thresholds of $q$-ary SC-LDPC code ensembles on the BIAWGNC are similar to those made for the BEC. As a result, only a few examples are given here.

\begin{figure}[!t]
    \centering
    \subfigure[FSD thresholds of the $(2,4,m)$ and $(3,6,m)$ ensembles]
    {
        \includegraphics[width=0.475\columnwidth]{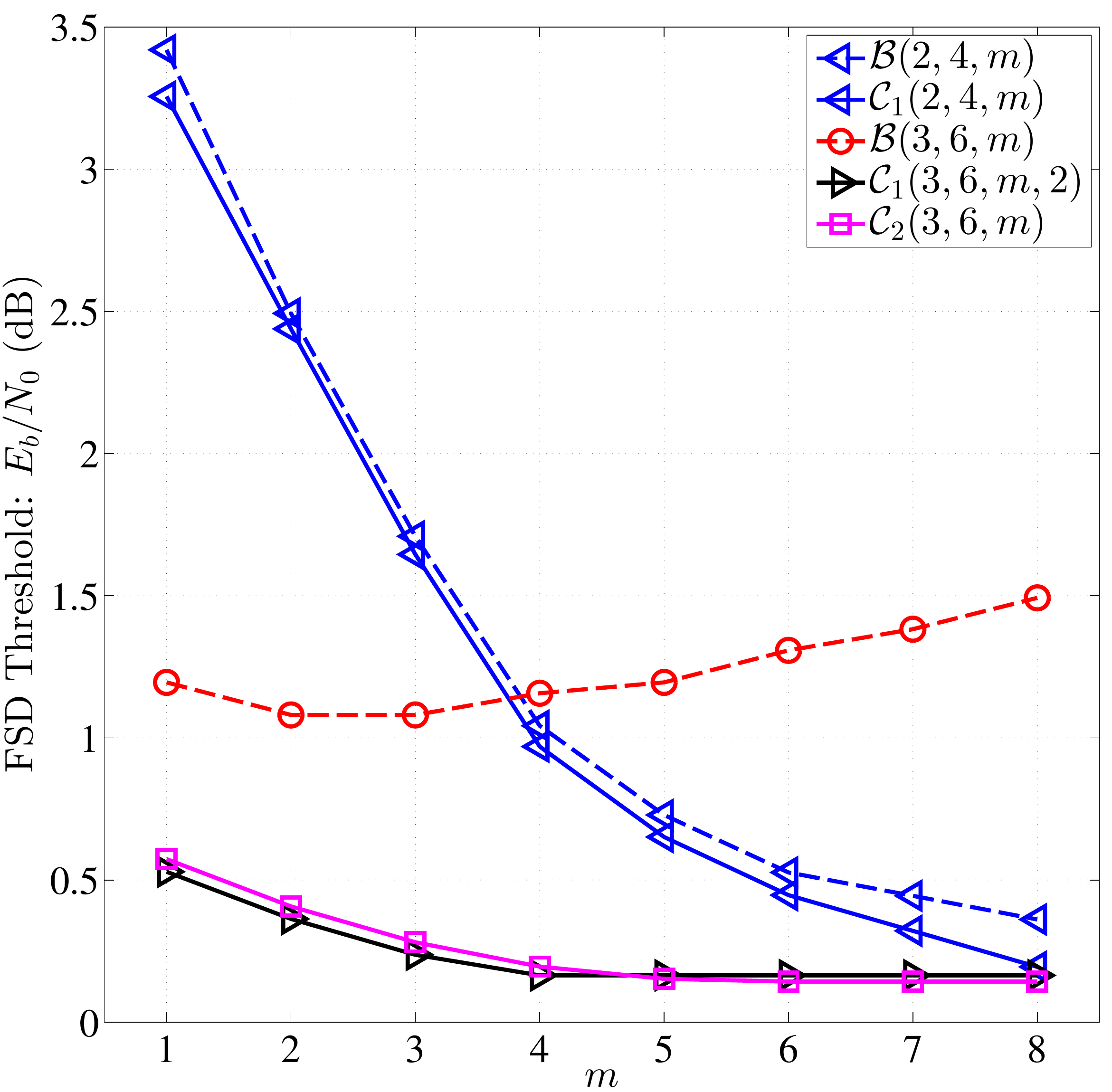}
        \label{Fig: TH BIAWGNC FSD}
    }
    \subfigure[WD thresholds of $\Cthreesixmsonetypetwo$]
    {
        \includegraphics[width=0.475\columnwidth]{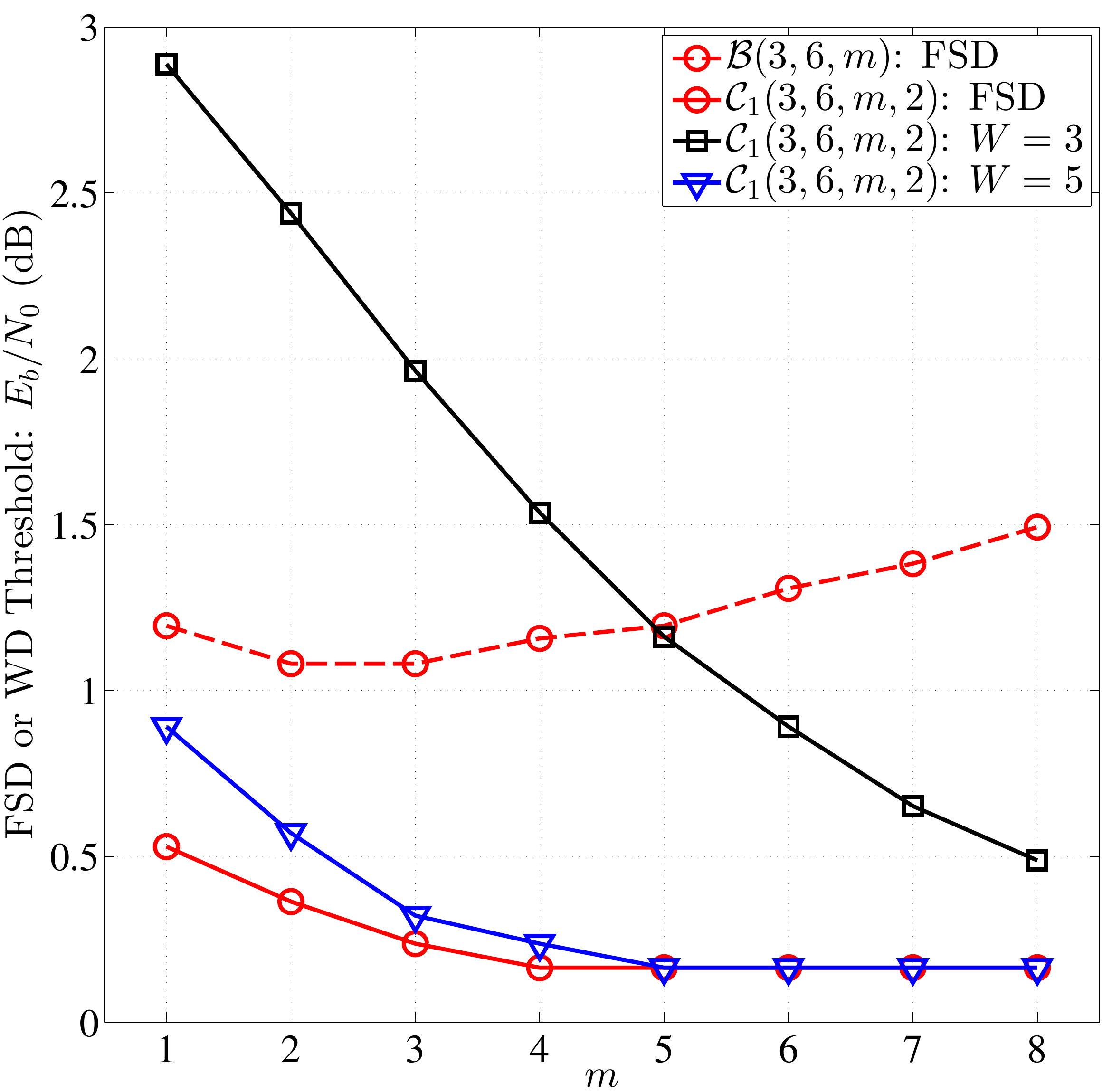}
        \label{Fig: TH BIAWGNC WD 3 6 ms1 type2}
    }
    \caption{FSD thresholds of the $(2,4,m)$ and $(3,6,m)$ ensembles and WD thresholds of $\Cthreesixmsonetypetwo$ on the BIAWGNC.}
    \label{Fig: TH BIAWGNC}
\end{figure}

Fig.~\ref{Fig: TH BIAWGNC FSD} compares the FSD thresholds of the $(2,4,m)$ and $(3,6,m)$ ensembles on the BIAWGNC.\footnote{Due to computational limitations, the BIAWGNC thresholds were calculated only up to $m=8$. However, as stated by Uchikawa~\emph{et al.} in~\cite{Uchikawa11-RC-NB-LDPCC}, it is reasonable to assume that the BIAWGNC thresholds for $m=9$ and $10$ are consistent with the corresponding BEC results.} $\Cthreesixmsonetypethree$ and $\Cthreesixmsonetypeone$ have the same FSD thresholds for all $m$, which are almost identical to those of $\Cthreesixmsonetypetwo$, so only $\Cthreesixmsonetypetwo$ is shown to represent the $w=1$ code ensembles and to compare with $\Cthreesixmstwo$. Fig.~\ref{Fig: TH BIAWGNC WD 3 6 ms1 type2} shows the WD thresholds of $\Cthreesixmsonetypetwo$ when $W=3$ and $5$. Both subfigures illustrate behavior similar to the BEC results presented in Section~\ref{Sec: TH BEC; Subsec: k=2}. To summarize, small gains are observed for $\Ctwofourmsone$ compared to $\Btwofour$ until the finite field size gets large, whereas (numerically) capacity achieving WD thresholds that are significantly better than the corresponding block code thresholds are observed for both $\Cthreesixmsonetypetwo$ and $\Cthreesixmstwo$. Again, $\Cthreesixmsonetypetwo$ turns out to be particularly well suited for WD; for $m=5$ and $W=5$, the WD threshold is essentially at capacity.

\section{Decoding Latency and Decoding Complexity}
\label{Sec: Dec cmplx}
This section considers the tradeoff between two critical decoding properties of $q$-ary spatially coupled LDPC code ensembles:
\begin{enumerate}
    \item \emph{Latency}: measured as the number of bits that must be received before decoding can begin, and
    \item \emph{Complexity}: measured as the number of decoding operations required per information bit.
\end{enumerate}

Our focus is the ensemble average behavior on the BEC when windowed decoding is used; different cases are compared on the same BEC, so that the tradeoff between decoding latency and decoding complexity can be examined. We use the $q$-ary WD-PDE algorithm (for WD) and the $q$-ary PDE algorithm (for FSD) in order to obtain our results, i.e., we consider an infinite lifting factor $M$ used for the ensemble construction, thereby removing the effect of $M$ from the latency-complexity tradeoffs. This allows us to get a general picture of the latency-complexity tradeoffs associated with a particular code ensemble, rather than analyzing specific codes, which can then be used to guide the design of practical, finite-length protograph-based $q$-ary SC-LDPC codes, especially when there is a limit on decoding latency.

In the remainder of this section, we focus on the $(J,2J)$ SC-LDPC code ensembles with $k=2$ and coupling width $w=1$ previously discussed in Section~\ref{Sec: TH BEC; Subsec: k=2}; however, similar results can be obtained for other code ensembles as well.

\subsection{Decoding Latency}
\label{Sec: Dec cmplx; Subsec: Latency}
For a $q$-ary SC-LDPC code constructed as described in Section~\ref{Sec: Ensembles; Subsec: Construction}, the decoding latency (normalized by $M$) of WD is given by $kmW$ measured in \emph{bits}, where we assume that the binary image of a codeword is transmitted, so each GF$(q)$ symbol contains $m$ bits. For $k=2$, the latency is proportional to the product of $m$ and $W$. In the numerical results presented in this section, we use
\begin{eqnarray}
\label{Eq: SC WD dec latency}
    \TSC = 2 m W
\end{eqnarray}
to represent the latency of WD for a $q$-ary SC-LDPC code ensemble. Also, FSD can be viewed as a special case of WD for which $W = L + w$, where $L$ is the coupling length, so the corresponding latency can also be obtained using~\eqref{Eq: SC WD dec latency}.

\subsection{Decoding Complexity}
\label{Sec: Dec cmplx; Subsec: Complexity}
As stated in~\cite{Voicila10-EMS} and the references therein, the decoding complexity of $q$-ary LDPC codes using the sum-product algorithm based on the fast Fourier transform can be summarized as follows:
\begin{itemize}
    \item One check-to-variable (c-to-v) update requires $\O\left( q \log q \right)$ operations, and
    \item One variable-to-check (v-to-c) update requires $\O\left( q \right)$ operations.
\end{itemize}

We define the order of decoding complexity as the number of operations required per \emph{information} bit, which is a fraction $1/\left(R_{L}mkML\right)$ of the total number of operations for all the c-to-v and v-to-c updates during the decoding process, where $R_L$ is the design rate. That is,
\begin{eqnarray}
\label{Eq: SC dec cmplx}
    \begin{aligned}
        \text{Order of Decoding complexity}
        &=
        \O\left(
            \dfrac{J \left( q + q \log_2 q \right) k M \sum\limits_{t=1}^{L} l_t}{R_L m k M L}
        \right)\\
        &=
        \O\left(
            \dfrac{J 2^m (m + 1) \sum\limits_{t=1}^{L} l_t}{R_L m L}
        \right),
    \end{aligned}
\end{eqnarray}
where $l_t$ is the number of iterations involving updates of variables at time instant $t$ ($1 \leq t \leq L$), which can be easily tracked during the decoding process. As previously mentioned, we let $L=100$.

Although~\eqref{Eq: SC dec cmplx} is derived for BP decoding of finite-length SC-LDPC codes, we use it for our ensemble-average complexity analysis as well. For a particular $q$-ary SC-LDPC code ensemble, the erasure rate of the BEC is chosen to be no greater than the WD threshold of the ensemble, so $q$-ary WD-PDE is guaranteed to decode successfully. As the algorithm iterates, the number of c-to-v and v-to-c updates at each time instant is tracked via $l_t$, and then the order of decoding complexity is calculated.

\subsection{Numerical Results}
\label{Sec: Dec cmplx; Subsec: Sim}
\subsubsection{WD vs. FSD, for the same decoding threshold}
\begin{figure}[!t]
    \centering
    \subfigure[Complexity]
    {
        \includegraphics[width=0.475\columnwidth]{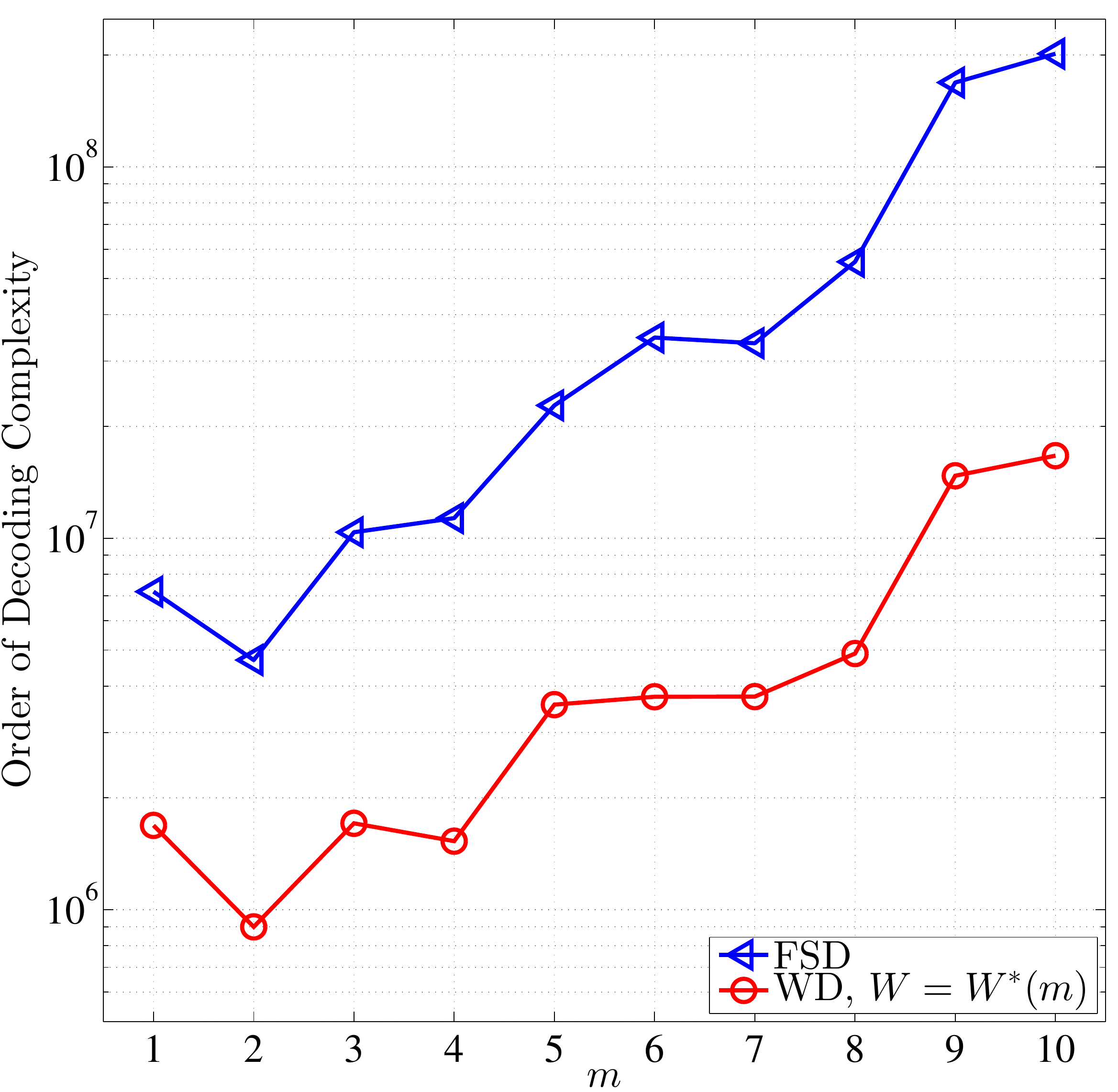}
        \label{Fig: dec cmplx FSD vs WD}
    }
    \subfigure[Latency]
    {
        \includegraphics[width=0.475\columnwidth]{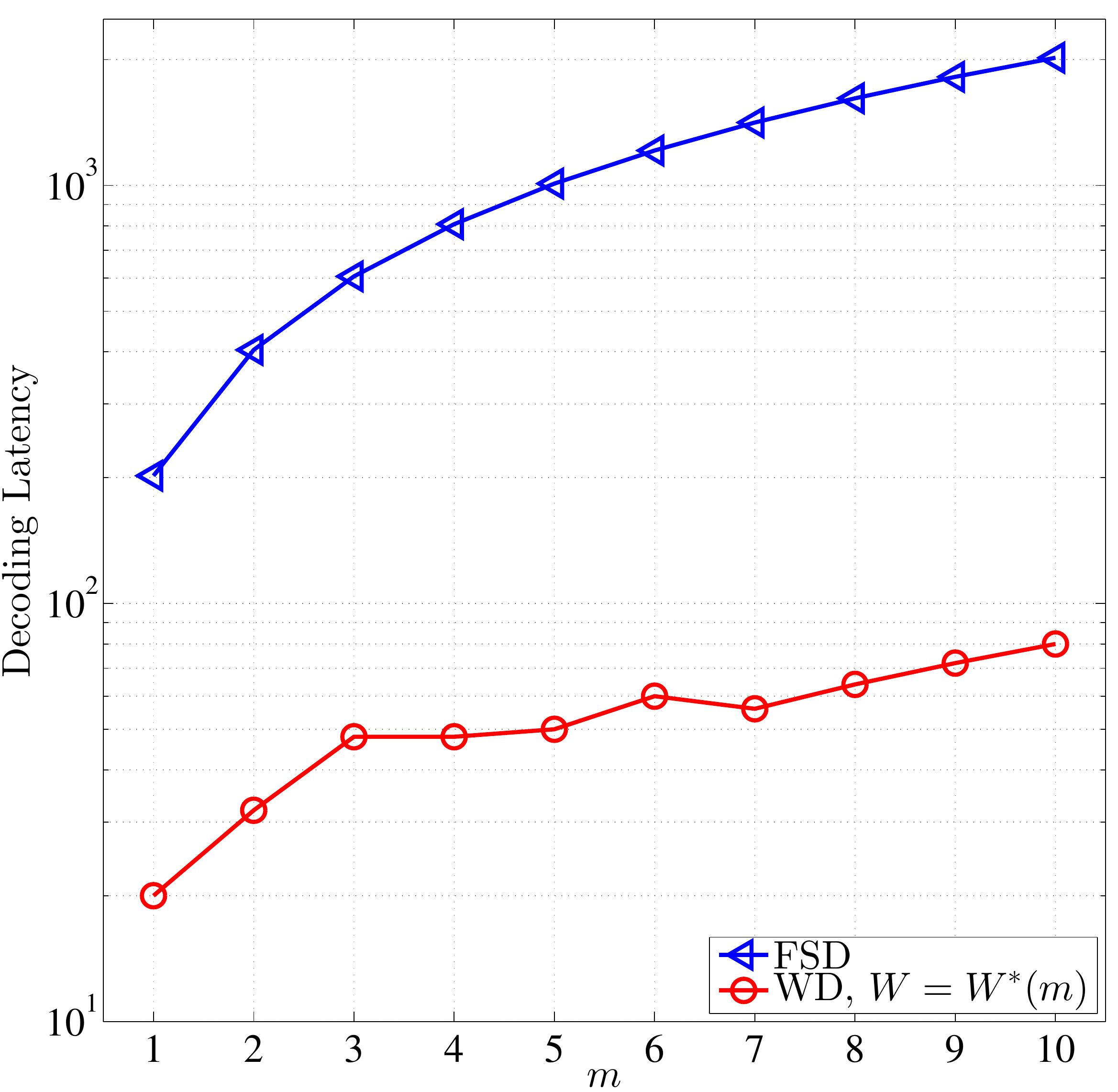}
        \label{Fig: dec latency FSD vs WD}
    }
    \caption{WD vs. FSD for $\Cthreesixmsonetypetwo$: comparison of (a) decoding complexity, and (b) decoding latency.}
    \label{Fig: dec cmplx latency FSD vs WD}
\end{figure}

Fig.~\ref{Fig: dec cmplx latency FSD vs WD} uses $\Cthreesixmsonetypetwo$ as an example to illustrate why WD is preferred to FSD for $q$-ary SC-LDPC code ensembles.

For each $m$, FSD is compared to WD with $W=\Wm$, where we recall that $\Wm$ is the minimum window size that provides the best achievable WD threshold $\ewdmaxm$. Here, $\Wm=10$, $8$, $8$, $6$, $5$, $5$, $4$, $4$, $4$, and $4$ for $m=1$ to $10$. Here also the WD threshold $\ewdmaxm$ equals the FSD threshold $\efsdm$ for $\Cthreesixmsonetypetwo$ for all $m$, i.e., these two cases have the same decoding threshold, and we set the channel erasure rate to this threshold, i.e., $\epsilon=\efsdm$. From Fig.~\ref{Fig: dec cmplx FSD vs WD}, we see that WD saves approximately $75\%$ to $90\%$ in decoding complexity compared to FSD, as $m$ ranges from $1$ to $10$; the larger the finite field size, the more the savings in complexity.

Moreover, as shown in Fig.~\ref{Fig: dec latency FSD vs WD}, WD also has a significant advantage in reducing decoding latency: the decoding latency of WD is only about $10\%$ of FSD when $m=1$ ($q=2$) and only about $4\%$ of FSD when $m=10$ ($q=1024$), i.e., the larger the finite field size, the more decoding latency is saved.

To summarize, WD is preferred to FSD for decoding $q$-ary SC-LDPC codes because the former provides large savings in both decoding complexity and decoding latency, due to the fact that, unlike FSD, WD is localized to include only a small portion of the parity-check matrix. Also, by choosing the window size appropriately, these savings incur no loss in threshold.

\subsubsection{WD complexity as a function of $m$ and $W$, with equal latency}

\begin{table}[!t]
    \caption{Order of Decoding Complexity of $\Cthreesixmsonetypetwo$}
    \label{Table: dec cmplx}
    \centering
    \begin{tabular}{| C{2.5cm} | C{2.5cm} | C{2.5cm} | C{2.5cm} |}
    \hline
    Decoding latency      & \multirow{2}{*}{$(m,W)$} & \multicolumn{2}{c|}{Order of decoding complexity} \\
    \hhline{~~--}
    $2mW$                 &                          & $\epsilon=0.488$ & $\epsilon=0.44$ \\
    \hline\hline
    \multirow{2}{*}{$24$} & $(1,12)$ & $4.55\times 10^{5}$ & $1.14\times 10^{3}$ \\
    \hhline{~---}
                          & $(2,6)$ & $\mathbf{9.25\times 10^{3}}$ & $\mathbf{1.03\times 10^{3}}$ \\
    \hline
    \multirow{4}{*}{$40$} & $(1,20)$ & $7.18\times 10^{5}$ & $1.77\times 10^{3}$ \\
    \hhline{~---}
                          & $(2,10)$ & $\mathbf{1.32\times 10^{4}}$ & $\mathbf{1.37\times 10^{3}}$ \\
    \hhline{~---}
                          & $(4,5)$ & $1.54\times 10^{4}$ & $2.77\times 10^{3}$ \\
    \hhline{~---}
                          & $(5,4)$ & $2.62\times 10^{4}$ & $4.92\times 10^{3}$ \\
    \hline
    \multirow{6}{*}{$48$} & $(1,24)$ & $8.38\times 10^{5}$ & $2.07\times 10^{3}$ \\
    \hhline{~---}
                          & $(2,12)$ & $1.57\times 10^{4}$ & $\mathbf{1.62\times 10^{3}}$ \\
    \hhline{~---}
                          & $(3,8)$ & $\mathbf{1.36\times 10^{4}}$ & $2.05\times 10^{3}$ \\
    \hhline{~---}
                          & $(4,6)$ & $1.76\times 10^{4}$ & $3.07\times 10^{3}$ \\
    \hhline{~---}
                          & $(6,4)$ & $4.43\times 10^{4}$ & $8.96\times 10^{3}$ \\
    \hhline{~---}
                          & $(8,3)$ & $2.06\times 10^{5}$ & $3.12\times 10^{4}$ \\
    \hline
    \multirow{6}{*}{$60$} & $(1,30)$ & $1.00\times 10^{6}$ & $2.49\times 10^{3}$ \\
    \hhline{~---}
                          & $(2,15)$ & $1.92\times 10^{4}$ & $\mathbf{1.98\times 10^{3}}$ \\
    \hhline{~---}
                          & $(3,10)$ & $\mathbf{1.68\times 10^{4}}$ & $2.51\times 10^{3}$ \\
    \hhline{~---}
                          & $(5,6)$ & $3.23\times 10^{4}$ & $5.91\times 10^{3}$ \\
    \hhline{~---}
                          & $(6,5)$ & $5.15\times 10^{4}$ & $9.77\times 10^{3}$ \\
    \hhline{~---}
                          & $(10,3)$ & $5.15\times 10^{5}$ & $1.12\times 10^{5}$ \\
    \hline
     \multirow{6}{*}{$80$} & $(1,40)$ & $1.23\times 10^{6}$ & $3.09\times 10^{3}$ \\
    \hhline{~---}
                          & $(2,20)$ & $\mathbf{2.48\times 10^{4}}$ & $\mathbf{2.55\times 10^{3}}$ \\
    \hhline{~---}
                          & $(4,10)$ & $2.84\times 10^{4}$ & $4.88\times 10^{3}$ \\
    \hhline{~---}
                          & $(5,8)$ & $4.24\times 10^{4}$ & $7.75\times 10^{3}$ \\
    \hhline{~---}
                          & $(8,5)$ & $1.91\times 10^{5}$ & $3.67\times 10^{4}$ \\
    \hhline{~---}
                          & $(10,4)$ & $5.85\times 10^{5}$ & $1.13\times 10^{5}$ \\
    \hline
    \multirow{7}{*}{$120$}& $(1,60)$ & $1.54\times 10^{6}$ & $3.94\times 10^{3}$ \\
    \hhline{~---}
                          & $(2,30)$ & $3.46\times 10^{4}$ & $\mathbf{3.58\times 10^{3}}$ \\
    \hhline{~---}
                          & $(3,20)$ & $\mathbf{3.16\times 10^{4}}$ & $4.71\times 10^{3}$ \\
    \hhline{~---}
                          & $(4,15)$ & $4.13\times 10^{4}$ & $7.10\times 10^{3}$ \\
    \hhline{~---}
                          & $(5,12)$ & $6.21\times 10^{4}$ & $1.13\times 10^{4}$ \\
    \hhline{~---}
                          & $(6,10)$ & $9.95\times 10^{4}$ & $1.88\times 10^{4}$ \\
    \hhline{~---}
                          & $(10,6)$ & $8.63\times 10^{5}$ & $1.66\times 10^{5}$ \\
    \hline
    \end{tabular}
\end{table}

Decoding latency is calculated by $2mW$, so if $mW$ is fixed, there can be multiple $(m,W)$ pairs that satisfy a latency constraint. Again, using the $\Cthreesixmsonetypetwo$ ensemble as an example, Table~\ref{Table: dec cmplx} shows the order of decoding complexity of different $(m,W)$ pairs, when $2mW$ is fixed at $24$, $40$, $48$, $60$, $80$, and $120$; the third column is for a BEC with $\epsilon = 0.488$, while the fourth column is for $\epsilon = 0.44$. For each $\epsilon$, the smallest decoding complexity for a particular $mW$ value is marked in boldface, corresponding to the most attractive $(m,W)$ pair for that particular decoding latency.

The channel erasure rate $\epsilon=0.488$ is within approximately $0.1\%$ of the best-achievable binary WD threshold of $\C_1(3,6,m=1,2)$ and $2.5\%$  from channel capacity.\footnote{For a fixed value of $mW$, not all possible $(m,W)$ pairs can guarantee decoding success for this channel erasure rate. For example, when $mW=12$, $m=3$ and $W=4$ results in a threshold below $\epsilon=0.488$.} As a result, when $m=1$, a large number of iterations is required to achieve decoding success using the $q$-ary WD-PDE algorithm. On the other hand, larger values of $m$ (and as a result, smaller values of $W$), for example, $m=2$, $3$, and $4$, show significant reductions in decoding complexity (one to two orders of magnitude), since the WD thresholds for the corresponding $(m,W)$ pairs are larger; in fact, the smallest decoding complexity is achieved when $m$ is either $2$ or $3$ for all the decoding latencies examined in Table~\ref{Table: dec cmplx}.

$q$-ary SC-LDPC codes may still provide benefits compared to their binary counterparts even at lower channel erasure rates. For example, in the fourth column of Table~\ref{Table: dec cmplx}, $\epsilon = 0.44$ is approximately $10\%$ from the best achievable binary WD threshold of $\Cthreesixmsonetypetwo$ and $12\%$ from channel capacity. Here, we see that $m=2$ has lower decoding complexity than $m=1$ for all decoding latencies and achieves the smallest complexity in all cases, although the gains compared to the $\epsilon = 0.488$ case are not large. Eventually, the advantage of $4$-ary codes compared to binary codes disappears as $\epsilon$ decreases further; nevertheless, Table~\ref{Table: dec cmplx} suggests that, for near-capacity performance requirements with a constraint on decoding latency, one should consider $q$-ary SC-LDPC codes as alternatives to binary codes.

\subsubsection{$\Bjj$ vs. $\Cjjmsonetypethree$, with equal latency}
We now compare the WD complexity of the $\Cjjmsonetypethree$ ensembles
(with $\bfB_0 = \begin{bmatrix} J-1 & J-1 \end{bmatrix}$
and $\bfB_1 = \begin{bmatrix} 1 & 1 \end{bmatrix}$)
with $W = 2 = w+1$ to the FSD complexity of the $\Bjj$ ensembles defined by
\begin{eqnarray}
    \bfB =
    \begin{bmatrix}
        \bfB_0 & \bfB_1 \\
        \bfB_1 & \bfB_0 \\
    \end{bmatrix}.
\end{eqnarray}
Similar to the derivation of~\eqref{Eq: SC WD dec latency}, the FSD (normalized) latency of $\Bjj$ is
\begin{eqnarray}
    \TBC = 4 m,
\end{eqnarray}
the same as $\TSC = 2mW = 4m$ for $\Cjjmsonetypethree$ with $W=2$, i.e., the decoding latencies are equal.

\begin{figure}[!t]
    \centering
    \includegraphics[width=0.475\columnwidth]{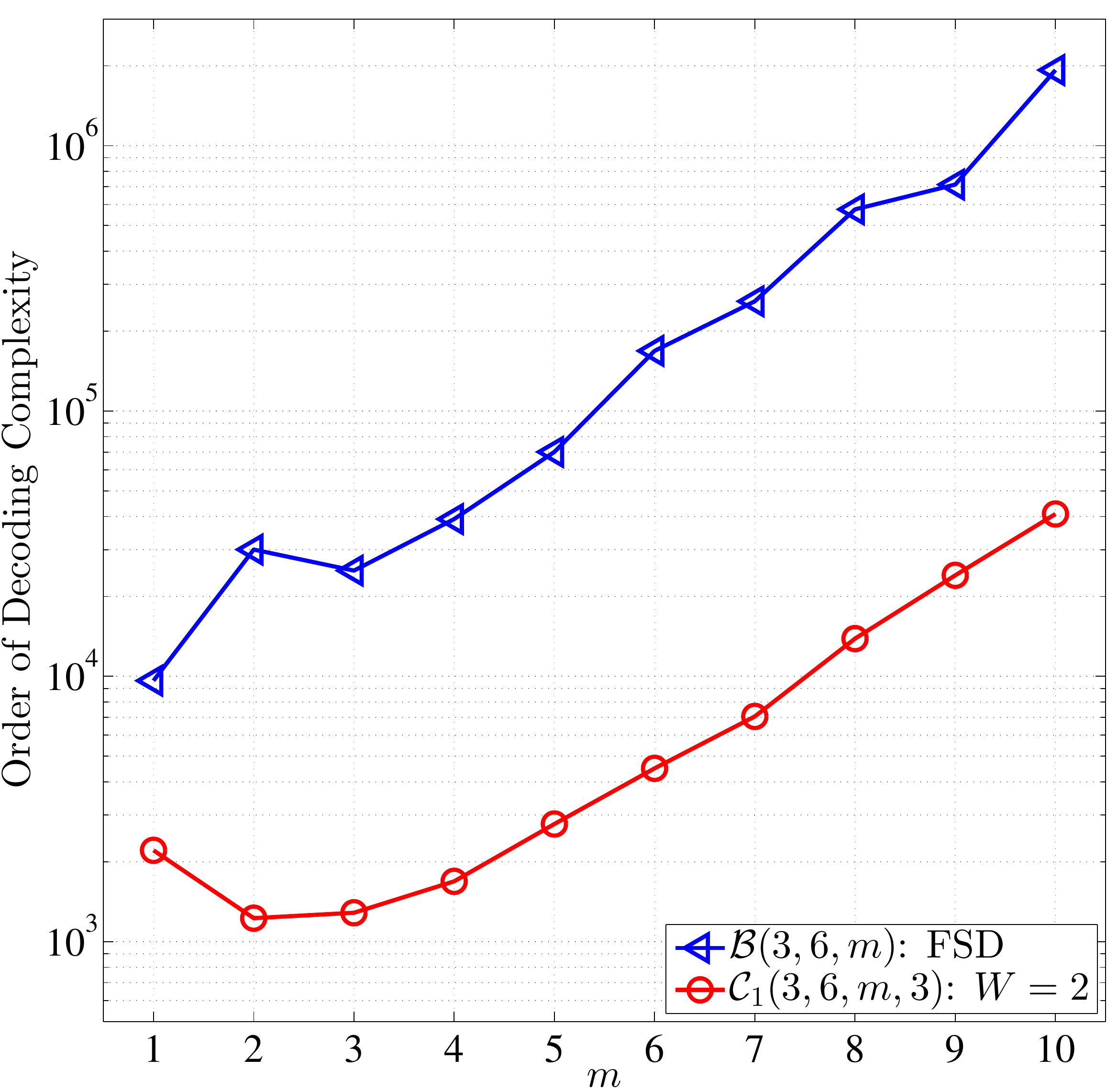}
    \caption{Comparison of decoding complexity: $\Bthreesix$ using FSD and $\Cthreesixmsonetypethree$ using WD with $W=2$.}
    \label{Fig: dec cmplx blk vs SC}
\end{figure}

The orders of decoding complexity for $J=3$ are illustrated in Fig.~\ref{Fig: dec cmplx blk vs SC}. For each $m$, the channel erasure rate is chosen as the FSD threshold $\efsdm$ of $\Bthreesix$, which is smaller than the WD threshold $\ewdmw$ of $\Cthreesixmsonetypethree$ with $W=2$. Similar results can be obtained for $J=4$ and $J=5$ as well. For $w=1$, using $W=2$ results in the smallest possible decoding latency, so Fig.~\ref{Fig: dec cmplx blk vs SC} suggests that, even under a very tight latency constraint, $q$-ary SC-LDPC code ensembles with type $3$ spreading still provide a significant reduction in decoding complexity compared to their block code counterparts. For a comparison of finite-length $q$-ary SC-LDPC codes and $q$-ary LDPC-BCs, where the lifting factor $M$ can be varied to achieve various tradeoffs between error probability, complexity, and latency, we refer the reader to~\cite{Huang14-NB-SC-LDPC}.

\section{Conclusions}
\label{Sec: Conclusions}
This paper proposes design rules for $q$-ary spatially coupled LDPC codes suitable for latency-constrained applications. The design rules are based on an analysis of the windowed decoding thresholds of various protograph-based $(J,K)$-regular $q$-ary SC-LDPC code ensembles for both the binary erasure channel and the BPSK-modulated additive white Gaussian noise channel. In particular, we show that mixing $\bfEA$ and $\bfEB$ edge spreadings to construct $q$-ary SC-LDPC code ensembles results in near-capacity WD thresholds when both the finite field size $q$ and the window size $W$ are relatively small, and that the balance between these two types of spreading depends on the degree distribution and the threshold requirements.

By tracking the number of density evolution update operations needed for decoding success of a $q$-ary SC-LDPC code ensemble for fixed channel conditions, we also demonstrate that WD is superior to FSD in both decoding complexity and decoding latency. Finally, when operation close to the binary SC-LDPC code ensemble threshold is required, we show that codes from $q$-ary SC-LDPC code ensembles provide significant reductions in decoding complexity compared to binary codes for the same decoding latency.



\end{document}